\documentclass[12pt,a4paper]{article}
\usepackage{amsmath,amsfonts,amsthm,graphicx,lscape}
\usepackage[dvips]{psfrag}
\usepackage{cite}
\usepackage[normalem]{ulem}

\usepackage{textcomp}
\usepackage{amsmath,amsthm,amssymb,mathrsfs,cases}
\usepackage{amsfonts,graphicx,lscape}

\usepackage{accents}
\makeatletter
\def\widebar{\accentset{{\cc@style\underline{\mskip10mu}}}}
\makeatother

\numberwithin{equation}{section}

\theoremstyle{definition}
\newtheorem{theorem}{Theorem}[section]

\newtheorem{proposition}[theorem]{Proposition}

\newcommand{\vt}[1]{\mbox{\boldmath$#1$}}

\newcommand{\dprod}[2]{\prod_{\stackrel{\scriptstyle #1}{#2}}}

\newcommand{\sca}[2]{\langle #1, #2 \rangle}

\begin{document}
\title{
%\vspace{-7mm}
\vspace{-9mm}
%A systematic method for generating (constructing?) new integrable systems 
%I.\ Method based on inverse Miura maps (: continuous and discrete)
%A 
%Refinements of 
%On a
Exact 
%Dark/bright 
%Soliton 
solutions of 
multicomponent 
nonlinear Schr\"odinger 
%NLS 
%systems 
equations 
under general 
plane-wave 
%nonvanishing 
%nondecaying 
boundary conditions 
}
\author{Takayuki \textsc{Tsuchida}
%\footnote{
%%E-mail address: 
%%E-mail:\ \{surname of the author\}@ms.u-tokyo.ac.jp
%E-mail:\ surname at 
%%atmark 
%poisson.ms.u-tokyo.ac.jp
%}
%\\
}
%\date{}
\maketitle
\begin{abstract} 
We 
%develop a method to 
construct 
%consider 
%describe 
%study 
%investigate 
%a variety of 
%nontrivial 
exact 
%explicit 
%expressions for various 
soliton 
solutions 
%for 
%to 
%integrable systems 
of 
%coupled 
integrable 
multicomponent 
%cubic 
nonlinear Schr\"odinger 
(NLS) 
equations 
under general nonvanishing boundary conditions. 
%on a general plane-wave background 
%using B\"acklund--Darboux transformations.  
%which can be considered as a perturbation of a general plane-wave solution. 
%All the three cases, 
%i.e.,  
%The NLS equations with 
%under  
%, 
%We consider 
%under 
%In contrast to most of the recent publications on the same subject, 
%We consider 
%%assume 
%The 
%most 
%The solutions are 
%general nonvanishing boundary conditions 
%at spatial infinity 
%, 
%are considered, 
%wherein 
%that is, 
%such that 
%different 
%%%the 
%The 
%In the 
%this 
%general setting, 
%At spatial infinity, 
Different 
components 
%rows 
%columns 
of 
the vector 
(or matrix)
%vector/matrix 
%multicomponent 
dependent variable 
%with different row indices 
%NLS field 
%generally 
can 
%are required to 
%may 
%approaches   
%is allowed 
%are allowed to 
%can 
approach 
%involve 
%different 
%a 
%%linear 
%combination of 
%more than one 
%an 
%a suitable 
%array of 
%totally 
%different 
plane waves 
%plane-wave solutions 
with 
%%involving 
different 
%%amplitudes, 
wavenumbers
and 
%different 
frequencies 
%as \mbox{$x \to \pm \infty$}. 
at spatial infinity. 
%and similar for the matrix dependent variable. 
%The method is applicable 
We apply B\"acklund--Darboux transformations
to 
the 
%multicomponent 
%vector 
cubic 
NLS 
equations 
%systems 
%obtained as the complex conjugation reduction. 
%of  
%and the matrix NLS equation
with a  
self-focusing nonlinearity, 
%a 
a self-defocusing nonlinearity 
or 
%and 
%and 
a mixed focusing-defocusing nonlinearity. 
%are considered. 
%The NLS equations 
%with a 
%self-
%focusing nonlinearity, 
%self-
%a defocusing nonlinearity or
%and 
%a mixed focusing-defocusing nonlinearity 
%are considered under  
%turn out to 
%They possess 
%
%Using elementary and binary 
%three different versions of 
%the   
%B\"acklund--Darboux transformations, 
%We 
%%and 
%obtain 
Both 
%the 
%bright-type 
bright-soliton solutions 
%solitons 
and 
%the 
dark-soliton
%the 
%dark-type soliton 
solutions are obtained, 
depending on the signs of the nonlinear terms 
%nonlinearity 
and the 
%values 
%choice of the soliton parameters. 
type of B\"acklund--Darboux transformation. 
%In particular, 
%In most cases, 
%those 
The 
multicomponent solitons 
%in the self-focusing 
%%%nonlinearity, 
%%%a 
%%%a self-defocusing nonlinearity 
%or 
%%and 
%%and 
%mixed focusing-defocusing case 
%nonlinearity.
generally 
%have 
%admit 
possess 
%allow 
%can 
%have 
%essential 
internal degrees of freedom 
%(a polarization)
%, 
%as 
%%in 
%the bright soliton solitons under 
%the zero 
%%decaying 
%boundary conditions, 
%so 
and 
%they 
%indeed provide 
%are indeed nontrivial multicomponent 
%are 
%indeed 
provide 
%essentially 
highly 
nontrivial generalizations of the 
scalar NLS solitons. 
%soliton solutions of the scalar NLS equation. 
%solutions. 
The main step in 
the construction of the multicomponent solitons 
%solutions 
is 
%equivalent to 
%reduced to 
%the computation of 
to compute the matrix exponential 
of 
a constant 
non-diagonal matrix arising from the Lax pair. 
%Lax-pair 
%representation. 
With a suitable re-parametrization of the non-diagonal matrix, 
%soliton parameters,
the 
%this 
matrix exponential can be computed explicitly 
in closed form 
for the most interesting cases such as 
the two-component vector NLS equation. 
%, as well as 
%and 
%and the \mbox{$2 \times 2$} matrix NLS equation
%, 
%without 
%%using the complicated 
%recourse to 
In particular, we do not resort to 
Cardano's 
formula 
%for 
in diagonalizing a \mbox{$3 \times 3$} matrix,  
%Thus, 
so our expressions for the 
multicomponent solitons 
%of the soliton solutions 
%results 
are in 
%a
some 
sense 
%respects 
more explicit and 
%practically 
useful than  
%compared with the pioneering work 
those obtained in 
%by Park and Shin in 
[Q-H.~Park and H.~J.~Shin,
%:\ 
Phys.\ Rev.\ E {\bf 61} (2000) 3093]. 
\end{abstract}
%%%%%%%%%%%%%%%%%%%%%%%%%%%%%%%%%%%%%%
%{\it MSC:} 82B23; 45G15; 82B20; 82B80 \\
%{\it PACS:} 75.10.Jm, 02.30.Ik, 05.70.-a, 05.30.-d \\
%{\it Key words:}
%
\vspace{15mm}
%{10mm}
\begin{minipage}{13cm}
{\it Keywords: }
%$\hspace{3mm}$
%system, 
%nonlinear Schr\"odinger (NLS), 
%cubic 
%multicomponent 
vector/matrix NLS, 
Lax pair, 
%discrete 
%multi
%$N$-
bright solitons, 
%solutions, 
dark solitons, 
%nonlinear Schr\"odinger (NLS) 
%vector/matrix 
multicomponent 
solitons, 
%with internal degrees of freedom, 
%vector/matrix NLS, 
%equation, 
B\"acklund--Darboux transformations
\end{minipage}

%lattices
%Ablowitz--Ladik lattice, 
%Calogero--Moser--Sutherland models
%systems 
%NLS, 
%nonlinear integral equation;  quantum transfer matrix
%\vspace{5mm}
%\\
%%{\it PACS: }75.10.Jm, 02.30.Ik, 05.70.-a, 05.30.-d \\
%{\it PACS numbers: }02.30.Ik, 05.45.Yv, 45.50.Jf 
%
%
%02.30.Jr Partial differential equations
%02.60.-x Numerical approximation and analysis
%02.60.Cb Numerical simulation; solution of equations
%02.60.Jh Numerical differentiation and integration
%02.60.Lj Ordinary and partial differential equations; boundary value problems
%02.70.-c Computational techniques in mathematical methods in physics
%02.70.Bf Finite-difference methods
%02.70.Dh Finite-element and Galerkin methods
%05.45.-a Nonlinear dynamics, 
%05.50.+q Lattice theory and statistics, 
%42.65.Tg Optical solitons; nonlinear guided waves
%42.81.Dp Propagation, scattering, and losses; solitons
%75.10.Hk Classical spin models
%
%{\it Report-no:} {\bf OIQP-08-??}  \\
%{\bf to appear in Physica A}
%%%%%%%%%%%%%%%%%%%%%%%%%%%%%%%%%%%
\newpage
\noindent
\tableofcontents

\newpage
\section{Introduction}
%The integrability of 
The cubic 
%scalar 
nonlinear Schr\"odinger (NLS) equation 
in 
%%\mbox{$(1+1)$}-dimensional 
\mbox{$1+1$} 
%space-time 
dimensions
%~\cite{ZS1,ZS2} 
%with a focusing or defocusing cubic nonlinearity~\cite{ZS1,ZS2} 
%is 
%the second example of 
%%an integrable 
%a partial differential equation 
%(PDE) 
%that can be 
%solved 
%shown to be 
%was 
%shown to be 
%integrable 
%first 
%solved 
%integrated 
%is 
was first solved 
%solvable 
by the inverse scattering method\footnote{The inverse scattering 
method was originally 
%first 
%applied to 
devised to solve 
the KdV equation\cite{GGKM}.} 
%in~\cite{GGKM}.}
%and known to be 
under 
%the 
%decaying 
vanishing boundary conditions at spatial infinity~\cite{ZS1,AKNS73,AKNS74}. 
%The inverse scattering method 
%It 
The method 
was 
soon 
generalized 
%extended 
to 
%solve 
integrate the NLS equation 
%the case of 
%solve the NLS equation 
%under 
%deal with
under 
%the more general case of 
nonvanishing (or plane-wave\footnote{
%Note
%%, however, 
%that 
The Galilean invariance of 
the NLS equation~\cite{SY74} 
%[Yajima--Satsuma] 
%is invariant under Galilean transformations [Yajima--Satsuma], so 
%in the scalar NLS case 
%and 
%we can 
allows us to 
set 
the wavenumber of the background plane wave 
%boundary conditions 
%can be set equal to 
as 
%equal to 
zero without 
%any 
loss of generality. 
%Note
%, however, 
%that 
However, 
%there can be a 
%shift 
%in the 
%complex phase 
%shift 
%of the physical field 
the 
%constant 
complex 
phase 
of the dependent variable 
%shift 
%between 
can 
%be different 
take different values 
for 
%as 
\mbox{$x \to - \infty$} and \mbox{$x \to + \infty$}~\cite{ZS2,Kawa78}. 
%can be different.
})  
%plane-wave 
boundary conditions at 
%spatial 
infinity 
%at spatial infinity
and 
various interesting 
%a richer class of 
solutions 
%of the NLS equation 
were 
%obtained
%derived
%found
obtained~\cite{ZS2,Kuz77,Kawa78,Ma79} 
%,Peregrine,Akh85}
%[Y-C.Ma] 
(also see~\cite{Peregrine,Akh85,Ste86}). 
%and references citing these papers).
%Peregrine, Akhmediev]. 
%(Akhmediev N N, Eleonskii V M and Kulagin N E 1985 Sov. Phys.-JETP 89 894--899).
%various interesting solutions of the NLS equation 
%were obtained~\cite{ZS2}[Kawata,Ma]. 
In particular, 
%we mention 
%the most relevant 
dark-soliton solutions for 
%of 
the 
%defocusing 
NLS equation with a self-defocusing nonlinearity~\cite{ZS2,Tsuzuki,Hase73-2,Hiro76} 
%possesses 
%admits a dip in a plane wave, 
%has 
%so-called 
%has dark-soliton solutions, 
%that is a dip, 
%hile 
and bright-soliton solutions on a plane-wave background 
%on a finite background 
%of 
for the 
%focusing 
NLS equation with a self-focusing nonlinearity~\cite{Kuz77,Kawa78,Ma79} 
%[Y-C.Ma]
%on a plane-wave background 
are 
the 
most 
%fundamental and  
relevant and fundamental.  
%to the subject of this paper. 
%admits 
%allows a hump, called 
%admits 
%has 
%bright-soliton solutions  
%solutions 
%moving 
%on a 
%finite 
%background 
%plane-wave background. 
%Note that the Galilean invariance of 
%the NLS equation~\cite{SY74} 
%[Yajima--Satsuma] 
%is invariant under Galilean transformations [Yajima--Satsuma], so 
%in the scalar NLS case 
%and 
%we can 
%allows us to 
%set 
%the wavenumber of the background plane wave 
%boundary conditions 
%can be set equal to 
%as 
%equal to 
%zero without 
%any 
%loss of generality. 
%, which could 
%can 
%be simply called 
%nonzero (or nonvanishing) boundary conditions. 
%reduced to 
%the nonzero boundary conditions 

The integrability of the NLS equation is based on 
%the 
its Lax-pair\footnote{The term 
%It 
%The term 
``Lax pair"
%``Lax representation"
%was 
%coined 
is named after 
%he 
%Lax's 
%formulation of 
the work of Peter Lax 
on 
%for 
the KdV hierarchy~\cite{Lax}.}
%\footnote{The Lax pair was first found for the KdV equation by Lax [Lax]. } 
(or zero-curvature) representation~\cite{ZS1,ZS2,AKNS73,AKNS74}, 
which 
%also 
provides a starting point for applying the inverse scattering method. 
By generalizing the 
%\mbox{$2 \times 2$} matrix 
Lax pair, 
%for the scalar NLS equation, 
we 
can straightforwardly
obtain  
%The NLS equation admits straightforward 
%its 
%the 
a vector/matrix 
%and matrix 
%generalization 
analog of the 
%scalar 
NLS equation~\cite{Mana74,ZS74} (also see~\cite{KuSk81}),
%preserving the integrability 
%[Manakov, ZS74], 
%and 
%Moreover, 
%Thus, 
%we 
%can solve 
%It is not a difficult to task 
%to generalize 
%the vector/matrix NLS equation 
%by 
which 
can be solved by 
the inverse scattering method 
%can be applied 
under the 
%decaying 
vanishing boundary conditions. 
%Thus, 
Thus, 
it is natural to 
%consider 
expect 
that the inverse scattering method 
%can be used to solve 
is also applicable to 
%solve 
%one is tempted to apply the inverse scattering method to 
the vector/matrix NLS equation 
%can 
%also 
%should be solved 
%be 
%is also 
%solvable 
%can also be solved 
%integrated 
%by the inverse scattering method 
under general nonvanishing 
%plane-wave 
boundary conditions. 
%as in the scalar case. 
%;  
%because they  
However, 
%it turns out to be not the case. 
in fact this is not the case. 
%it is an extremely difficult task and 
%it is extremely difficult to 
%appears to be impossible. 
%remains as an open problem. 
%in general. 
The main difficulty 
%in the case of the general nonvanishing 
%plane-wave 
%boundary conditions 
is that 
%we need 
%have 
%to consider 
%the components of 
%the vector/matrix dependent variable 
%generally 
%approach 
%different background 
%plane waves, i.e., 
the background plane waves 
%with 
can 
have different 
wavenumbers and frequencies 
in 
%each 
%for 
different components of 
%which correspond to 
%for 
%corresponding to 
%different 
%the 
%each 
%component of 
%because 
the 
vector/matrix 
dependent variable~\cite{Mak81,Mak82,Dub88,Mak89}. 
%[Mak89]. 
%is a vector/matrix. 
%The spectral properties 
%of such a Lax  
%Only a few special cases 
%The 
%In 
Such a 
%this 
general case is not 
%tractable by 
amenable to 
%any known version of 
the usual formulation of 
the inverse scattering method~\cite{Dub88}, 
%cannot be applied, 
%to such a general 
%the generic 
%case; 
%and 
%indeed, 
so 
only 
%a few 
%special cases 
a degenerate case in which 
%wherein 
%there is only one plane-wave background 
all the components 
%become 
%are 
%asymptotically 
%proportional to 
approach 
the same 
%a single 
plane wave at spatial infinity, 
%as \mbox{$x \to \pm \infty$} 
up to (possibly zero) 
proportionality factors, 
%have 
has been considered~\cite{Mak83,MPS83,Mak84,Ieda07} 
(also see~\cite{GerKu85}).  
%[MakPasha81, 82, 
%[MakPasha83, 84, 
%[Ieda1]. 
%in the existing 
%literature. 
%all the components 
%a 
%special degenerate case such that 
%all the components have the 
%there is only one 
%same plane wave as a background, up to proportionality factors, 
%has been solved by the inverse scattering method.  
Although some 
%the 
soliton solutions 
%obtained 
in 
%such a 
%special 
the degenerate 
case 
%cases 
%a non-generic case  
are certainly 
%often 
%can be 
%in general, 
%sometimes 
nontrivial and 
%exhibit interesting behaviors 
interesting~\cite{Mak83,MMP81,Kuro07,Ieda06}, 
% [Mak81, 82, Mak84, 
%[MakPasha83, 
%[Ieda2], 
they 
%belong to 
form 
%are 
%merely 
%just a limited 
%special 
%only 
a 
rather 
restricted 
%degenerate 
%case 
subclass of the general soliton solutions 
%obtained 
%that satisfy 
%under 
%existing 
%in 
satisfying 
%in the more 
%most 
the 
%more 
%general 
%case of the 
plane-wave
%case of the 
%nonvanishing 
boundary conditions. 

The main objective of this paper is to construct 
%present 
%a class of exact 
new 
%the general 
soliton solutions of the vector/matrix NLS equation, which 
%under the nonvanishing boundary conditions 
%that 
have not been 
%not 
%explicitly 
presented explicitly 
in the 
%existing 
literature. 
%explicitly. 
To 
%investigate 
deal with 
%consider 
the 
%most 
%general 
non-degenerate case 
%where 
%the components of the vector/matrix dependent variable can approach different plane waves, 
%the background plane waves 
of 
the 
%plane-wave 
nonvanishing 
boundary conditions 
involving more than one background plane wave, 
we use B\"acklund--Darboux transformations~\cite{Sall82,MatSall91} 
%[Mat--Salle91] 
instead of the inverse scattering method. 
The main advantage of the B\"acklund--Darboux transformations 
is that 
%we can generate 
a new solution of the nonlinear equation considered can be 
constructed 
from a known solution 
%through  
%need only linear operations 
using only linear operations. 
More specifically, we need to 
%solve 
find 
%obtain 
a 
linear 
%wavefunction that satisfies 
eigenfunction
%s 
of 
%satisfying 
the 
%Lax pair 
Lax-pair representation 
%with 
for 
%with 
a given seed solution of the nonlinear equation. 
%to obtain a linear wavefunction. 
%but  
%eigenfunction 
%Broadly speaking, we need only solve 
%the pair of linear 
In essence, this task is 
%essentially 
equivalent to solving 
%a pair of 
a linear system of ordinary differential equations 
in the spatial 
%(or 
%/
%and 
%temporal) 
variable. 
%, respectively. 
%by exponentiating a Lax matrix. 
%In our case of 
For the vector/matrix 
%multicomponent 
NLS equation 
under the general 
plane-wave 
%nonvanishing 
boundary conditions, 
%we only have to compute 
%it 
the construction of soliton solutions 
reduces to the computation of 
%requires to 
%explicitly 
%compute 
%explicitly 
the 
%matrix 
exponential of a 
%certain 
constant non-diagonal 
%constant 
matrix 
%of a certain specific form 
arising from the Lax pair~\cite{Park00,Dega09}; 
this 
%the same 
%approach 
%construction 
also 
applies 
%approach is also valid for 
%the 
%note that 
%naturally, 
%soliton solutions 
%of 
for the higher 
%commuting 
flows 
of the vector/matrix NLS hierarchy 
%can be constructed 
%in the same manner
%~\cite{Park-OC}. 
(cf.~\cite{Park-OC}).
%[Dega--Lomb]. 
A straightforward way to compute 
%this 
the 
matrix exponential 
is to diagonalize the non-diagonal matrix 
by solving its 
%the 
characteristic equation.\footnote{In this paper, 
%In 
we do not separately discuss 
%consider separately 
the non-generic case wherein this 
%the 
non-diagonal matrix 
%is not diagonalizable, 
cannot be diagonalized, 
i.e., 
%its 
it has 
%the case of 
a non-diagonal Jordan normal form 
%is 
%not a diagonal matrix 
%not diago 
%a non-diagonal matrix 
(cf.~\cite{Dega13}).}
However, 
even in the simplest nontrivial case 
of the two-component vector NLS equation, 
%this 
it requires to 
%diagonalize 
%it reduces to the problem of the diagonalization of 
%a \mbox{$3 \times 3$} matrix 
solve a cubic equation with 
generally 
complex
%-valued 
coefficients 
%by solving the characteristic equation, which can be achieved 
%by 
using Cardano's formula\footnote{The
%As
%{The 
%As 
%is well known, 
%Cardano's formula 
%A 
%The 
%the
formula 
for 
finding 
the 
%three 
roots of 
a cubic equation has a rather 
complicated history
%, 
%but we do not present it here. 
%he history of 
(see, {\it e.g.}, 
{\tt 
%\mbox{
http://en.wikipedia.org/wiki/Cubic\_function}).
%~.
}~\cite{Park00,Wright} 
%[Wright00-1] 
(also see~\cite{Dub88}). 
%The cubic equation generally involves complex-valued 
%coefficients, so 
%Thus, 
Unfortunately, 
the obtained expression for the matrix exponential 
is too cumbersome
%too 
%complicated 
and 
not suitable
%so 
%useful 
for investigating 
%further 
the 
asymptotic 
behaviors 
%behavior 
of the 
%soliton 
solutions. 
%solution. 
%analysis, 
%because 
%%we 
%one 
%cannot 
%%tell which term becomes 
%%dominant 
%identify 
%%the dominant 
%%leading 
%which terms 
%become dominant 
%in 
%each 
%which 
%the 
%considered 
%some 
%region of the $(x,t)$-plane. 
%In particular, 
%Indeed, 
%Note that 
Moreover, such an expression intrinsically 
%it 
has 
%some 
an 
%inherent 
ambiguity,  
because Cardano's formula generally 
%for 
%has the shortcoming that it 
involves the square and cubic roots  
of a complex number, 
%which is 
%that 
which are 
%cannot 
%is 
not uniquely determined. 
%this ambiguity should be 
%; 
%%so 
%thus, 
%it is impossible to 
%identify the dominant 
%%leading 
%terms 
%in each 
%%which 
%%the 
%considered 
%some 
%region of the $(x,t)$-plane. 
%
%under the general nonvanishing 
%general plane-wave 
%boundary conditions. 
To fix this shortcoming, 
%To 
%and to 
%obtain 
%a more useful 
%another expression that is 
%appropriate 
%a more 
%explicit and 
%useful expression, 
%for 
%the asymptotic analysis etc., 
we 
%can 
re-parametrize the non-diagonal matrix so that 
the matrix exponential can be computed 
%more easily. 
in a more easy-to-read 
%``easy-to-read" 
form without any ambiguity. 
%In particular, 
Consequently, 
%As a result, 
for 
%the case of 
the two-component vector NLS equation, 
%and the \mbox{$2 \times 2$} matrix NLS equation, 
we can express the general 
%bright 
one-soliton solution 
%solutions 
explicitly 
%using only 
%the quadratic formula in
by 
%solving 
using the solutions of 
a quadratic equation with real
%-valued 
coefficients. 
%In addition, this procedure can, in principle, be performed 
%%applied 
%%repeated 
%repeatedly 
%%to construct 
%so that we can obtain 
%%applied re
%%given 
%two-soliton solutions, etc. 

%To be precise, 
The nature and the solution space
of the integrable 
NLS equation 
%changes drastically depending on 
depend critically on the sign of the cubic term. 
In 
%the case of 
the scalar case, 
%NLS equation, 
%there are only 
there exist only 
%the 
two choices 
%of the sign 
%are called 
corresponding 
to 
the self-focusing NLS 
%case 
%NLS 
and 
the 
self-defocusing NLS, 
%case, 
%NLS,  
%case, 
respectively; these 
names 
%which 
can be justified by 
%looking into 
considering 
%the 
%%its 
%canonical 
%Hamiltonian structure
%, 
%and the sign 
%and, 
%in particular, 
the 
%sign 
physical 
%effect 
meaning 
of the potential 
energy 
term in the canonical Hamiltonian formalism~\cite{ZM74}. 
%[ZakMan74]. 
%structure. 
%energy. 
%which become evident if we consider its Hamiltonian structure. 
%If the dependent variable is a vector 
%For 
In the multicomponent case, 
%NLS equations, 
{\it e.g.}, 
%\ }
when 
%vector/matrix NLS equation, 
the dependent variable is 
%a vector or 
%a 
%matrix
vector- or matrix-valued~\cite{Mana74,ZS74}, 
%[Manakov, ZS74], 
%there 
%are 
%exists 
we 
%also 
%need to examine 
%take into account 
have 
%need to consider 
%the third possibility 
%the case 
%have 
the third possibility:\
%, i.e., 
%choice 
%corresponding to 
%of mixed 
%of 
a mixed focusing-defocusing nonlinearity 
%[Newell78, Mak81?, ZakShul82] 
%; 
%that is, 
%wherein 
%some 
%nonlinear 
that contains both terms 
%are
%have 
with a plus 
%positive 
sign 
%definite 
%definite, while 
and 
%the 
%some 
%others have 
terms with a minus sign~\cite{YO2,Ab78,New79,Mak81,MMP81,Mak82,ZakShul82}.
%are negative 
%[Mak81, 82, ZakShul82].  
%definite. 
%definite and some other terms are negative definite. 

In the scalar case, 
the self-defocusing NLS equation admits 
dark-soliton solutions. 
A 
%single 
dark soliton is a dip 
of the 
%envelope 
density 
in 
%on a plane wave 
the background plane wave, which 
%they 
can be 
obtained by applying an elementary B\"acklund--Darboux 
transformation~\cite{Kono82,Calo84}
to the plane-wave solution. 
%It 
Each dark soliton 
is associated with a square-integrable eigenfunction 
(i.e., a bound state) of the Lax pair 
%representation 
%at 
with 
a real eigenvalue, 
%with a real value of the spectral parameter
%of 
because the spatial Lax operator for the self-defocusing NLS equation   
%which is  
%real value of the spectral parameter 
%because 
%the underlying spectral problem 
is self-adjoint~\cite{ZS2,AKNS73,AKNS74}. 
%[ZS73]. 
%; the spectral parameter is restricted to a real number 
In contrast, 
%The scalar 
the self-focusing NLS equation admits 
%has 
%allows 
bright-soliton solutions. A 
%single 
bright soliton is 
%, in general, 
generally 
a hump of the density, 
which can be obtained 
%generated 
by applying 
%to the seed solution 
a binary B\"acklund--Darboux transformation~\cite{Sall82,MatSall91} 
(also see~\cite{Chen1,Lamb74,KW75}) 
%[Lamb74]) 
%[Salle83]
%, 
%to the seed solution, 
%; 
%it 
%the latter 
that 
%which 
is 
%a 
%suitable 
equivalent to a suitable 
composition of two 
elementary B\"acklund--Darboux transformations~\cite{Kono82,Calo84}. 
%Naturally, 
Each bright soliton is associated with 
%a pair of 
%a 
two 
bound states 
%at two 
%different 
%complex
%-valued 
% values of 
%the 
%eigenvalues of the spatial Lax operator, 
%which 
of the Lax pair, 
%with a 
%pair of 
%complex conjugate eigenvalues, 
%with a complex conjugate pair of eigenvalues, 
%at 
%%two different 
%complex-valued eigenvalues, 
%at two distinct complex values of the eigenvalue, 
wherein the spatial Lax operator is not self-adjoint 
and the bound-state 
%discrete 
eigenvalues form 
%occur in 
a complex conjugate pair~\cite{ZS1,AKNS74}. 
%and 
%%thus 
%%the eigenvalue takes 
%%complex values.  
%has 
%a pair of complex conjugate eigenvalues. 
%with two complex-valued eigenvalues 
%%at complex values of the spectral parameter.  
%Note that 
%both the dark soliton and the bright soliton are 
%associated with 
%
%Each soliton a square-integrable eigenfunction 
%of the Lax-pair representation, i.e., a bound state. 
%the bound states of 

%In the multicomponent case
%Similarly, 
In 
a similar way, 
%a way similar to the 
%the same way as in 
%Similarly to the 
%scalar case,  
%for the multicomponent NLS equations, 
%systems, 
we can 
obtain multicomponent dark-soliton solutions in 
%for 
the self-defocusing case and multicomponent 
bright-soliton solutions 
%for 
in the self-focusing case
%\footnote{In [Park--Shin00, 02], Park and Shin used 
%a binary B\"acklund--Darboux transformation to 
%obtain a ``general dark-dark soliton". However, 
%their soliton solution is } 
%in 
%a 
%essentially the same way 
%similar to 
%as in the scalar NLS equation. 
using an elementary B\"acklund--Darboux transformation 
and a binary B\"acklund--Darboux transformation, respectively. 
%This is natural 
%possible 
%Note that 
Indeed, 
the general 
%formalism 
formulation 
of the B\"acklund--Darboux transformations 
%does not depend on 
is 
%in
%not 
%insensitive to the 
valid irrespective of the
%valid for any 
matrix 
%dimension 
size 
of the Lax-pair representation~\cite{Steu88,Park00,Park02,Dega09}, 
%[Wright00-1, 00-2],  
%Park--Shin01], 
%and thus 
%that 
which is related to 
%determined by 
the number of 
components of the vector/matrix 
dependent variable. 
%is irrelevant.  
%It turns out that 
%The 
A vector 
%multicomponent 
dark soliton~\cite{Dub88} 
in the self-defocusing case 
%generally 
does not 
%contain 
%have 
admit 
any essential 
%any 
%free 
parameter 
%for 
representing 
the 
internal degrees of freedom,
% [Dub88], 
%[Ohta10], 
although it provides a nontrivial generalization 
of the scalar dark soliton. 
%[Dub88].
%under the general plane-wave boundary conditions,  
%while 
%whereas 
Contrastingly, 
%it turns out that 
%but 
a matrix dark soliton 
in the self-defocusing case 
and 
a vector/matrix 
%multicomponent 
bright soliton in the self-focusing case 
%possess 
%admit 
%involve 
contain 
%essential 
free parameters 
%corresponding to 
representing 
the internal degrees of freedom. 
%i.e.,  
%which can be considered as an analog of 
%%or, equivalently, 
%a polarization~\cite{Mana74}. 
%[Manakov]. 
The 
case of 
%multicomponent case with 
a mixed focusing-defocusing nonlinearity 
is 
%the most 
%interesting 
%more complicated 
more complicated, because 
%if one is interested 
%in the soliton solutions; 
it admits both a dark soliton~\cite{Mak81,Mak82,Dub88} 
%the 
%dark-soliton solutions  
%solitons 
%bright 
and 
%dark 
%the 
%bright-soliton solutions;
a bright soliton~\cite{Dub88}; 
%solitons; 
%a variety of soliton solutions 
%Thus, we 
%can 
%apply 
%both 
they can be obtained by 
%successively 
%either 
applying an elementary B\"acklund--Darboux transformation 
%or 
and 
a binary B\"acklund--Darboux transformation, respectively. 
%to the 
%%focusing-defocusing 
%multicomponent NLS equations 
%with both focusing and defocusing components. 
%with a mixed focusing-defocusing nonlinearity. 
Moreover, 
in this case, 
we can also consider 
%use 
%Using, a special 
a 
limiting case 
%coalescence 
%%certain confluent
%limit 
%limiting form 
%case 
%form 
of the 
%a 
binary B\"acklund--Darboux transformation 
%such that 
in such a way that the associated 
%corresponding 
two 
%bound states 
bound-state eigenvalues coalesce into a 
%single bound state
%single 
real eigenvalue 
%corresponding to the coalescence of two bound states 
%eigenvalues  
(cf.~\S 2.4 in~\cite{MatSall91} and 
\cite{Dega09}); 
%and apply it to multicomponent NLS equations 
%with both focusing and defocusing components;
%which can used to produce 
%to obtain 
%so that 
%Thus 
%thus, 
as a result of its application, 
%consequently, 
%we can 
%also obtain 
a 
%construct 
vector dark soliton with internal degrees of freedom 
can 
%also be derived. 
be constructed. 
%obtained. 
%for multicomponent NLS equations 
%with both focusing and defocusing components. 
%(polarization) 
%can be obtained.  
In particular, 
%it can be shown that 
the three-component vector NLS equation with 
one focusing and two defocusing components 
admits a 
non-stationary 
%vector 
dark soliton
%, 
%with 
%internal degrees of freedom. 
%having 
%with 
%a polarization 
%vector. 
%that has a different 
%exhibits 
%which 
that exhibits a soliton mutation phenomenon; 
that is, 
%it 
%has 
%exhibits 
%different asymptotic behaviors 
%shapes 
%as 
the shape and 
%, 
velocity 
%and polarization 
of the dark soliton 
can change spontaneously 
%in the passage of time. 
%as time goes from $-\infty$ to $+\infty$. 
in the time evolution. 
%passes 
%be different for 
%forms 
%\mbox{$t \to - \infty$} and for \mbox{$t \to +\infty$}. 
%the soliton can have different shapes, velocities and polarizations.

In the multicomponent case, 
it is
%, in general, not so 
%not always 
%meaningful 
%easy 
%possible 
generally impossible 
to distinguish 
%clearly 
between 
a dark soliton and a bright soliton 
in a rigorous manner 
%or to identify the number of solitons 
just 
by 
%taking a look 
%%identifying 
looking 
at 
%obsreving
%the wave profile 
%from 
%the 
their wave profiles. 
%at a fixed moment of time. 
%of a soliton 
%in one component. 
%as a dip or a hump from the background. 
Indeed, 
%even a one-soliton solution 
%it 
each soliton 
%can 
may exhibit a rather complicated behavior 
%profile 
in 
%each 
%the 
individual 
components~\cite{Kuro07,Park00,Park02,Ieda06}, 
%of the vetor/matrix dependent variable 
%[Ieda2], 
which cannot be identified as a simple dip or 
%a 
hump. 
In addition, there exists 
%the invariance group 
a group of 
%invariance 
linear 
transformations, called the symmetry group, which 
%that 
mixes the components of the vector/matrix 
dependent variable 
but leaves the equation of motion 
%form-
invariant~\cite{MMP81,Mak82,Konop3};
% [Kono81];
%, MakPash81, 82]; 
%as 
%{\it e.g.}, 
for example, 
the self-focusing/defocusing vector NLS equation is 
invariant under the group of (special) 
unitary transformations. 
Such 
%symmetry group 
linear 
%invariance 
transformations 
%may 
can 
drastically 
change 
%not only 
both the boundary conditions and 
the wave profile of 
%each 
%the 
a soliton 
observed 
in 
%each component. 
individual components. 
%drastically. 
%considerably. 
%For instance, 
%%example, 
%a non-generic 
%%special 
%dark-dark soliton solution of the 
%%self-defocusing 
%two-component vector NLS equation with a self-defocusing nonlinearity 
%can be 
%identified with 
%%rewritten as 
%%reduced 
%%transformed 
%%to 
%a simpler 
%dark-bright soliton 
%solution 
%using 
%%by 
%%a linear transformation 
%an $SU(2)$ rotation 
%[Park-Shin]. 
%in some special cases. 
Thus, 
%in this paper, 
%it is not 
%so 
%meaningful to use 
instead of 
%using the terms 
distinguishing between a dark soliton and a bright soliton
%``dark soliton" and 
%%or 
%``bright soliton"  
%such as ``dark-dark" or ``dark-bright" solitons. 
intuitively, in this paper 
%and thus, 
%in this paper 
we 
%distinguish between them 
%the dark soliton and the bright soliton 
%determine the type of a soliton 
%simply 
%by 
use 
%them 
these terms 
in correspondence with 
the type of 
%a 
B\"acklund--Darboux transformation 
%used to generate the 
generating 
%used to generate 
each 
%such 
%these 
soliton 
solution. 
%solutions. 
%, or equivalently, 
%More specifically, 
That is, in analogy with the case of 
the scalar NLS equation, 
%case, 
%of the scalar NLS equation, 
a dark soliton is associated with a 
%single 
%bound state 
square-integrable eigenfunction 
of the Lax pair 
%at 
with a real eigenvalue,  
%a bound state 
%of the spatial Lax operator, 
%pair,  
%at 
%with a real eigenvalue, 
while a bright soliton entails 
%is related to 
%with 
%we compare 
%%consider 
%the positions of 
%if 
%the bound-state eigenvalue 
%of the Lax pair 
%is real, 
%then we 
%in the complex plane of the spectral parameter 
%and compare them 
%with those in the case of the scalar NLS equation. 
%Broadly speaking, 
%a 
%dark soliton should be accociated with 
%soliton giving 
%producing 
%a bound state 
%with a real eigenvalue is a dark soliton, 
%while a soliton 
%is associated with 
a pair of 
%related to 
%two 
bound states with 
%a 
%complex conjugate 
%pair of 
%complex conjugate 
%a pair of complex conjugate 
eigenvalues 
%they 
that are complex
conjugates of each other. 
%that occur in complexd conjugate pairs of eigenvalues 
%of .  
%is a bright soliton. 
%identifying 
%by comparing 
%looking at 
%into 
%the positions of the bound-state eigenvalues of the Lax-pair representation 
%with those in the scalar NLS case. 
Note that this definition is more stringent than 
%not exactly the same as 
that commonly 
used 
in the 
%existing 
literature. In particular, the spatial Lax operator 
for the vector/matrix NLS equation with a self-defocusing nonlinearity 
is self-adjoint~\cite{Mak84,Ieda07}, 
%[Ieda1], 
so 
%its bound-state eigenvalues 
it does not admit any 
bright-soliton solution. 
%are restricted to real numbers. 
%%be real-valued. 
%associated with bound states with complex eigenvalues. 
%at any non-real 
This is in contrast to the work of Park and Shin~\cite{Park00,Park02}, 
wherein they applied 
a binary B\"acklund--Darboux transformation 
%%is applied 
to the 
%self-defocusing 
%vector 
two-component 
vector 
NLS equation 
with a self-defocusing nonlinearity 
%to construct 
%and obtained 
and constructed 
%because some of their 
``dark-bright" and ``dark-dark" soliton solutions 
(cf.~``double solitons" in~\cite{Dub88}). 
%Some of 
Their solutions 
%are not purely solitons because they 
are not associated with 
square-integrable eigenfunctions 
%bound states 
of the Lax pair, 
so we call 
%their 
%such 
%solutions
them 
%, which
%that 
%can be generated by applying a B\"acklund--Darboux transformation, 
``soliton-like" 
%soliton-like 
solutions, 
%; 
which 
%they 
are not genuine 
%rather than 
%pure 
soliton solutions in our definition. 
%; they are stable and spacially localized structures. 
%
%We call solitons corresponding to a pair of complex conjugate eigenvalues 
%the bright-type 
%and solitons corresponding to a real eigenvalue the dark-type, 
%
%
%To construct soliton solutions 
%for the multicomponent NLS equations 
%with a focusing nonlinearity, defocusing nonlinearity and 
%mixed  focusing-defocusing nonlinearity, 
%we 
%%need to 
%use appropriate 
%B\"acklund--Darboux transformations. 
%That is, depending on the three-types of nonlinearity, 
%elementary B\"acklund--Darboux transformations, 
%binary B\"acklund--Darboux transformation [Sall] which is a composition of 
%two kinds of elementary B\"acklund--Darboux transformations, 
%and the limiting case of binary B\"acklund--Darboux transformation [DegaLom] 
%
%Solitons 
%in the multicomponent case 
%wherein the dependent variable 
%comprises 
%contains more than one component, 
%wherein 
%such that the dependent 
%variable is a vector or a matrix, 
%
%there are three types of nonlinearity: 
%self-focusing [Manakov], self-defocusing [ZS74] 
%and mixed focusing-defocusing [Newell78, Mak81?, ZakShul82].  
%
%For the vanishing boundary conditions, the vector/matrix solitons 
%have internal degrees of freedom 
%
%carrier frequency (Park--Shin)

The remainder of this paper is organized as follows. 
In section 2, we introduce step by step 
%step-by-step 
%consider 
three types of 
B\"acklund--Darboux transformations for the vector/matrix NLS system. 
%on a step-by-step basis.  
%using a step-by-step approach. 
%step by step. 
%consider the Lax-pair representation 
%and 
%its 
%the 
%adjoint representation 
%for the vector/matrix NLS system 
%and then 
%define 
%Firstly, 
First, we introduce 
two 
%different 
%kinds 
%types 
%of 
%an 
elementary B\"acklund--Darboux transformations 
%based 
on the basis of the 
%(adjoint) 
Lax-pair representation  
and 
its 
%the 
%the 
adjoint problem. 
%representation. 
%for the vector/matrix NLS system. 
%on the basis of the Lax-pair representation 
%and 
%its 
%the 
%adjoint representation. 
%and its 
%the 
%adjoint representation.  
%In section 3, 
%Then, 
%Secondly, 
Second, we 
%discuss 
%consider 
%a nonlinear superposition 
%the composition of 
%compose 
%superimpose
%superpose
combine the 
two 
%kinds of 
elementary B\"acklund--Darboux transformations 
%into 
to define 
%give 
%and define 
a binary B\"acklund--Darboux transformation. 
%In section 4, 
%Moreover, 
%Thirdly, 
Third, we consider 
%discuss 
a 
%certain 
limiting case 
of the binary B\"acklund--Darboux transformation 
in such a way that 
%such that 
the associated 
%two 
%complex conjugate pair of 
%two associated 
%bound-state 
%complex 
two 
%complex 
eigenvalues 
%coalesce into 
merge into 
a single 
%real 
%value; 
eigenvalue. 
%(cf.~\S 2.4 in~\cite{MatSall91}). 
%which 
%it can be effectively 
%applied  
%%and apply it 
%to the multicomponent NLS equations 
%%on a general plane-wave background. 
%with a 
%%self-defocusing or 
%mixed focusing-defocusing nonlinearity. 
%certain 
In section 3, 
by applying an 
%Using 
%either 
%of the 
elementary B\"acklund--Darboux transformation 
to a 
%the 
general plane-wave solution, 
%and 
%choosing 
%%tuning 
%the 
%arbitrary 
%soliton 
%parameters 
%therein 
%appropriately, 
we 
%can realize the (pseudo-)Hermitian conjugation relation 
%reduction 
%between the two vector/matrix dependent variables. 
%in the NLS system. 
%Thus, 
%we 
%can 
obtain 
%derive 
dark-soliton solutions 
of the multicomponent NLS equations 
%on a general plane-wave background. 
with a self-defocusing or mixed focusing-defocusing nonlinearity. 
%and repeated application results  
In section 4, we apply 
the limiting case 
of the binary B\"acklund--Darboux transformation 
%in such a way 
%such that the associated 
%two 
%complex conjugate pair of 
%two associated 
%bound-state 
%complex 
%two complex eigenvalues 
%coalesce into 
%merge into 
%a real 
%value; 
%eigenvalue. 
%; 
%which 
%it can be effectively 
%applied  
%and apply it 
to the 
%multicomponent NLS equations 
vector NLS equation 
%on a general plane-wave background. 
with a 
%self-defocusing or 
mixed focusing-defocusing nonlinearity. 
Thus, we 
%%can 
%This enables 
%allows 
%us to 
%and 
obtain a
%its their 
dark-soliton solution with 
%essential 
internal degrees of 
freedom, which 
is 
%are 
more general 
%and interesting 
than 
%cannot be obtained
%those 
the 
%corresponding 
vector dark-soliton 
solution
%that 
obtained 
%given 
%is not obtained 
in section 3. 
In section 5, 
using the binary B\"acklund--Darboux transformation, 
we 
%can 
construct bright-soliton solutions 
on a general plane-wave background 
%of 
for the multicomponent NLS equations 
%on a general plane-wave background. 
with a self-focusing or mixed focusing-defocusing nonlinearity; 
%They can be expressed explicitly 
%with a suitable choice of the 
%The problem of how to express the matrix exponential of a 
%non-diagonal matrix is discussed. 
%We can also 
%characterize 
%derive 
a ``soliton-like" 
%soliton-like 
%solutions 
solution 
%of Park and Shin 
in the self-defocusing case 
given 
in~\cite{Park00} (also see~\cite{Park02}) 
can also be obtained 
%derived 
in 
%with 
a more explicit 
%expression. 
form.
%are also 
%in our setting. 
The last section, section 6, is devoted to concluding remarks.

%\section
%{Two 
%{Elementary 
\section{General formulation of 
B\"acklund--Darboux transformations}

In this section, we formulate three types of 
B\"acklund--Darboux transformations for the vector/matrix 
NLS system 
using 
%on the basis of 
the 
%its 
Lax-pair representation and 
its adjoint 
problem. 
%problem. 
%the adjoint Lax--pair 
%representation. 
%In this section, 
%First, we formulate two 
%elementary B\"acklund--Darboux transformations 
%%for 
%based on the Lax-pair representation for 
%the vector/matrix NLS system. 
%The Hermitian conjugation reduction between the two 
%dependent variables is realized by suitable choice 
%of the parameters 
%The application of 
%each 
%either elementary B\"acklund--Darboux transformation 
%to 
%%the 
%a plane-wave solution provides 
%%gives 
%a solution, which can be reduced 
%%reduces 
%to a dark-soliton solution 
%of the multicomponent NLS equations 
%by a suitable choice of 
%restriction on 
%the parameters. 

\subsection{Multicomponent NLS systems}
%{General formulation}
\label{sec2.1}

%We consider 
%start with 
The 
%vector/
matrix generalization~\cite{
%Mana74,
ZS74} 
of the 
nonreduced 
%matrix 
NLS system~\cite{AKNS73,AKNS74}
%: 
is given by
%\cite{
%%Mana74,
%ZS74} 
%
%\begin{equation}
%\left\{
%\begin{split}
%& \mathrm{i} Q_t + Q_{xx} - 2Q R Q = O, \\
%& \mathrm{i} R_t - R_{xx} + 2R Q R = O, 
%\end{split}
%\right.
%\label{}
%\end{equation}
%
\begin{subnumcases}{\label{mNLS}}
{}
%\begin{equation} \left\{ \begin{split}
\label{mNLS1}
 \mathrm{i} Q_t + Q_{xx} - 2Q R Q = O, 
%\hspace{13mm}
\\[0.5mm]
%[1mm]
\label{mNLS2}
 \mathrm{i} R_t - R_{xx} + 2R Q R = O.
%\hspace{13mm}
%\notag \\
%\end{split} \right. \end{equation}
\end{subnumcases}
%where 
Here, $Q$ and $R$ are 
%is an 
\mbox{$l \times m$} 
%\mbox{$l \times m$} 
%matrix 
%is an 
and \mbox{$m \times l$} matrices, respectively, and 
the subscripts $t$ and $x$ denote the partial
differentiation; 
%with respect to these variables, 
%and 
we use 
the symbol
%italic
$O$ 
%is used 
instead of $0$ to stress 
%when dependent variables in the equation considered can take their
%values in matrices.
%is used to
%%imply that the left-hand side is matrix-valued.
%stress that this is a system of matrix equations.
%Note that
%%the symbol italic
%$O$
%on the right-hand side
%%s
%of the
%equations
%implies 
that the dependent variables 
%$Q$ and $R$ 
can take
their
values in
%a matrix algebra.
matrices. 
%The nonreduced matrix NLS system (\ref{mNLS}) allows 
%By imposing 
%There exist too many references on matrix generalization
A 
%useful 
%informative 
list of references on the 
%nonreduced 
%vector/
matrix NLS system 
(\ref{mNLS}) and related equations 
%systems 
%and its reductions 
can be found in~\cite{DM2010}. 
%
%A general plane-wave solution of (\ref{mNLS}) 
%that will be used later is 
%\begin{align}
%Q = P_1 
%\mathrm{e}^{-2\mathrm{i} t BC} B  
%\mathrm{e}^{\mathrm{i} x A - \mathrm{i} t A^2} P_2, \hspace{5mm} 
%R= P_2^{-1} \mathrm{e}^{-\mathrm{i} x A + \mathrm{i} t A^2}
%	 C \mathrm{e}^{2\mathrm{i} t BC} P_1^{-1}, 
%\label{plane1}
%\end{align}
%where 
%%$P_1$, $P_2$, 
%$A$, $B$, $C$, $P_1$ and $P_2$ are constant matrices; 
%$P_1$ is an \mbox{$l \times l$} matrix,
%$A$ and $P_2$ are \mbox{$m \times m$} matrices, 
%$B$ is an \mbox{$l \times m$} matrix 
%and $C$ is an \mbox{$m \times l$} matrix. 
%Actually, 
%%In fact, 
%%we can remove 
%$P_1$ and $P_2$ can be removed 
%%are unnecessary, because they can be absorbed 
%by redefining $A$, $B$ and $C$. 
%However, 
%it is more convenient to leave them 
%%$P_1$ and $P_2$ 
%as in (\ref{plane1}), 
%because they can be used to 
%%assume 
%%write 
%put 
%$A$ and $BC$ in Jordan 
%%their 
%%canonical 
%normal form. 
%%forms. 

%For the nonreduced 
%vector/
%matrix 
%NLS system (\ref{mNLS}), 
%we can impose 
%The matrix NLS system 
Note that (\ref{mNLS}) allows 
%the 
a 
%(pseudo-)
Hermitian conjugation 
reduction~\cite{Mak82} (also see~\cite{YO2,Ab78,New79,MMP81,ZakShul82}):
%~\cite{Mak82,Zak82}
\begin{equation}
R = \Sigma \hspace{1pt} Q^\dagger \hspace{1pt} \Omega.  
%\Lambda_2 
\label{Herm}
\end{equation}
%on the nonreduced 
%vector/
%matrix 
%NLS system (\ref{mNLS}), 
%where 
Here, $\Sigma$ and 
%$\Lambda_2$ 
$\Omega$ are \mbox{$m \times m$} 
and \mbox{$l \times l$} 
constant Hermitian matrices, respectively, 
and the dagger denotes the Hermitian 
%conjugate. 
conjugation. 
%which reduces 
%simplifies 
%Thus, 
The reduction (\ref{Herm}) simplifies 
the nonreduced matrix NLS system 
(\ref{mNLS}) 
%is simplified 
%reduced 
to the matrix NLS equation: 
\begin{equation}
\label{redNLS}
 \mathrm{i} Q_t + Q_{xx} - 2Q \Sigma \hspace{1pt} Q^\dagger 
	\hspace{1pt} \Omega Q = O.
\end{equation}
%The freedom of 
%By applying 
Considering 
%unitary 
a 
%suitable 
linear transformation  
%to $Q$, 
\mbox{$Q 
\to 
%\mapsto 
U_1 Q U_2$}
with 
nonsingular 
constant 
%unitary 
matrices
%, where 
$U_1$ and $U_2$, 
%are 
%unitary 
%invertible 
%constant 
%nonsingular 
%constant 
%matrices, 
%enables us to reduce 
%set 
we can recast 
%rewrite 
$\Sigma$ and $\Omega$ in their canonical forms, 
i.e., diagonal matrices whose diagonal 
%elements 
entries 
are $+1$, $-1$ or $0$. 
%However, the case in which some of these diagonal elements 
%are zero is 
%%more or less 
%trivial and less interesting. 
Because we are 
%only 
more 
%not 
interested in 
%The case 
%integrable systems of 
truly coupled systems 
%multicomponent equations 
than 
%and not 
in triangular systems, 
we require 
%assume 
that 
%do not consider the special case in which 
%the 
%some 
%all 
the diagonal entries 
%elements 
of 
(the canonical forms of) 
$\Sigma$ and $\Omega$ are 
%$\pm1$. 
%either 
$+1$ or $-1$. 
%In this paper, 
In particular, 
in the subsequent sections, 
%3--5, 
we 
%are interested in 
mainly consider 
%focus on 
the following two 
%special 
cases. 
%subcases. 
%of (\ref{redNLS}) 
%corresponding to the vector NLS equation and 
%the square matrix NLS equation, respectively. 
\begin{itemize}
\item 
%\\
The 
%the 
case \mbox{$l=1$}: 
%(generalized) 
vector NLS equation~\cite{ 
%in the general form~\cite{
%Mana74,
YO2,Ab78,New79,MMP81,
Mak82,
ZakShul82}, 
\begin{equation}
\label{rvNLS}
 \mathrm{i} \vt{q}_t + \vt{q}_{xx} - 2\sca{\vt{q}\Sigma}{\vt{q}^\ast}
	\vt{q} = \vt{0},
\end{equation}
which generalizes the Manakov model~\cite{Mana74}. 
Here, \mbox{$\vt{q} := (q_1, q_2, \ldots, q_m)$} is an $m$-component row vector
%,  
and \mbox{$\Sigma := \mathrm{diag} (\sigma_1, \sigma_2, \ldots, \sigma_m)$} is a diagonal 
matrix with 
entries 
%each diagonal entry $\sigma_j$ equal to 
\mbox{$\sigma_j = +1$} or $-1$. 
%, 
Thus, after a suitable re-numbering of the components, 
the scalar product can be written explicitly 
%is defined, 
%with 
%after a suitable re-numbering of the components, 
as
\[
%\Sigma := \left[
\sca{\vt{q}\Sigma}{\vt{q}^\ast} := \sum_{j=1}^n |q_j|^2 - \sum_{k=n+1}^m |q_k|^2, 
\]
where \mbox{$n \; (0 \le n \le m)$} is the number of defocusing components and \mbox{$m-n$} 
is the number of focusing components. 
%Note that (\ref{rvNLS}) is invariant under the action of 
%the (peudo-)unitary group. 
\item 
%\\
The case \mbox{$l=m$}: square matrix NLS equation~\cite{ZS74}, 
\begin{equation}
\label{rmNLS}
 \mathrm{i} Q_t + Q_{xx} - 2 \sigma Q Q^\dagger Q = O, 
 \hspace{5mm} \sigma = +1 \;\, \mbox{or} \;\, \mbox{$-1$}.  
\end{equation}
Here, the choice of 
%the sign 
\mbox{$\sigma=+1$} 
and the choice of \mbox{$\sigma=-1$} correspond to the self-defocusing case 
and the self-focusing case, respectively.  
\end{itemize}

\subsection{Lax-pair representation 
and 
%inverse 
Miura maps}
\label{subs2.2}

%Next, we consider 
%introduce 
%%formulate 
%two 
%elementary B\"acklund--Darboux transformations 
%%The Lax-pair representation 
%for the nonreduced matrix NLS system (\ref{mNLS}), 
%Recall that 
%which admits 
%ossesses 
The 
%following 
%The 
Lax-pair representation~\cite{Lax} 
for the nonreduced matrix NLS system (\ref{mNLS}) is 
given by~\cite{Zakh,Konop1} 
%as 
%: 
% for (\ref{mNLS}) 
%is given by the pair of linear 
%%partial 
%differential equations~\cite{Zakh,Konop1}: 
%The Lax representation 
%%of 
%for the matrix NLS system (\ref{mNLS})
%is given by~\cite{Zakh,Konop1}
%
\begin{subequations}
\label{NLS-UV}
%\begin{equation}
%\left\{ \begin{array}{l}
\begin{align}
& \left[
\begin{array}{c}
 \Psi_1  \\
 \Psi_2 \\
\end{array}
\right]_x 
= \left[
\begin{array}{cc}
-\mathrm{i}\zeta I_l & Q \\
 R & \mathrm{i}\zeta I_m\\
\end{array}
\right] 
\left[
\begin{array}{c}
 \Psi_1  \\
 \Psi_2 \\
\end{array}
\right],
\label{NLS-U}
\\[1.5mm]
& \left[
\begin{array}{c}
 \Psi_1  \\
 \Psi_2 \\
\end{array}
\right]_t 
= \left[
\begin{array}{cc}
-2\mathrm{i}\zeta^2 I_l -\mathrm{i} QR & 2 \zeta Q + \mathrm{i} Q_x \\
 2 \zeta R - \mathrm{i} R_x & 2\mathrm{i}\zeta^2 I_m +\mathrm{i} RQ \\
\end{array}
\right]
\left[
\begin{array}{c}
 \Psi_1  \\
 \Psi_2 \\
\end{array}
\right]. 
\label{NLS-V}
\end{align}
%\end{array}\right.
%\end{equation}
\end{subequations}
%as well as the adjoint Lax-pair representation: 
%
Here, 
$\zeta$ is the spectral parameter, which is
%%assumed to be
an arbitrary constant, 
%%, namely,
%independent of $x$ and $t$, 
and 
%;
$I_l$ and $I_m$ 
%is 
are the \mbox{$l \times l $} and \mbox{$m \times m$} 
%unit 
identity matrices, respectively. 
%$Q$ is a potential function 
%which takes its value in \mbox{$m \times m$} matrices. 
The compatibility condition for 
%of 
the overdetermined linear system (\ref{NLS-UV})
%, \mbox{$\partial_x \partial_t \Psi_j = \partial_t \partial_x \Psi_j$}, 
%is 
indeed 
%provides 
%equivalent to 
implies 
(\ref{mNLS}). 
%, which justifies the use of the term ``Lax pair"~\cite{Lax}. 
Note that (\ref{NLS-U}) can be rewritten 
%more 
explicitly 
as 
%in the form of 
an eigenvalue problem~\cite{ZS1,ZS2}: 
\[
\left[
\begin{array}{cc}
 \mathrm{i} \partial_x & - \mathrm{i} Q \\
 \mathrm{i} R & - \mathrm{i} \partial_x \\
\end{array}
\right] 
\left[
\begin{array}{c}
 \Psi_1  \\
 \Psi_2 \\
\end{array}
\right] = \zeta 
\left[
\begin{array}{c}
 \Psi_1  \\
 \Psi_2 \\
\end{array}
\right]. 
\]
%where $\zeta$ is the eigenvalue of the spatial Lax operator. 
In the self-defocusing case 
%of 
\mbox{$R=Q^\dagger$}, 
%the 
this eigenvalue problem is self-adjoint, 
%and 
%; 
so the eigenvalue $\zeta$ 
is restricted to be real-valued 
%must be real 
for any square-integrable 
(i.e., bound-state) eigenfunction; note that 
%, 
%because 
%This can be easily seen using the relation: 
\[
 \int_{-\infty}^\infty
 \left[
\begin{array}{cc}
\! \Psi_1^\dagger \! & \! \Psi_2^\dagger \!
\end{array}
\right]
\left[
\begin{array}{cc}
 \mathrm{i} \partial_x & - \mathrm{i} Q \\
 \mathrm{i} Q^\dagger & - \mathrm{i} \partial_x \\
\end{array}
\right] 
\left[
\begin{array}{c}
 \Psi_1  \\
 \Psi_2 \\
\end{array}
\right] \mathrm{d} x = \zeta \int_{-\infty}^\infty
\left[
\begin{array}{cc}
\! \Psi_1^\dagger \! & \! \Psi_2^\dagger \!
\end{array}
\right]
\left[
\begin{array}{c}
 \Psi_1  \\
 \Psi_2 \\
\end{array}
\right] \mathrm{d} x. 
\]

We also 
%need 
%crucial 
%important 
%to 
consider 
the 
%``adjoint" eigenvalue problem
adjoint Lax-pair 
%representation (see, {\it e.g.}, [])
representation:\footnote{If 
%Note that if 
we have 
%know consider 
a square-matrix solution 
$\Psi$ of 
%to 
the linear system 
%eigenvalue problem 
\mbox{$\Psi_x = U \Psi,\; \Psi_t = V \Psi$}, 
then its 
%the 
inverse 
%matrix 
\mbox{$\Phi := \Psi^{-1}$} solves 
%satisfies 
the adjoint system 
%eigenvalue problem 
\mbox{$\Phi_x = -\Phi U, \; \Phi_t = -\Phi V$}~\cite{New79,Agra79}.}
%(see, {\it e.g.}, [AG79])
%(without transposing $Q_n$ and $R_n$),
%
%For later convenience, we also consider 
\begin{subequations}
\label{NLS-adUV}
%\begin{equation}
%\left\{ \begin{array}{l}
\begin{align}
& \left[
\begin{array}{cc}
%\Psi_{1,n}^{\mathrm{ad}} \! & \! \Psi_{2,n}^{\mathrm{ad}} 
\! \Phi_1 \! & \! \Phi_2 \!
\end{array}
\right]_x 
= - \left[
\begin{array}{cc}
%\Psi_{1,n}^{\mathrm{ad}} \! & \! \Psi_{2,n}^{\mathrm{ad}} 
\! \Phi_1 \! & \! \Phi_2 \!
\end{array}
\right]
\left[
\begin{array}{cc}
-\mathrm{i}\zeta I_l & Q \\
 R & \mathrm{i}\zeta I_m\\
\end{array}
\right],
\label{NLS-adU}
\\[1.5mm]
& \left[
\begin{array}{cc}
%\Psi_{1,n}^{\mathrm{ad}} \! & \! \Psi_{2,n}^{\mathrm{ad}} 
\! \Phi_1 \! & \! \Phi_2 \!
\end{array}
\right]_t 
= - \left[
\begin{array}{cc}
%\Psi_{1,n}^{\mathrm{ad}} \! & \! \Psi_{2,n}^{\mathrm{ad}} 
\! \Phi_1 \! & \! \Phi_2 \!
\end{array}
\right]
\left[
\begin{array}{cc}
-2\mathrm{i}\zeta^2 I_l -\mathrm{i} QR & 2 \zeta Q + \mathrm{i} Q_x \\
 2 \zeta R - \mathrm{i} R_x & 2\mathrm{i}\zeta^2 I_m +\mathrm{i} RQ \\
\end{array}
\right]. 
\label{NLS-adV}
\end{align}
%\end{array}\right.
%\end{equation}
\end{subequations}
%The compatibility condition 
Indeed, 
%the use of the expression 
%%word 
%``adjoint equation" 
%can be justified by 
%the following identity, 
%a discrete analog of 
%Lagrange's 
a Lagrange-like
identity, 
%guarantees 
%the use of 
%the word 
%``adjoint equation":
\begin{align}
& \left[
\begin{array}{cc}
\! \Phi_1 \! & \! \Phi_2 \!
\end{array}
\right] 
\left\{
\left[
\begin{array}{c}
 \Psi_1 \\
 \Psi_2 \\
\end{array}
\right]_x 
-
\left[
\begin{array}{cc}
-\mathrm{i}\zeta I_l & Q \\
 R & \mathrm{i}\zeta I_m\\
\end{array}
\right]
\left[
\begin{array}{c}
 \Psi_1 \\
 \Psi_2 \\
\end{array}
\right] \right\}
\nonumber \\
& \mbox{} + \left\{
\left[
\begin{array}{cc}
\!
\Phi_1
%\;\, 
 \! & \!  
\Phi_2 \!
\end{array}
\right]_x 
+
\left[
\begin{array}{cc}
\! \Phi_1 \! & \! \Phi_2 \!
\end{array}
\right]
\left[
\begin{array}{cc}
-\mathrm{i}\zeta I_l & Q \\
 R & \mathrm{i}\zeta I_m\\
\end{array}
\right]
\right\} \left[
\begin{array}{c}
 \Psi_1 \\
 \Psi_2 \\
\end{array}
\right]
\nonumber \\[1mm] 
& = 
%\partial_x
\left\{ 
\left[
\begin{array}{cc}
\! \Phi_1 \! & \! \Phi_2 \!
\end{array}
\right]
\left[
\begin{array}{c}
 \Psi_1 \\
 \Psi_2 \\
\end{array}
\right]
\right\}_x,
\nonumber 
\end{align}
and a similar relation 
%with respect to 
for 
the $t$-differentiation 
%show 
imply  
%: 
%justifies 
%implies 
that 
%(\ref{NLS-UV}) and 
(\ref{NLS-adUV}) 
%(\ref{mAL0}) and (\ref{mAL2}) 
%can be called 
%are 
%can be 
%considered as 
%said to be 
is 
the adjoint problem of (\ref{NLS-UV}). 
%to each other. 
%
%Here, $z$ is 
%%the 
%a constant spectral parameter, 
%$I$ is the $l \times l$ unit matrix, and 
%$Q_n$ and $R_n$ are $l \times l$ matrix potentials. 
%The eigenvalue problem (\ref{mAL0}) is equivalent, up to 
%a trivial gauge transformation, 
%to the 
%% straightforward 
%standard matrix generalization of the 
%%that by 
%Ablowitz--Ladik eigenvalue problem [AL] 
%%considered 
%studied in \cite{GI82}, etc. 
%Then, 
In particular, 
%for any pair of 
%Note that 
the product of 
%any 
%the 
solutions 
%a solution 
of 
%(\ref{NLS-UV}) 
%the adjoint linear eigenfunction satisfying  
(\ref{NLS-adUV}) and 
%the linear eigenfunction satisfying 
%a solution of 
%the solution of 
%(\ref{NLS-adUV}) 
(\ref{NLS-UV})
%, we have the important relation: 
%does not depend on $x$ and $t$: 
is \mbox{$(x,t)$}-independent: 
%the quantity 
\begin{equation}
\left[
\begin{array}{cc}
\! \Phi_1 \! & \! \Phi_2 \!
\end{array}
\right]
\left[
\begin{array}{c}
 \Psi_1 \\
 \Psi_2 \\
\end{array}
\right] = \mathrm{const.}
\label{inner}
\end{equation}

We consider an \mbox{$(l+m) \times l$} 
%block-matrix 
matrix-valued 
%solution 
%of 
%to 
%for
%the pair of linear equations 
linear eigenfunction of the Lax-pair representation 
%the Lax-pair representation 
(\ref{NLS-UV}) 
%so that 
such that 
%wherein 
$\Psi_1$ is an \mbox{$l \times l$} 
%%square 
%%regular 
nonsingular
%invertible 
matrix 
%%nonsingular for generic values of $x$, $t$, and $\zeta$. 
%Then, 
%in terms of 
and define 
%the 
an \mbox{$m \times l$} matrix $\widetilde{R}$ 
as \mbox{$\widetilde{R} := \Psi_2 \Psi_1^{-1}$}. 
%the Lax representation 
Then, 
%the Lax-pair representation 
%(\ref{NLS-U}) 
%the Lax-pair representation 
(\ref{NLS-UV}) 
%can be rewritten as 
%a pair of matrix Riccati equations 
%(see
implies the 
%important 
relations 
(cf.~\cite{Chen1,
%Chen2,
WSK,KW75}):
%,
% for the scalar case), 
%
\begin{subequations}
\label{NLS-R}
\begin{align}
& R = \widetilde{R}_x - 2 \mathrm{i} \zeta \widetilde{R} 
	+ \widetilde{R} Q \widetilde{R}, 
\label{NLS-R1}
\\
%[1mm]
& \widetilde{R}_t = 2 \zeta R- \mathrm{i}R_x + 4 \mathrm{i}\zeta^2 \widetilde{R} 
 + \mathrm{i}RQ  \widetilde{R}
	+ \mathrm{i} \widetilde{R}QR - 2\zeta \widetilde{R}Q\widetilde{R} 
	- \mathrm{i}\widetilde{R} Q_x \widetilde{R}.
\label{NLS-R2}
\end{align}
\end{subequations}
%
%the most remarkable
%of particular importance
%of particular note is
%The crucial observation is that 
%The important point is that
%Using 
Relation (\ref{NLS-R1}) 
%allows 
enables us 
to  
%one 
%we can 
%rewrite 
express 
$R$ %in (\ref{NLS1}) and (\ref{NLS-R2}) 
%in terms of $Q$ and $\widetilde{R}$ 
%and $Q$ 
%as \mbox{$ -2\mathrm{i} \zeta P +P_x +PQP$}. 
in (\ref{mNLS1}) and (\ref{NLS-R2}) in terms of 
$Q$ and $\widetilde{R}$~\cite{TsuJMP11}.  
%to obtain 
Thus, 
%we obtain 
%(\ref{NLS1}) and (\ref{NLS-R2}) 
%form 
%now comprise 
we obtain 
a closed two-component 
system for 
%the pair of variables $(Q, \widetilde{R})$, 
$Q$ and $\widetilde{R}$:
%, 
%i.e., 
%as
%: 
\begin{subnumcases}{\label{mGI}}
{}
%\begin{equation} \left\{ \begin{split}
%\begin{subequations}
%\label{mGI}
%\begin{align}
 \mathrm{i} Q_t + Q_{xx} +4\mathrm{i} \zeta Q \widetilde{R}Q 
 - 2Q \widetilde{R}_x Q -2Q\widetilde{R} Q\widetilde{R} Q = O,
\label{GI1}
\\[0.5mm]
%[1mm]
\label{GI2}
 \mathrm{i} \widetilde{R}_t - \widetilde{R}_{xx} 
 -4\mathrm{i} \zeta \widetilde{R} Q \widetilde{R} 
 - 2 \widetilde{R} Q_x \widetilde{R} +2\widetilde{R} Q \widetilde{R}Q \widetilde{R} = O.
%\end{array}
%\end{align}
%\end{subequations}
\end{subnumcases}
%
%In the case of scalar depedent variables, 
%this system 
This is 
%(
a matrix generalization~\cite{Linden1,SviYami91,Olver2,TW3,Ad,Dimakis} of
%)
%essentially 
%intrinsically 
the derivative NLS system 
%and is 
%which is called the Gerdjikov--Ivanov 
%which was investigated 
%studied 
%by Ablowitz, Ramani, and Segur~\cite{ARS}
studied by Ablowitz, Ramani and Segur~\cite{ARS} 
%{\it et al.}~\cite{ARS}
%, 
(and 
later by 
%subsequently by 
Gerdjikov and Ivanov~\cite{GI} 
without 
%referring
%any 
%reference 
%to
citing the paper~\cite{ARS}); 
the cubic terms involving 
%with 
the parameter $\zeta$ can be 
%eliminated 
removed 
by a simple point 
%Galilean 
%transformation (cf.~\cite{IKWS,Kawata2,Clark87}). 
transformation (cf.~\cite{IKWS,Kawata2,Clark87}). 
%simple point transformation. 
%in the case of scalar dependent variables. 
%system 
%or Ablowitz--Ramani--Segur system. 
%, and Ablowitz, Ramani, and Segur 
%(also see Ablowitz, Ramani, and Segur~\cite{ARS}). 
%The matrix generalization (\ref{mGI}) was studied 
%%discussed 
%in~\cite{Linden1,Olver2,TW3,Ad,Dimakis}. 
%[Linden,OS99,TW,Adler00,DM]. 
Note that (\ref{NLS-R1}) defines a Miura map 
%\mbox{$(Q, \widetilde{R}) \to (Q,R)$} 
from 
(\ref{mGI}) to (\ref{mNLS}): \mbox{$(Q, \widetilde{R}) \mapsto (Q,R)$}. 

Similarly, 
using a linear eigenfunction 
of the adjoint Lax-pair representation (\ref{NLS-adUV}), 
we define
%can 
%introduce  
an \mbox{$l \times m$} matrix $\widehat{Q}$ 
%$\widetilde{Q}$ 
%as \mbox{$\widetilde{Q} := \Phi_1^{-1} \Phi_2$} 
%from 
%%using 
%%the solution 
%a linear eigenfunction 
%of the adjoint Lax-pair representation (\ref{NLS-adUV}) 
as \mbox{$\widehat{Q} 
%\widetilde{Q} 
:= \Phi_1^{-1} \Phi_2$}. 
Then, we obtain 
%(\ref{NLS-adUV}) implies 
%we obtain the two 
the relations: 
\begin{subequations}
\label{NLS-Q}
\begin{align}
& Q = -\widehat{Q}_x - 2 \mathrm{i} \zeta \widehat{Q} 
	+ \widehat{Q} R \widehat{Q}, 
\label{NLS-Q1}
\\
%[1mm]
& \widehat{Q}_t = -2 \zeta Q- \mathrm{i}Q_x - 4 \mathrm{i}\zeta^2 \widehat{Q} 
 - \mathrm{i}QR  \widehat{Q}
	- \mathrm{i} \widehat{Q}RQ + 2\zeta \widehat{Q}R\widehat{Q} 
	- \mathrm{i}\widehat{Q} R_x \widehat{Q}.
\label{NLS-Q2}
\end{align}
\end{subequations}
Relation (\ref{NLS-Q1}) 
allows us 
%enables us 
to  
%one 
%we can 
%rewrite 
express 
$Q$ %in (\ref{NLS1}) and (\ref{NLS-R2}) 
%in terms of $Q$ and $\widetilde{R}$ 
%and $Q$ 
%as \mbox{$ -2\mathrm{i} \zeta P +P_x +PQP$}. 
in (\ref{NLS-Q2}) and (\ref{mNLS2}) 
%and (\ref{NLS-Q2}) 
%using 
in terms of 
%and 
$\widehat{Q}$ and $R$. 
%~\cite{TsuJMP11}.  
%to obtain 
%Thus, 
%we obtain 
%(\ref{NLS1}) and (\ref{NLS-R2}) 
%form 
%now comprise 
%so 
Thus, 
%we obtain 
a closed 
%two-component 
%the same 
%derivative NLS 
system for 
%the pair of variables $(Q, \widetilde{R})$, 
$\widehat{Q}$ and $R$ is obtained:
%, 
%i.e., 
%as
%: 
\begin{subnumcases}{\label{mGI2}}
{}
%\begin{equation} \left\{ \begin{split}
%\begin{subequations}
%\label{mGI}
%\begin{align}
 \mathrm{i} \widehat{Q}_t + \widehat{Q}_{xx} +4\mathrm{i} \zeta 
	\widehat{Q} R \widehat{Q} 
 - 2\widehat{Q} R_x \widehat{Q} -2 \widehat{Q} R \widehat{Q} R \widehat{Q} = O,
\label{GI1-2}
\\[0.5mm]
%[1mm]
\label{GI2-2}
 \mathrm{i} R_t - R_{xx} 
 -4\mathrm{i} \zeta R \widehat{Q} R 
 - 2 R \widehat{Q}_x R +2 R \widehat{Q} R \widehat{Q} R = O.
%\end{array}
%\end{align}
%\end{subequations}
\end{subnumcases}
Note 
that 
(\ref{mGI}) and 
%the coincidence 
%equivalence 
%of 
(\ref{mGI2}) 
%is essentially 
%equivalent to 
%identical to 
%can be identified 
%with (\ref{mGI}). 
are identical, so 
%the same system, 
%; 
%is the same system as (\ref{mGI}). 
%``same". 
%Thus, 
%and 
we 
%now 
have 
%obtain 
another Miura map 
%and the same system for the pair of variables $(\widetilde{Q}, R)$,
%Note that 
(\ref{NLS-Q1}) 
%defines a Miura map 
%\mbox{$(Q, \widetilde{R}) \to (Q,R)$} 
from 
%the matrix derivative NLS system 
%the matrix Ablowitz--Ramani--Segur system 
(\ref{mGI2}) to 
%the matrix NLS system 
(\ref{mNLS}): \mbox{$(\widehat{Q}, R) \mapsto (Q,R)$}.

\subsection{Elementary B\"acklund--Darboux transformations}
%To sum up, 
%Hence, 
%Thus, 
We have 
obtained 
two distinct 
%different 
Miura maps from 
the 
matrix 
derivative NLS system 
%(\ref{mGI}) or ((\ref{mGI2}))
to the 
matrix 
NLS system~\cite{Clark87,SviYami91}; 
%(cf.~Adler, Shabat, Mikhailov, Yamilov); 
the inverse of each Miura map can be expressed in terms 
of a linear eigenfunction of the (adjoint) Lax-pair representation 
for the
matrix 
NLS system. 
Thus, combining the inverse of one Miura map 
%one of the two inverse Miura maps 
with 
%the inverse of 
the other Miura map, we can construct 
%(the spatial part of) 
an elementary B\"acklund--Darboux transformation 
for the matrix NLS system. 
Moreover, we can also identify 
the 
%corresponding 
%a 
gauge transformation 
%write down 
%trace 
%The action of such a B\"acklund--Darboux transformation 
%acting on 
%for 
of the linear eigenfunction 
%of the Lax pair, 
of the Lax-pair representation, 
%, 
%%of the Lax-pair representation
%which generates 
%induced by 
%corresponding to 
which generates 
%such 
%an 
each elementary 
%each 
B\"acklund--Darboux transformation~\cite{DJM83,Li87,Adler94}. 
%[Chinese87]. 
%can be easily identified as well.  
%however, 
In the following, 
%the value of 
we often fix 
the arbitrary 
%spectral 
parameter $\zeta$ 
in the B\"acklund--Darboux transformation 
%is 
%considered to be 
%fixed 
at some 
%a 
specific 
%some 
value, say \mbox{$\zeta=\mu$}, 
%or $\nu$, 
%\mbox{$\zeta=\zeta_1$},  
%and 
%which should not be confused with 
to distinguish it from the original 
%arbitrary 
spectral parameter $\zeta$
%fixed 
in the Lax-pair representation. 

\begin{proposition}
\label{prop2.1}
The spatial 
%(i.e., $x$-) 
part of 
an elementary auto-B\"acklund transformation 
%\mbox{$(Q,R) \mapsto (\widetilde{Q}, \widetilde{R})$}
%of
for 
%between two solutions, $(Q,R)$ and $(\widetilde{Q}, \widetilde{R})$, of 
the matrix NLS system (\ref{mNLS}), which connects 
%connecting 
%%relates 
two solutions $(Q,R)$ and $(\widetilde{Q}, \widetilde{R})$, 
%of 
%for 
%the matrix NLS system (\ref{mNLS})
%, \mbox{$(Q,R) \mapsto (\widetilde{Q}, \widetilde{R})$}, 
%can be 
%written 
%defined 
%expressed 
%described 
%as
is given by
%read
%~\cite{Kono82,Calo84} 
(see~\cite{Kono82,Calo84} for the scalar case 
and~\cite{SviYami91,Adler94} for the matrix case) 
%\begin{subequations}
%\label{eBT1}
\begin{align}
& \widetilde{Q} = -Q_x - 2 \mathrm{i} \mu Q + Q \widetilde{R} Q, 
\nonumber 
\\
& R = \widetilde{R}_x - 2 \mathrm{i} 
%\zeta_1 
 \mu \widetilde{R} 
 + \widetilde{R} Q \widetilde{R}. 
\nonumber
\end{align}
%\end{subequations}
%which connects 
%%%relates 
%two solutions $(Q,R)$ and $(\widetilde{Q}, \widetilde{R})$. 
%\mbox{$(\widetilde{Q}, \widetilde{R}) \mapsto (Q,R)$}
%It 
The 
transformation 
%map 
\mbox{$(Q,R) \mapsto (\widetilde{Q}, \widetilde{R})$} 
can be expressed 
explicitly 
%as an explicit mapping 
%rewritten 
in terms of a linear eigenfunction 
of the Lax-pair representation (\ref{NLS-UV}) at \mbox{$\zeta=\mu$} 
as~\cite{Kono82,Calo84} 
%[Chinese87] 
\begin{equation}
\widetilde{R} = \left. \Psi_2 \Psi_1^{-1}\right|_{\zeta= \mu}, 
\hspace{5mm} \widetilde{Q} = -Q_x - 2 \mathrm{i} \mu Q + Q \widetilde{R} Q. 
\label{eBD1}
\end{equation}
The corresponding gauge transformation~\cite{DJM83,Li87,Adler94}, 
%[Chinese87], 
%given by~\cite{Adler94} [Chinese87], 
\begin{align}
& \left[
\begin{array}{c}
\widetilde{\Psi}_1 \\ 
\widetilde{\Psi}_2 \\
\end{array}
\right] := G (Q, \widetilde{R}; \mu)
 \left[
\begin{array}{c}
 \Psi_1  \\
 \Psi_2 \\
\end{array}
\right], \hspace{5mm}
G(Q, \widetilde{R}; \mu):= g(\zeta,\mu) 
\left[
\begin{array}{cc}
2 \mathrm{i} (\zeta-\mu)  I_l + Q \widetilde{R}  & - Q \\
- \widetilde{R} & I_m \\
\end{array}
\right], 
%\nonumber
\label{eBT-g1}
\end{align}
where 
%$g(\zeta,\mu)$ 
$g$ is an arbitrary (but 
%nonzero) 
not identically zero) 
scalar function 
of $\zeta$ and $\mu$,  
%of 
changes the original Lax-pair representation 
(\ref{NLS-UV}) with 
\mbox{$R =  \widetilde{R}_x - 2 \mathrm{i} \mu \widetilde{R} 
 + \widetilde{R} Q \widetilde{R}$} 
%is given 
%can be described 
%as~\cite{Adler94} [Chinese87]
%by rewriting 
into 
%the new one 
a new Lax-pair representation of the same form. 
%;
%, 
That is, 
%. That is, 
%\begin{subequations}
%\begin{align}
%& \left[
%\begin{array}{c}
% \Psi_1  \\
% \Psi_2 \\
%\end{array}
%\right]_x 
%= \left[
%\begin{array}{cc}
%-\mathrm{i}\zeta I_l & Q \\
% R & \mathrm{i}\zeta I_m\\
%\end{array}
%\right] 
%\left[
%\begin{array}{c}
% \Psi_1  \\
% \Psi_2 \\
%\end{array}
%\right], 
%\hspace{5mm} 
%R =  \widetilde{R}_x - 2 \mathrm{i} \mu \widetilde{R} 
% + \widetilde{R} Q \widetilde{R}, 
%\label{}
%\end{align}
%as~\cite{Adler94} [Chinese87]
\begin{align}
%\\[2mm]
&  \left[
\begin{array}{c}
 \widetilde{\Psi}_1 \\
 \widetilde{\Psi}_2 \\
\end{array}
\right]_x 
= \left[
\begin{array}{cc}
-\mathrm{i}\zeta I_l & \widetilde{Q} \\
 \widetilde{R} & \mathrm{i}\zeta I_m\\
\end{array}
\right] 
\left[
\begin{array}{c}
 \widetilde{\Psi}_1 \\
 \widetilde{\Psi}_2 \\
\end{array}
\right], \hspace{5mm}  
\widetilde{Q} = -Q_x - 2 \mathrm{i} \mu Q + Q \widetilde{R} Q, 
%\\
%& \widetilde{\Psi}_1 = 2 \mathrm{i} (\zeta-\mu) \Psi_1 + Q \widetilde{R} \Psi_1 - Q \Psi_2, 
%  \hspace{5mm}
%   \widetilde{\Psi}_2 = \Psi_2 - \widetilde{R} \Psi_1, 
\nonumber
\end{align}
%\end{subequations}
and 
similar 
%similarly 
%a similar result 
for the $t$-differentiation. 
%time part. 
%
%Let 
%Then, an elementary B\"acklund--Darboux transformation 
%for the nonreduced matrix NLS system is given by 
%Naturally, 
In the same manner, 
%way, 
the gauge transformation
%, 
%for 
%linear eigenfunction of 
%that 
%which preserves 
preserving the form of 
the adjoint Lax-pair representation 
(\ref{NLS-adUV})
%, 
is given by 
%reads 
%transforms as 
\begin{equation}
\left[
\begin{array}{cc}
\! \widetilde{\Phi}_1 \! & \! \widetilde{\Phi}_2 \!
\end{array}
\right] 
:= 
\left[
\begin{array}{cc}
\! \Phi_1 \! & \! \Phi_2 \!
\end{array}
\right] H (Q, \widetilde{R}; \mu), \hspace{5mm}
H(Q, \widetilde{R}; \mu):= h(\zeta, \mu) 
\left[
\begin{array}{cc}
 I_l  &  Q \\
 \widetilde{R} & 2 \mathrm{i} (\zeta-\mu) I_m + \widetilde{R} Q \\
\end{array}
\right], 
%\label{}
\nonumber
\end{equation}
where 
%$h(\zeta,\mu)$ 
$h$ is 
an arbitrary (but 
%nonzero) 
not identically zero) 
%a nonzero 
scalar function of $\zeta$ and $\mu$. 
%, 
%a scalar function and 
Note that 
\mbox{$
%\[
G(Q, \widetilde{R}; \mu) H(Q, \widetilde{R}; \mu) 
 = 2 \mathrm{i} (\zeta-\mu) g(\zeta,\mu) h(\zeta,\mu) I_{l+m}
%\]
$}. 
\end{proposition}

%Note that
%(\ref{eBD1}) 
%can  
%define the map 
%Actually, 
%Strictly speaking, 
The elementary B\"acklund--Darboux transformation 
\mbox{$(Q,R) \mapsto (\widetilde{Q}, \widetilde{R})$} 
defined by (\ref{eBD1}) 
is not 
%really 
%a unique map, 
%does not 
%cannot determine 
a map determining 
%that can determine 
\mbox{$(\widetilde{Q}, \widetilde{R})$} uniquely; indeed, 
%in the strict sense, 
%mapping, 
%because 
it 
%is not 
%explicit but 
%not 
%non-
%unique and 
%contains 
involves 
%some
additional 
%new 
%some arbitrary parameters
%arbitrary constants
%arbitrary 
free parameters, 
%; 
%indeed, 
%note that 
because 
each column 
%vector 
of the linear eigenfunction 
%of the Lax pair 
used in (\ref{eBD1}) 
%representation  
%we can take 
%can be 
%any 
is 
%, in general, 
an arbitrary 
linear combination of a fundamental set 
of 
%column-vector 
solutions. 
%as the linear eigenfunction considered. 
%Because 
%It is easy to see that 
The transformation 
%allows 
admits the following 
conservation law: 
\begin{align}
\widetilde{Q} \widetilde{R} - QR &=  
	- \left( Q \widetilde{R} \right)_x 
\nonumber \\ &= 
%QR 
	- \left( Q \left. \Psi_2 \Psi_1^{-1}\right|_{\zeta= \mu} \right)_x 
\nonumber \\ &= 
%QR 
	- \left( \left. \Psi_{1,x} \Psi_1^{-1}\right|_{\zeta= \mu} \right)_x. 
\nonumber 
\end{align}
Thus, if we can choose the linear eigenfunction 
%in such a way that 
such that \mbox{$\lim_{x \to - \infty} \Psi_{1,x} \Psi_1^{-1}$} and 
\mbox{$\lim_{x \to + \infty} \Psi_{1,x} \Psi_1^{-1}$} exist 
and $\Psi_{1,x} \Psi_1^{-1}$ does not oscillate extraordinarily 
rapidly as \mbox{$x \to \pm \infty$}, 
then $\widetilde{Q} \widetilde{R}$ 
%has the same 
and $QR$ 
exhibit 
%have 
the same asymptotic behavior  
%as \mbox{$x \to \pm \infty$}. 
at spatial infinity. 
%as $QR$. 
%Hence, 
In addition, there exists another conservation law:
\begin{align}
 \widetilde{R} \widetilde{Q}- R Q &=  
	- \left( \widetilde{R} Q \right)_x 
\nonumber \\ &= 
%QR 
	- \left( \left. \Psi_2 \Psi_1^{-1}\right|_{\zeta= \mu} Q \right)_x. 
\nonumber 
\end{align}
%This 
These conservation laws 
%provide 
%which also 
%provides 
give 
a useful hint 
%information 
on 
%the choice of $Q$ and $R$ 
%$QR$ 
%in order 
how to realize 
a 
%complex/
Hermitian conjugation reduction between 
$\widetilde{Q}$ and $\widetilde{R}$. 

%uniquely only if 
%unless 
%we 
%strictly 
%rigorously 
%specify 
%the boundary conditions on 
%because we have some freedom in choosing 
%the linear 
%eigenfunction; 
%considered is strictly specified; 
%;
% appropriately; 
%otherwise, 
%the \mbox{$(l+m) \times l$} 
%linear eigenfunction involves 
%$l$ 
%\mbox{$(l+m) \times l$} 
%arbitrary constants, so 
%so
%thus, 
%otherwise, 
%the quantity 
%ratio 
%\mbox{$\widetilde{R} = \left. \Psi_2 \Psi_1^{-1}\right|_{\zeta= \mu}$}
%contains 
%generally 
%involves 
%new 
%some 
%\mbox{$l-1$} 
%arbitrary parameters. 

%Considering 
%Taking the composition of 
%an 
%one inverse 
Combining the inverse of one 
Miura map 
%and 
%a 
with the other Miura map 
%in a different way, 
%an opposite way, 
%in the opposite direction, 
in the other way, 
%the other way around, 
we 
obtain 
%can 
%also 
%define 
%state 
another 
elementary 
%auto-
B\"acklund--Darboux transformation;
%the other 
by construction, 
it provides the inverse 
of the 
%previous 
elementary 
%auto-
B\"acklund--Darboux transformation 
defined in Proposition~\ref{prop2.1}
if appropriate boundary conditions are 
%assumed. 
imposed (cf.~\cite{Kono82}). 

\begin{proposition}
\label{prop2.2}
The spatial 
%(i.e., $x$-) 
part of 
an elementary auto-B\"acklund transformation 
%\mbox{$(Q,R) \mapsto (\widetilde{Q}, \widetilde{R})$}
%of
for 
%between two solutions, $(Q,R)$ and $(\widetilde{Q}, \widetilde{R})$, of 
the matrix NLS system (\ref{mNLS}), which connects 
%connecting 
%%relates 
two solutions $(Q,R)$ and $(\widehat{Q}, \widehat{R})$, 
%of 
%for 
%the matrix NLS system (\ref{mNLS})
%, \mbox{$(Q,R) \mapsto (\widetilde{Q}, \widetilde{R})$}, 
%can be 
%written 
%defined 
%expressed 
%described 
%as
is given by
%read
%~\cite{Kono82,Calo84} 
(see~\cite{Kono82,Calo84} for the scalar case 
and~\cite{SviYami91,Adler94} for the matrix case) 
%\begin{subequations}
%\label{eBT2}
\begin{align}
& Q = -\widehat{Q}_x - 2 \mathrm{i} \nu \widehat{Q} + \widehat{Q} R \widehat{Q}, 
\nonumber 
\\
& \widehat{R} = R_x - 2 \mathrm{i} 
%\zeta_1 
 \nu R 
 + R \widehat{Q} R. 
\nonumber
\end{align}
%\end{subequations}
%which connects 
%%%relates 
%two solutions $(Q,R)$ and $(\widetilde{Q}, \widetilde{R})$. 
%\mbox{$(\widetilde{Q}, \widetilde{R}) \mapsto (Q,R)$}
%It 
The 
transformation 
%map 
\mbox{$(Q,R) \mapsto (\widehat{Q}, \widehat{R})$} 
can be expressed 
explicitly 
%as an explicit mapping 
%rewritten 
in terms of a linear eigenfunction 
of the adjoint Lax-pair representation (\ref{NLS-adUV}) at \mbox{$\zeta=\nu$} 
as
%~\cite{Kono82,Calo84} 
%[Chinese87] 
\begin{equation}
\widehat{Q} = \left. \Phi_1^{-1} \Phi_2 \right|_{\zeta= \nu}, 
\hspace{5mm} \widehat{R} = R_x - 2 \mathrm{i} \nu R + R \widehat{Q} R. 
%\label{eBD2}
\nonumber
\end{equation}
The corresponding gauge transformation~\cite{Li87,Adler94,DJM83}, 
%[Chinese87], 
%given by~\cite{Adler94} [Chinese87], 
\begin{align}
& \left[
\begin{array}{c}
\widehat{\Psi}_1 \\ 
\widehat{\Psi}_2 \\
\end{array}
\right] := H (\widehat{Q}, R; \nu)
 \left[
\begin{array}{c}
 \Psi_1  \\
 \Psi_2 \\
\end{array}
\right], \hspace{5mm}
H(\widehat{Q}, R; \nu)= h(\zeta, \nu) 
\left[
\begin{array}{cc}
 I_l  &  \widehat{Q} \\
 R & 2 \mathrm{i} (\zeta-\nu) I_m + R \widehat{Q} \\
\end{array}
\right],
\nonumber
%\label{eBT-g2}
\end{align}
%of 
changes the original Lax-pair representation 
(\ref{NLS-UV}) with 
\mbox{$Q = -\widehat{Q}_x - 2 \mathrm{i} \nu \widehat{Q} + \widehat{Q} R \widehat{Q}$} 
%is given 
%can be described 
%as~\cite{Adler94} [Chinese87]
%by rewriting 
into 
%the new one 
a new Lax-pair representation of the same form. 
%, 
That is, 
\begin{align}
%\\[2mm]
&  \left[
\begin{array}{c}
 \widehat{\Psi}_1 \\
 \widehat{\Psi}_2 \\
\end{array}
\right]_x 
= \left[
\begin{array}{cc}
-\mathrm{i}\zeta I_l & \widehat{Q} \\
 \widehat{R} & \mathrm{i}\zeta I_m\\
\end{array}
\right] 
\left[
\begin{array}{c}
 \widehat{\Psi}_1 \\
 \widehat{\Psi}_2 
\\
\end{array}
\right], \hspace{5mm}  
\widehat{R} = R_x - 2 \mathrm{i} \nu R 
 + R \widehat{Q} R,
%\widetilde{Q} = -Q_x - 2 \mathrm{i} \mu Q + Q \widetilde{R} Q, 
%\\
%& \widetilde{\Psi}_1 = 2 \mathrm{i} (\zeta-\mu) \Psi_1 + Q \widetilde{R} \Psi_1 - Q \Psi_2, 
%  \hspace{5mm}
%   \widetilde{\Psi}_2 = \Psi_2 - \widetilde{R} \Psi_1, 
\nonumber
\end{align}
%\end{subequations}
and similar 
%similarly 
%a similar result 
for the $t$-differentiation. 
%time part. 
%
%Let 
%Then, an elementary B\"acklund--Darboux transformation 
%for the nonreduced matrix NLS system is given by 
%Naturally, 
%In the same manner, 
%way, 
The gauge transformation
%, 
%for 
%linear eigenfunction of 
%that 
%which preserves 
preserving the form of 
the adjoint Lax-pair representation 
(\ref{NLS-adUV})
%, 
is given by 
%reads 
%transforms as 
\begin{equation}
\left[
\begin{array}{cc}
\! \widehat{\Phi}_1 \! & \! \widehat{\Phi}_2 \!
\end{array}
\right] 
:= 
\left[
\begin{array}{cc}
\! \Phi_1 \! & \! \Phi_2 \!
\end{array}
\right] G (\widehat{Q}, R; \nu), \hspace{5mm}
G(\widehat{Q}, R; \nu)= g(\zeta,\nu) \left[
\begin{array}{cc}
2 \mathrm{i} (\zeta-\nu)  I_l + \widehat{Q} R & - \widehat{Q} \\
- R & I_m \\
\end{array}
\right].
%\label{}
\nonumber
\end{equation}
%where \mbox{$g(Q, \widetilde{R}; \mu) h(Q, \widetilde{R}; \mu) 
% = 2 \mathrm{i} (\zeta-\mu) I_{l+m}$}. 
%Let 
%Then, an elementary B\"acklund--Darboux transformation 
%for the nonreduced matrix NLS system is given by 
%\\
\hphantom{aa}
%\\
\end{proposition}

%Note that 
The elementary B\"acklund--Darboux transformation 
\mbox{$(Q,R) \mapsto (\widehat{Q}, \widehat{R})$} 
admits the following 
conservation laws: 
\begin{align}
\widehat{Q} \widehat{R} - QR &=  
	\left( \widehat{Q} R \right)_x 
\nonumber \\ &= 
%QR 
	- \left( \left. \Phi_1^{-1} \Phi_{1,x} \right|_{\zeta= \nu} \right)_x, 
\nonumber \\[2mm]
\widehat{R} \widehat{Q} - R Q &=  
	\left( R \widehat{Q} \right)_x 
\nonumber \\ &= 
%QR 
	\left( R \left. \Phi_1^{-1} \Phi_2 \right|_{\zeta= \nu} \right)_x. 
\nonumber 
\end{align}

\subsection{Binary B\"acklund--Darboux transformation}

We can 
%consider 
%compose 
further combine 
%the composition of 
the 
two 
elementary B\"acklund--Darboux transformations 
%given 
%introduced 
defined in Propositions~\ref{prop2.1} and 
%that in Proposition~
\ref{prop2.2} 
%at 
%(generally) 
%different 
%the 
%values $\mu$ and $\nu$ 
%of 
%with 
involving the B\"acklund parameters $\mu$ and $\nu$, respectively. 
%, $\mu$ and $\nu$, 
%as
%in the order
%We can consider 
%either 
%It can be shown 
%%checked directly 
%that 
In the following, we consider 
the composition 
in the order 
%\mbox{$
\begin{equation} 
(Q,R) \mapsto (\widetilde{Q}, \widetilde{R}) \mapsto 
(\skew{3}\widehat{\widetilde{Q}}, \skew{3}\widehat{\widetilde{R}})
%, 
\label{path1}
\end{equation}
%or equivalently, 
%which turns out to be equivalent to 
%It 
%Actually, it 
%is also possible to consider 
%it can be checked directly that 
%or 
%and 
because 
the composition 
in 
%%a 
%%different order, i.e., 
the opposite order
%, 
%: 
%as 
\begin{equation}
(Q,R) \mapsto (\widehat{Q}, \widehat{R}) \mapsto 
(\skew{3}\widetilde{\widehat{Q}}, \skew{3}\widetilde{\widehat{R}}) 
%. 
%\label{path2}
\nonumber
\end{equation}
%$}. 
%Thus, we obtain 
%Both 
%also 
provides the same final result; 
%as given below; 
%but 
%However, it can be shown 
%checked directly 
%that 
%the final result is 
%will be 
%the same~\cite{Kono82} [Kono79]. 
that is, 
%That is, 
the 
two 
elementary B\"acklund--Darboux transformations
%are commutative
commute with each other~\cite{Kono82,Konop3,Kono79}. 
%as long as 
%%we employ 
%a generic linear eigenfunction is used in 
%%defining 
%applying 
%%to defined 
%the second B\"acklund--Darboux transformation. 
%[Kono79]. 

Thus, 
%considering the composition 
%using 
by applying Propositions~\ref{prop2.1} and 
%that in Proposition~
\ref{prop2.2} 
%as 
%of 
in the order 
%as in 
(\ref{path1}), we obtain 
\begin{subequations}
\label{}
\begin{align}
\skew{3}\widehat{\widetilde{Q}} &= 
\left[ \Phi_1 (\nu) + \Phi_2 (\nu) \widetilde{R} \right]^{-1}
 \left[ \Phi_1 (\nu) Q + 2 \mathrm{i} (\nu - \mu ) \Phi_2 (\nu) 
 + \Phi_2 (\nu) \widetilde{R} Q  \right] 
 \nonumber \\ 
 &= Q +  2 \mathrm{i} (\nu -\mu) \left[ \Phi_1 (\nu) 
 + \Phi_2 (\nu) \Psi_2 (\mu) \Psi_1 (\mu)^{-1} \right]^{-1} \Phi_2 (\nu)
 \nonumber \\
 &= Q +   2 \mathrm{i} (\nu -\mu) \Psi_1 (\mu)  \left[ \Phi_1 (\nu) \Psi_1 (\mu)
 + \Phi_2 (\nu) \Psi_2 (\mu)  \right]^{-1} \Phi_2 (\nu), 
%\nonumber
\end{align}
and 
\begin{align}
\skew{3}\widehat{\widetilde{R}} &= \widetilde{R}_x - 2 \mathrm{i} \nu \widetilde{R} 
	+ \widetilde{R} \skew{3}\widehat{\widetilde{Q}} \widetilde{R}
 \nonumber \\
 &= R - 2 \mathrm{i} (\nu-\mu) \widetilde{R} 
	+ \widetilde{R} 
%\left( 
%\bigl( 
 \Bigl( \skew{3}\widehat{\widetilde{Q}} - Q \Bigr)
%\bigr)
%\right) 
 \widetilde{R}
 \nonumber \\
  &= R - 2 \mathrm{i} (\nu-\mu) \Psi_2 (\mu) \Psi_1 (\mu)^{-1} 
  \nonumber \\ 
 & \hphantom{=} \;\,
 + 2 \mathrm{i} (\nu -\mu) \Psi_2 (\mu)  \left[ \Phi_1 (\nu) \Psi_1 (\mu)
 + \Phi_2 (\nu) \Psi_2 (\mu)  \right]^{-1} \Phi_2 (\nu) \Psi_2 (\mu) \Psi_1 (\mu)^{-1} 
 \nonumber \\
    &= R - 2 \mathrm{i} (\nu-\mu) \Psi_2 (\mu) \left[ \Phi_1 (\nu) \Psi_1 (\mu)
 + \Phi_2 (\nu) \Psi_2 (\mu)  \right]^{-1} \Phi_1 (\nu).
%\nonumber 
\end{align}
\end{subequations}
%
%With the aid of 
%Using (\ref{eBT-g1}) and (\ref{eBT-g2}), 
The corresponding gauge transformation 
of the linear eigenfunction 
%that preserves the form of the Lax-pair representation 
%is 
%given 
%defined 
can be written 
as 
\begin{align}
\left[
\begin{array}{c}
\widehat{\widetilde{\Psi}}_1 \\ 
\widehat{\widetilde{\Psi}}_2 \\
\end{array}
\right] :=& \hspace{3pt} H(\skew{3}\widehat{\widetilde{Q}}, \widetilde{R}; \nu)
G (Q, \widetilde{R}; \mu)
\left[
\begin{array}{c}
 \Psi_1  \\
 \Psi_2 \\
\end{array}
\right] 
\nonumber \\
=& \hspace{3pt}
h(\zeta, \nu) g(\zeta,\mu) 
\left[
\begin{array}{cc}
 I_l  &  \skew{3}\widehat{\widetilde{Q}} \\
 \widetilde{R} & 2 \mathrm{i} (\zeta-\nu) I_m 
	+ \widetilde{R} \skew{3}\widehat{\widetilde{Q}} \\
\end{array}
\right]
%g(\zeta,\mu) 
\left[
\begin{array}{cc}
2 \mathrm{i} (\zeta-\mu)  I_l + Q \widetilde{R}  & - Q \\
- \widetilde{R} & I_m \\
\end{array}
\right] 
\left[
\begin{array}{c}
 \Psi_1  \\
 \Psi_2 \\
\end{array}
\right] 
\nonumber \\
=& \hspace{3pt} h(\zeta, \nu) g(\zeta,\mu) 
\left[
\begin{array}{cc}
2 \mathrm{i} (\zeta-\mu)  I_l + \Bigl( Q - \skew{3}\widehat{\widetilde{Q}} \Bigr) \widetilde{R}  
 & \skew{3}\widehat{\widetilde{Q}}  - Q \\
  2 \mathrm{i} (\nu-\mu) \widetilde{R} 
	+ \widetilde{R} \Bigl( Q-\skew{3}\widehat{\widetilde{Q}} \Bigr) \widetilde{R}
 & 2 \mathrm{i} (\zeta-\nu) I_m 
	+ \widetilde{R} \Bigl( \skew{3}\widehat{\widetilde{Q}} -Q \Bigr) \\
\end{array}
\right] 
\left[
\begin{array}{c}
 \Psi_1  \\
 \Psi_2 \\
\end{array}
\right]
\nonumber \\
\propto & \hspace{1pt}
%\hspace{3pt} 
%h(\zeta, \nu) g(\zeta,\mu) 
\left\{ I_{l+m} + \frac{\nu-\mu}{\zeta-\nu} \left[
\begin{array}{c}
 \Psi_1(\mu) \\
 \Psi_2 (\mu) \\
\end{array}
\right] \left[ \Phi_1 (\nu) \Psi_1 (\mu)
 + \Phi_2 (\nu) \Psi_2 (\mu)  \right]^{-1} 
\left[
\begin{array}{cc}
\! \Phi_1 (\nu) \! & \! \Phi_2 (\nu) \!
\end{array}
\right] 
\right\}
\left[
\begin{array}{c}
 \Psi_1  \\
 \Psi_2 \\
\end{array}
\right]. 
\nonumber 
\end{align}
%where 
Here, we 
%have omitted 
omit the unessential 
proportionality factor 
%overall 
%constant 
in the last line. 
%overall factor.
%Similarly, 
In 
%the same way, 
a similar way, 
%manner, 
the gauge transformation 
of the adjoint linear eigenfunction can be written as 
%is given as 
\begin{align}
& \hspace{1pt} \left[
\begin{array}{cc}
\! \widehat{\widetilde{\Phi}}_1 \! & \! \widehat{\widetilde{\Phi}}_2 \!
\end{array}
\right]
\nonumber \\
\propto & \hspace{1pt} 
\left[
\begin{array}{cc}
\! \Phi_1 \! & \! \Phi_2 \!
\end{array}
\right]
%h(\zeta, \nu) g(\zeta,\mu) 
\left\{ I_{l+m} + \frac{\mu-\nu}{\zeta-\mu} \left[
\begin{array}{c}
 \Psi_1(\mu) \\
 \Psi_2 (\mu) \\
\end{array}
\right] \left[ \Phi_1 (\nu) \Psi_1 (\mu)
 + \Phi_2 (\nu) \Psi_2 (\mu)  \right]^{-1} 
\left[
\begin{array}{cc}
\! \Phi_1 (\nu) \! & \! \Phi_2 (\nu) \!
\end{array}
\right] 
\right\}. 
\nonumber 
\end{align}

%
%Incidentally, 
Note that 
the commutativity 
%condition 
of the two 
elementary B\"acklund
%--Darboux 
transformations can be expressed 
%as 
%the commutativity of the two 
%gauge transformations
%the matrix exchange relation 
in the matrix product form (cf.~\cite{BPT83,QNCL,NCWQ}):
% [Adler94]):
%
%The commutativity conditions for the elementary B\"acklund--Darboux transformations 
%can be directly expressed as the matrix commutativity equations: 
%\begin{equation}
% h (\widehat{Q}, R; \nu) g(Q, \widetilde{R}; \mu) 
%=  g(Q, \widetilde{R}; \mu) h (\widehat{Q}, R; \nu), 
%\end{equation}
%: 
%equation
\begin{align}
& \left[
\begin{array}{cc}
 I_l  &  \skew{3}\widehat{\widetilde{Q}} \\
 \widetilde{R} & 2 \mathrm{i} (\zeta-\nu) I_m 
	+ \widetilde{R} \skew{3}\widehat{\widetilde{Q}} \\
\end{array}
\right]
%g(\zeta,\mu) 
\left[
\begin{array}{cc}
2 \mathrm{i} (\zeta-\mu)  I_l + Q \widetilde{R}  & - Q \\
- \widetilde{R} & I_m \\
\end{array}
\right] 
\nonumber \\
&= \left[
\begin{array}{cc}
2 \mathrm{i} (\zeta-\mu)  I_l + \widehat{Q} \skew{3}\widetilde{\widehat{R}}  & - \widehat{Q} \\
- \skew{3}\widetilde{\widehat{R}} & I_m \\
\end{array}
\right] 
\left[
\begin{array}{cc}
 I_l  &  \widehat{Q} \\
 R & 2 \mathrm{i} (\zeta-\nu) I_m + R \widehat{Q} \\
\end{array}
\right],
\nonumber 
\end{align}
which provides 
%a 
the nonlinear superposition principle 
%for 
%of 
%the B\"acklund
%%--Darboux 
%transformations 
for the solutions 
of 
the matrix NLS hierarchy. 
It 
%can be rewritten as 
%reduces to 
%is equivalent to 
can be described by 
the 
pair of 
relations~\cite{Kono82,Calo84,Adler94}:
%[Nijhoff86?, 
%[Adler94]: 
%equations: 
\begin{align}
&  \skew{3}\widehat{\widetilde{Q}}  = Q + 2 \mathrm{i} (\nu-\mu) \widehat{Q} 
	\left( I_m + \widetilde{R}  \widehat{Q}  \right)^{-1}, 
\nonumber \\
&  \skew{3}\widetilde{\widehat{R}} = R + 2 \mathrm{i} (\mu - \nu)
%\left( I_m + \widetilde{R}  \widehat{Q}  \right)^{-1} 
\widetilde{R}
\left( I_l + \widehat{Q} \widetilde{R} \right)^{-1}. 
\nonumber 
\end{align}
%
%If we interpret the hat and tilde as 
%In addition, 
%Actually, it is possible to interpret 
These relations 
%we can interpret this 
can be reinterpreted 
as a fully discrete integrable system defined 
on the two-dimensional 
lattice~\cite{DJM83,NQC83,QNCL,NCWQ}.
%of fully discrete 
%[NQC84]. 

The composition of the 
two 
elementary B\"acklund--Darboux transformations
%can be reformulated as 
%
%Summarizing the above results, 
%This provides 
%us with 
%can  
defines a binary B\"acklund--Darboux transformation~\cite{Sall82, MatSall91}
%,MatSall91}  
that can be 
%stated 
formulated as follows. 
%summarized as follows.  
%as stated in the following proposition. 

\begin{proposition}
\label{prop2.3}
%The 
%%binary B\"acklund--Darboux 
%transformation 
%map 
%\mbox{$(Q,R) \mapsto (\widehat{\widetilde{Q}}, \widehat{\widetilde{R}})$} 
%can be expressed 
%explicitly 
%as an explicit mapping 
%rewritten 
%in terms of 
%Using 
For 
%a pair of 
a 
%given 
linear eigenfunction 
satisfying 
%of the Lax-pair representation 
(\ref{NLS-UV}) at \mbox{$\zeta=\mu$} 
%as~\cite{Kono82,Calo84} 
%
%For a linear eigenfunction 
%of the Lax-pair representation (\ref{NLS-UV}) 
and 
%its adjoint 
%the 
an adjoint linear eigenfunction 
satisfying 
%of the Lax-pair representation 
(\ref{NLS-adUV}) at \mbox{$\zeta=\nu$}, 
% as 
we 
define a 
%the 
projection matrix $P(\mu,\nu)$ 
in the \mbox{$2 \times 2$} block-matrix 
form: 
%matrix $P$, respectively. 
%as 
%matrix 
\begin{align}
P(\mu,\nu) &:= 
\left[
\begin{array}{c}
 \Psi_1(\mu) \\
 \Psi_2 (\mu) \\
\end{array}
\right] \left[ \Phi_1 (\nu) \Psi_1 (\mu)
 + \Phi_2 (\nu) \Psi_2 (\mu)  \right]^{-1} 
\left[
\begin{array}{cc}
\! \Phi_1 (\nu) \! & \! \Phi_2 (\nu) \!
\end{array}
\right].
\nonumber 
%\\[1mm]
%&=: 
%\left[
%\begin{array}{cc}
%P_{11} & P_{12} \\
%P_{21} & P_{22} \\
%\end{array}
%\right] 
\end{align}
%
%which satisfies the 
%Note that 
Indeed, 
%it indeed satisfies 
\mbox{$P^2 = P$}. 
%\mbox{$\left[ P(\mu,\nu) \right]^2 = P(\mu,\nu)$}.
Then, 
the binary B\"acklund--Darboux transformation
%an elementary auto-B\"acklund transformation 
%\mbox{$(Q,R) \mapsto (\widetilde{Q}, \widetilde{R})$}
%of
\mbox{$(Q,R) \mapsto (\skew{3}\widehat{\widetilde{Q}}, \skew{3}\widehat{\widetilde{R}})$}
for 
%between two solutions, $(Q,R)$ and $(\widetilde{Q}, \widetilde{R})$, of 
the matrix NLS system (\ref{mNLS}) 
%, which 
%connects 
%connecting 
%%relates 
%two solutions $(Q,R)$ and $(\widehat{\widetilde{Q}}, \widehat{\widetilde{R}})$, 
%corresponds to the composition (\ref{path1}), 
can be expressed as 
(see~\cite{Chen1,
%Lamb74,
KW75} for the scalar case and 
\cite{Steu88,Park00,Park02,Forest,Wright} 
%[Wright00-01,02] 
for the vector case) 
%and [DegaLom] for the matrix case) 
%explicitly 
%as
%\begin{subequations}
%\label{}
\begin{align}
\skew{3}\widehat{\widetilde{Q}} &= 
%Q +  2 \mathrm{i} (\nu -\mu) 
%\Psi_1 (\mu)  \left[ \Phi_1 (\nu) \Psi_1 (\mu)
% + \Phi_2 (\nu) \Psi_2 (\mu)  \right]^{-1} \Phi_2 (\nu)
%\nonumber \\ &= 
Q +  2 \mathrm{i} (\nu -\mu) P_{12} (\mu,\nu), 
\nonumber \\
\skew{3}\widehat{\widetilde{R}} &= 
% R - 2 \mathrm{i} (\nu-\mu) \Psi_2 (\mu) \left[ \Phi_1 (\nu) \Psi_1 (\mu)
% + \Phi_2 (\nu) \Psi_2 (\mu)  \right]^{-1} \Phi_1 (\nu)
%\nonumber \\ &=
  R - 2 \mathrm{i} (\nu-\mu) P_{21} (\mu,\nu). 
\nonumber 
\end{align}
%\end{subequations}
Here, $P_{12}$ and $P_{21}$ denote 
the upper-right and lower-left block
%blocks 
of 
the 
%%\mbox{$2 \times 2$} block 
matrix 
$P$, 
respectively. 
The 
corresponding 
gauge transformation 
of the linear eigenfunction 
%that preserves the form of the Lax-pair representation 
%is 
%given 
%defined 
%can be written 
is given 
%in 
%using 
%in 
as 
\begin{align}
\left[
\begin{array}{c}
\widehat{\widetilde{\Psi}}_1 \\ 
\widehat{\widetilde{\Psi}}_2 \\
\end{array}
\right] 
\propto & 
%\hspace{1pt} 
%h(\zeta, \nu) g(\zeta,\mu) 
\left\{ I_{l+m} - \frac{\mu-\nu}{\zeta-\nu} P(\mu,\nu) 
\right\}
\left[
\begin{array}{c}
 \Psi_1  \\
 \Psi_2 \\
\end{array}
\right], 
\nonumber 
\end{align}
%where 
%Here, 
%we omit 
%have omitted 
%omit the unessential 
%the 
up to an overall constant, 
which
%this 
coincides with the form 
%{\em a la} 
%implied 
suggested by the Zakharov--Shabat dressing 
%operator
method~\cite{ZS79}. 
%form 
%form 
%(see also 
%, {\it e.g.},~
%also refrences in 
%[Caudrelier] and references therein).  
%factor.
% is unimportant and omitted.  
%in the last line. 
%overall factor.
%Similarly, 
%In 
%the same way, 
%a similar way, 
%manner, 
The gauge transformation 
of the adjoint linear eigenfunction 
%is 
%can be written as 
is given using the inverse of the 
above 
%ZS 
dressing operator 
as 
\begin{align}
%\hspace{3pt} 
\left[
\begin{array}{cc}
\! \widehat{\widetilde{\Phi}}_1 \! & \! \widehat{\widetilde{\Phi}}_2 \!
\end{array}
\right]
%\nonumber \\
\propto 
%& \hspace{3pt} 
\left[
\begin{array}{cc}
\! \Phi_1 \! & \! \Phi_2 \!
\end{array}
\right]
%h(\zeta, \nu) g(\zeta,\mu) 
\left\{ I_{l+m} - \frac{\nu-\mu}{\zeta-\mu} P(\mu,\nu) 
\right\}. 
\nonumber 
\end{align}
%\hphantom{a}
In the limit \mbox{$\nu \to \mu$}, 
%special case of \mbox{$\mu = \nu$}, 
the binary B\"acklund--Darboux transformation
reduces to 
%becomes 
the identity transformation 
%mapping 
%if 
as long as \mbox{$
%\left[ 
\Phi_1 (\mu) \Psi_1 (\mu)
 + \Phi_2 (\mu) \Psi_2 (\mu)  
%\right]^{-1} 
$} is 
%nonsingular. 
an invertible matrix (cf.~(\ref{inner})). 
%exists. 
%At 
\end{proposition}

%Note that 
Thus, under 
%a naive
a suitable 
%natural 
%the simplest 
choice of the (adjoint) linear eigenfunctions 
used in 
%the definitions of 
%to define 
the 
%B\"acklund--Darboux 
transformations, 
the two 
elementary B\"acklund--Darboux 
transformations 
defined in Propositions~\ref{prop2.1} and 
%that in Proposition~
\ref{prop2.2} 
are the inverse of each other 
at 
%if 
%the same value of the B\"acklund parameter 
\mbox{$\mu = \nu$}. 
%if we consider  
%%$\zeta$; 
%however, we can combine them at 
%different values of the parameter (section ), and then take 
%a coalescence limit to obtain a nontrivial result (section ). 
%it is possible to 
However, we can 
consider 
a more interesting 
%another 
%the 
choice of the (adjoint) linear eigenfunctions 
%case 
%where 
such that the binary B\"acklund--Darboux transformation 
reduces to a nontrivial transformation  
in the limit 
\mbox{$\nu \to \mu$}. 
%\mbox{$\mu \to \nu$}. 

\subsection{Coalescence 
%Nontrivial 
limit of the 
binary B\"acklund--Darboux transformation}
%For the product of 
%a pair of 
%For the pair of 
%an adjoint linear eigenfunction 
%satisfying 
%of the Lax-pair representation 
From 
%Using 
the Lax-pair representation (\ref{NLS-UV}) 
%at \mbox{$\zeta=\mu$},  
and 
%the adjoint Lax-pair representation 
its adjoint 
%problem 
(\ref{NLS-adUV}), 
%at \mbox{$\zeta=\nu$} 
%
%For a linear eigenfunction 
%of the Lax-pair representation (\ref{NLS-UV}) 
%and 
%its adjoint 
%the 
%a
%given 
%linear eigenfunction 
%satisfying 
%of the Lax-pair representation 
%the Lax-pair representation (\ref{NLS-UV}) at \mbox{$\zeta=\mu$},  
%as~\cite{Kono82,Calo84}
% as 
we 
%obtain the relations 
%satisfied by 
%find 
can show that the product of an adjoint linear eigenfunction 
and a linear eigenfunction satisfies 
(cf.~\cite{ZS1,AKNS74})
%he following relations: 
%: 
%
\begin{align}
%\left[
%\begin{array}{cc}
%\! \Phi_1 (\nu) \! & \! \Phi_2 (\nu) \!
%\end{array}
%\right]
%\left[
%\begin{array}{c}
% \Psi_1(\mu) \\
% \Psi_2 (\mu) \\
%\end{array}
%\right] = 
& \left\{ \frac{1}{\mathrm{i} (\nu-\mu)}\left[ \Phi_1 (\nu) \Psi_1 (\mu)
 + \Phi_2 (\nu) \Psi_2 (\mu)  \right] \right\}_x 
 = \Phi_1 (\nu) \Psi_1 (\mu) - \Phi_2 (\nu) \Psi_2 (\mu)
%, 
\nonumber 
\end{align}
and 
\begin{align}
& \left\{ \frac{1}{\mathrm{i} (\nu-\mu)}\left[ \Phi_1 (\nu) \Psi_1 (\mu)
 + \Phi_2 (\nu) \Psi_2 (\mu)  \right] \right\}_t  
\nonumber \\ =& \hspace{3pt} 2 (\mu + \nu) 
\left[ \Phi_1 (\nu) \Psi_1 (\mu) - \Phi_2 (\nu) \Psi_2 (\mu) \right] 
+ 2 \mathrm{i} \left[
 \Phi_1 (\nu) Q \Psi_2 (\mu) + \Phi_2 (\nu) R \Psi_1 (\mu) \right]. 
\nonumber
\end{align}
By construction, these two relations are compatible, i.e.
%, 
\begin{align}
& \left\{ \Phi_1 (\nu) \Psi_1 (\mu) - \Phi_2 (\nu) \Psi_2 (\mu) \right\}_t 
\nonumber \\ =& \hspace{3pt} \left\{ 2 (\mu + \nu) 
\left[ \Phi_1 (\nu) \Psi_1 (\mu) - \Phi_2 (\nu) \Psi_2 (\mu) \right] 
+ 2 \mathrm{i} \left[
 \Phi_1 (\nu) Q \Psi_2 (\mu) + \Phi_2 (\nu) R \Psi_1 (\mu) \right] \right\}_x. 
\nonumber
\end{align}
%Considering the limit, 
Thus, if we introduce 
%define 
an \mbox{$l \times l$} matrix $F$ as 
\begin{equation}
F(\mu) := 
\lim_{\nu \to \mu}
\frac{1}{\mathrm{i} (\nu-\mu)}\left[ \Phi_1 (\nu) \Psi_1 (\mu)
 + \Phi_2 (\nu) \Psi_2 (\mu)  \right], 
\label{F-def}
\end{equation}
it satisfies the
%compatible 
pair of relations: 
\begin{subequations}
\label{F-xt}
\begin{align}
& F_x = \Phi_1 (\mu) \Psi_1 (\mu) - \Phi_2 (\mu) \Psi_2 (\mu), 
%\nonumber 
\label{F-xt1}
\\[1mm] 
& F_t = 4 \mu F_x + 2 \mathrm{i} \left[
 \Phi_1 (\mu) Q \Psi_2 (\mu) + \Phi_2 (\mu) R \Psi_1 (\mu) \right]. 
%\nonumber
\label{F-xt2}
\end{align}
\end{subequations} 
%Note that 
Moreover, $F^{-1}$ should satisfy 
\[
F^{-1} 
%(\mu) 
\left[ \Phi_1 (\mu) \Psi_1 (\mu)
 + \Phi_2 (\mu) \Psi_2 (\mu)  \right] = 
\left[ \Phi_1 (\mu) \Psi_1 (\mu)
 + \Phi_2 (\mu) \Psi_2 (\mu)  \right] F^{-1} 
%(\mu) 
= O. 
\]
%
%In the following, 
In this paper, 
we 
%are not interested in rational solutions of the matrix NLS system 
%(\ref{mNLS}),  
%so we 
%only 
consider the simplest case 
%of 
where 
%of 
\mbox{$\Phi_1 (\mu) \Psi_1 (\mu) + \Phi_2 (\mu) \Psi_2 (\mu)
%=O
$} 
vanishes 
(cf.~(\ref{inner})). 
%, 
%%is zero, 
%because otherwise we need toconsider 
%%discuss, in general, 
%rational solutions of the matrix NLS system, in general. 
%in general. 
%, which are 
%%that are 
%outside the scope of this paper. 
Thus, 
%combining 
%in view of 
%the above results 
%with 
considering the
\mbox{$\nu \to \mu$} limit of 
Proposition~\ref{prop2.3}, 
%in the limit 
%\mbox{$\nu \to \mu$} provides 
%we can state 
we 
%obtain 
arrive at a nontrivial 
%an interesting 
%the 
limiting case of 
%next 
the binary B\"acklund--Darboux transformation.\footnote{Such a 
%A similar 
%This 
limiting procedure is 
already well-known 
%well known 
for the linear Schr\"odinger equation
%,  which is 
associated with the KdV equation 
(see
%, {\it e.g.}, 
\S 2.4 
%in
of~\cite{MatSall91}); 
%and 
%. 
%It 
%The 
the resulting transformation 
is sometimes 
%called 
referred to as 
%an 
the Abraham--Moses transformation~\cite{Ab80}
%after their work in 1980 
(see~\cite{RS13} for 
%more 
details).}
%following 
%proposition; 
Relavant results were obtained independently 
%previously 
by Degasperis and Lombardo~\cite{Dega09} 
using a different approach.  
%[DegaLom1,2]. 

\begin{proposition}
\label{prop2.4} 
%For 
We choose 
%a pair of 
a 
%given 
linear eigenfunction 
%satisfying 
%%of the Lax-pair representation 
%(\ref{NLS-UV}) 
%at \mbox{$\zeta=\mu$} 
%as~\cite{Kono82,Calo84} 
%
%For a linear eigenfunction 
%of the Lax-pair representation (\ref{NLS-UV}) 
and 
%its adjoint 
%the 
an adjoint linear eigenfunction 
satisfying 
(\ref{NLS-UV}) 
and
%of the Lax-pair representation 
(\ref{NLS-adUV}), respectively, 
%both 
at \mbox{$\zeta=\mu$}
%, we require that
in such a way that 
their \mbox{$(x,t)$}-independent 
product is 
equal to 
zero:  
%vanishes 
%(cf.~(\ref{inner})):
\[
\Phi_1 (\mu) \Psi_1 (\mu) + \Phi_2 (\mu) \Psi_2 (\mu)=O. 
\]
% as 
We 
can 
%define 
introduce 
%an \mbox{$l \times l$} matrix 
$F(\mu)$ 
that 
%solves 
satisfies 
%satisfying 
both
%as a simultaneous solution of 
%of 
%which satisfies 
%the pair of 
%compatible 
%relations 
%(\ref{F-xt}). 
(\ref{F-xt1}) and (\ref{F-xt2}), 
%which 
because they are 
%indeed 
compatible. 
Thus, $F(\mu)$ is defined up to the addition of 
an integration constant, 
which is an arbitrary \mbox{$l \times l$} 
%arbitrary 
matrix. 
%the addition of a constant matrix. 
%
%\mbox{$\left[ P(\mu,\nu) \right]^2 = P(\mu,\nu)$}.
Then, the limiting case of 
the binary B\"acklund--Darboux transformation, 
%an elementary auto-B\"acklund transformation 
%\mbox{$(Q,R) \mapsto (\widetilde{Q}, \widetilde{R})$}
%of
\mbox{$(Q,R) \mapsto (\breve{Q}, \breve{R})$}, 
for 
%between two solutions, $(Q,R)$ and $(\widetilde{Q}, \widetilde{R})$, of 
the matrix NLS system (\ref{mNLS}) 
%, which 
%connects 
%connecting 
%%relates 
%two solutions $(Q,R)$ and $(\widehat{\widetilde{Q}}, \widehat{\widetilde{R}})$, 
%corresponds to the composition (\ref{path1}), 
can be expressed as 
%(see~\cite{Chen1,
%%Lamb74,
%KW75} for the scalar case, 
%\cite{Steu88,Park00,Park02} [Wright00-01,02] for the vector case 
%and [DegaLom] for the matrix case) 
%explicitly 
%as
\begin{subequations}
\label{}
\begin{align}
%\skew{3}\widehat{\widetilde{Q}} 
\breve{Q} 
&= 
%Q +  2 \mathrm{i} (\nu -\mu) 
%\Psi_1 (\mu)  \left[ \Phi_1 (\nu) \Psi_1 (\mu)
% + \Phi_2 (\nu) \Psi_2 (\mu)  \right]^{-1} \Phi_2 (\nu)
%\nonumber \\ &= 
Q +  2  {\cal P}_{12} (\mu), 
%\Psi_1(\mu) F^{-1} (\mu) \Phi_2 (\mu), 
\\[0.5mm]
%\skew{3}\widehat{\widetilde{R}}
\breve{R} &= 
% R - 2 \mathrm{i} (\nu-\mu) \Psi_2 (\mu) \left[ \Phi_1 (\nu) \Psi_1 (\mu)
% + \Phi_2 (\nu) \Psi_2 (\mu)  \right]^{-1} \Phi_1 (\nu)
%\nonumber \\ &=
  R - 2  {\cal P}_{21} (\mu). 
%\Psi_2 (\mu) F^{-1} (\mu) \Phi_1 (\mu). 
%\nonumber 
\end{align}
\end{subequations}
Here, ${\cal P}_{12}$ and ${\cal P}_{21}$ are
%,  
%respectively
%, 
%denote 
the upper-right and lower-left block, respectively, 
%blocks 
of 
the nilpotent 
%%\mbox{$2 \times 2$} block 
matrix 
%${\cak P}$, 
%respectively,  
%A nilpotent matrix
%the 
%projection matrix 
${\cal P}(\mu)$ defined as 
%in the \mbox{$2 \times 2$} block-matrix 
%form: 
%matrix $P$, respectively. 
%as 
%matrix 
\begin{align}
{\cal P}(\mu) &:= 
\left[
\begin{array}{c}
 \Psi_1(\mu) \\
 \Psi_2 (\mu) \\
\end{array}
\right] F (\mu)^{-1}
\left[
\begin{array}{cc}
\! \Phi_1 (\mu) \! & \! \Phi_2 (\mu) \!
\end{array}
\right].
\nonumber 
%\\[1mm]
%&=: 
%\left[
%\begin{array}{cc}
%P_{11} & P_{12} \\
%P_{21} & P_{22} \\
%\end{array}
%\right] 
\end{align}
%
%which satisfies the 
Note that 
%Indeed, 
%it indeed satisfies 
\mbox{${\cal P}^2 = O$}. 
The 
corresponding 
gauge transformation 
%of the linear eigenfunction 
that preserves the form of the Lax-pair representation (\ref{NLS-UV})
%is 
%given 
%defined 
%can be written 
is given 
%in 
%using 
%in 
as 
\begin{align}
\left[
\begin{array}{c}
\breve{\Psi}_1 \\ 
\breve{\Psi}_2 \\
\end{array}
\right] 
\propto & 
%\hspace{3pt} 
%h(\zeta, \nu) g(\zeta,\mu) 
\left\{ I_{l+m} - \frac{\mathrm{i}}{\zeta-\mu} {\cal P}(\mu) 
\right\}
\left[
\begin{array}{c}
 \Psi_1  \\
 \Psi_2 \\
\end{array}
\right], 
\nonumber 
\end{align}
%where 
%Here, 
%we omit 
%have omitted 
%omit the unessential 
%the 
up to 
an overall 
%a proportionality 
constant. 
The gauge transformation 
that preserves the form of the adjoint Lax-pair representation (\ref{NLS-adUV})
%of the adjoint linear eigenfunction 
%proportionality constant
%is 
%can be written as 
is given using the inverse of the 
above 
%ZS 
%dressing 
operator 
as 
\begin{align}
%\hspace{3pt} 
\left[
\begin{array}{cc}
\! \breve{\Phi}_1 \! & \! \breve{\Phi}_2 \!
\end{array}
\right]
%\nonumber \\
\propto 
%& \hspace{3pt} 
\left[
\begin{array}{cc}
\! \Phi_1 \! & \! \Phi_2 \!
\end{array}
\right]
%h(\zeta, \nu) g(\zeta,\mu) 
\left\{ I_{l+m} + \frac{\mathrm{i}}{\zeta-\mu} {\cal P}(\mu) 
\right\}. 
\nonumber 
\end{align}
%\vphantom{At}
\hphantom{At}
%At 
\end{proposition}
The validity of Proposition~\ref{prop2.4} 
can be 
checked 
%confirmed 
by a direct calculation. 
%computation. 
Note that 
%%In the above proposition, 
%we have introduced 
$F(\mu)$ 
%as a 
%%the 
%solution of (\ref{F-xt}), 
satisfying (\ref{F-xt}) 
%%instead of the limit (\ref{F-def}), 
%which 
can be informally written as 
\begin{align}
F(\mu) &= 2C+ \int
%^x 
\left[ \Phi_1 (\mu) \Psi_1 (\mu) - \Phi_2 (\mu) \Psi_2 (\mu)
	\right] \mathrm{d} x
	\nonumber \\
	&= 2 \left[ C + \int
%^x 
\Phi_1 (\mu) \Psi_1 (\mu) \hspace{1pt} \mathrm{d} x \right], 
	\nonumber
\end{align}
where $C$ is 
%with 
an  
arbitrary 
\mbox{$l \times l$} constant matrix. 
%; 
%$C$; 
More precisely, the time independence of $C$ should be 
checked by using (\ref{F-xt2}); 
%As a matter of fact, 
%In practice, we need only 
%%have to 
%choose 
however, this process can be skipped by choosing 
the linear eigenfunction and the 
adjoint linear eigenfunction 
so that 
%in such a way that 
the 
right-hand sides of 
%both 
(\ref{F-xt1}) and (\ref{F-xt2}) 
%(\ref{F-xt}) 
decay rapidly 
as \mbox{$x \to -\infty$} (or \mbox{$
%x \to 
+\infty$}). 
This expression involves one integration, 
but it 
is generally 
%, in general, 
%more useful 
much 
easier to compute 
%for practical computations 
than 
%na\"ively 
taking the limit 
%as in 
%according to 
%in 
%na\"ively 
following 
the original definition (\ref{F-def}) 
(cf.~\cite{Park00,Zeng05}). 
%based on 
%that requires us 
%%one 
%to take 
%%consider 
%compute 
%through 
%using 
%the limit. 
%(\ref{F-def}). 
%The validity of Proposition~\ref{prop2.4} 
%can be checked 
%%confirmed 
%by a direct computation. 
%without any recourse to the limiting 
%definition of $F(\mu)$ (\ref{F-def}). 
%
%\newpage

%\subsection{Vector dark soliton} 
%dark-soliton solutions of vector NLS}

%\subsection{Matrix dark soliton}
%Dark-soliton solutions of matrix NLS}

\subsection{Seed solution}
\label{sec2.6}

A general plane-wave solution of the nonreduced 
%vector/
matrix NLS system (\ref{mNLS}) 
%that will be used later 
is given by 
\begin{align}
Q = P_1 
\mathrm{e}^{-2\mathrm{i} t AB} A  
\mathrm{e}^{\mathrm{i} x \Gamma - \mathrm{i} t \Gamma^2} P_2, \hspace{5mm} 
R= P_2^{-1} \mathrm{e}^{-\mathrm{i} x \Gamma + \mathrm{i} t \Gamma^2}
	 B \mathrm{e}^{2\mathrm{i} t AB} P_1^{-1}
, 
\label{plane2}
\end{align}
where 
%$P_1$, $P_2$, 
$\Gamma$, $A$, $B$, $P_1$ and $P_2$ are constant matrices; 
$P_1$ is an \mbox{$l \times l$} matrix,
$\Gamma$ and $P_2$ are \mbox{$m \times m$} matrices, 
$A$ is an \mbox{$l \times m$} matrix 
and $B$ is an \mbox{$m \times l$} matrix. 
Actually, 
%In fact, 
%we can remove 
$P_1$ and $P_2$ can be 
%absorbed
removed 
%are unnecessary, because they can be absorbed 
%by redefining  
by redefining 
%by a suitable redefinition of 
$\Gamma$, $A$ and $B$; 
%however, 
%However, 
%it is more convenient to 
%we leave them as they are, 
%%$P_1$ and $P_2$ 
%as in (\ref{plane2}), 
%because 
%Note, however, that 
note, however, that 
%$P_1$ and $P_2$ 
they 
can be used to 
%assume 
%write 
%put 
cast 
$\Gamma$ and $AB$ in Jordan 
%their 
%canonical 
normal form. 
%forms. 
%In particular, we are interested in the Hermitian conjugation reduction (\ref{Herm}); 
%The 
%In addition, 
A general plane-wave solution 
of the matrix NLS equation (\ref{redNLS}) 
%in this case 
%satisfying the Hermitian conjugation reduction (\ref{Herm}) 
can be 
%is 
obtained by 
%satisfying the Hermitian conjugation reduction (\ref{Herm})
imposing the reduction conditions 
\[
P_1^\dagger \Omega P_1 = \Omega, \hspace{5mm}
P_2 \Sigma P_2^\dagger = \Sigma, 
\hspace{5mm} \Sigma \Gamma^\dagger \Sigma^{-1} = \Gamma, \hspace{5mm}
B = \Sigma A^\dagger \Omega
%, 
\]
on (\ref{plane2}) so that 
%satisfying 
the Hermitian conjugation reduction (\ref{Herm}) is realized. 
%To satisfy the third condition, 
The third condition is satisfied 
for 
%the diagonal form of 
a diagonal matrix $\Sigma$ 
%if we simply set 
by setting 
\begin{equation}
\Gamma = \mathrm{diag} (\gamma_1, \gamma_2, \ldots, \gamma_m), 
\nonumber
\end{equation}
where $\gamma_j$ are real constants. 
%We will impose 
%use 
%As seed solutions, 
%In section~\ref{sec3}, we 
%use 
%apply an elementary B\"acklund--Darboux transformation to  
%the nonreduced plane-wave solution; 
%, while 
In 
%
%In 
sections~\ref{sec5}
%--\ref{sec4}, 
and~\ref{sec4}, 
we 
%will assume 
%%consider 
%these reduction conditions 
%will be 
%%assumed 
%used 
%in sections~\ref{sec4} and~\ref{sec5}
%, 
%but 
%wherein 
%so that 
%to 
apply a binary B\"acklund--Darboux transformation 
and its limiting version 
%in an effective manner. 
to 
%this 
the 
reduced plane-wave solution, 
%preserving 
which can maintain 
%a 
the 
Hermitian conjugation reduction. 
%way 
%can preserve the Hermitian conjugation reduction. 
%case 
%will be applied. 
%considered. 
%, but in the next section~\ref{sec3}, 
%we need to consider 
%In this section and the next section, we do not use 
%assume 
%these conditions directly; 
%they will be used 
%in sections~\ref{sec4} and~\ref{sec5}. 

%%The starting point is 
%Recall that 
%a
%%the 
%general plane-wave solution 
%%(\ref{plane1}) 
%of the nonreduced 
%%vector/
%matrix NLS system (\ref{mNLS}) is 
%given by (\ref{plane1}). 
To obtain the linear eigenfunction of 
the Lax-pair representation 
(\ref{NLS-UV}) 
%for 
%the case of 
%in the case where the potentials $Q$ and $R$ are given by 
with $Q$ and $R$ given by
the general plane-wave solution (\ref{plane2}), 
we consider a gauge transformation, 
\begin{align}
%\mathscr{\Psi} 
%{\mathit \Psi}
\Psi_1 = P_1 \mathrm{e}^{-2\mathrm{i} t AB} \boldsymbol{\mathit \Psi}_1, 
\hspace{5mm} 
\Psi_2 = \mathrm{i} P_2^{-1} \mathrm{e}^{-\mathrm{i} x \Gamma + \mathrm{i} t \Gamma^2} 
\boldsymbol{\mathit \Psi}_2. 
%\mathscr{ABC}
%{\mathcal \Psi} {\mathbf \Psi} \boldsymbol{\mathit \Psi}
\label{gauge1}
\end{align}
%to change 
%make 
%so that 
Then, 
%the Lax matrices into constant matrices 
the Lax-pair representation 
at \mbox{$\zeta = \mu$} 
%in this case 
%can be rewritten in 
reduces to 
the 
%simple 
%constant matrix 
form: 
%as 
%
\begin{subequations}
\label{NLS-UV2}
%\begin{equation}
%\left\{ \begin{array}{l}
\begin{align}
& \left[
\begin{array}{c}
 \boldsymbol{\mathit \Psi}_1  \\
 \boldsymbol{\mathit \Psi}_2 \\
\end{array}
\right]_x 
= \mathrm{i} \left[
\begin{array}{cc}
-\mu I_l & A \\
-B & \mu I_m + \Gamma \\
\end{array}
\right] 
\left[
\begin{array}{c}
 \boldsymbol{\mathit \Psi}_1  \\
 \boldsymbol{\mathit \Psi}_2 \\
\end{array}
\right],
\label{NLS-U2}
\\[1.5mm]
& \left[
\begin{array}{c}
 \boldsymbol{\mathit \Psi}_1  \\
 \boldsymbol{\mathit \Psi}_2 \\
\end{array}
\right]_t 
= \mathrm{i} \left[
\begin{array}{cc}
-2\mu^2 I_l + AB & 2 \mu A -A\Gamma \\
 -2 \mu B + \Gamma B & 2 \mu^2 I_m + BA - \Gamma^2 \\
\end{array}
\right]
\left[
\begin{array}{c}
 \boldsymbol{\mathit \Psi}_1  \\
 \boldsymbol{\mathit \Psi}_2 \\
\end{array}
\right]. 
\label{NLS-V2}
\end{align}
%\end{array}\right.
%\end{equation}
\end{subequations}
%Here, no assumption is made on the relation between $A$ and $B$. 
%at this stage. 
%are 
%a priori 
%mutually 
%unrelated constant matrices.
%
The general solution of this linear system 
%is 
can be 
%written 
expressed 
using the matrix exponential 
of an \mbox{$(l+m) \times (l+m)$} matrix 
as 
\begin{align}
\left[
\begin{array}{c}
 \boldsymbol{\mathit \Psi}_1 \\
 \boldsymbol{\mathit \Psi}_2 
%(\zeta) 
\\
\end{array}
\right] = \mathrm{e}^{
%\mathrm{i} x \boldsymbol{Z} 
%+ \mathrm{i}  t (-\boldsymbol{Z}^2 + 2 \zeta \boldsymbol{Z} + \zeta^2 I)
\mathrm{i} x (\boldsymbol{Z} + \mu I
%_{l+m}
%_{l+m} 
) + \mathrm{i}  t (-\boldsymbol{Z}^2 + 2 \mu^2 I
%_{l+m}
%_{l+m} 
)}
\left[
\begin{array}{c}
 {\mathscr C}_1 \\
 {\mathscr C}_2 \\
\end{array}
\right], 
%{Z x + 2 \zeta Z^2 t},
%\exp \left( Z x \right)
\hspace{5mm}
\boldsymbol{Z} (\mu) 
%Z
:= \left[
\begin{array}{cc}
-2 \mu I_l & A \\
 -B & \Gamma \\
\end{array}
\right].  
%\nonumber
\label{mat-ex1}
\end{align}
%
%where 
Here, 
$I$ denotes the 
%\mbox{$(l+m) \times (l+m)$}  
identity matrix 
%of size \mbox{$(l+m)$} 
$I_{l+m}$ and 
${\mathscr C}_1$ and ${\mathscr C}_2$ are 
\mbox{$l \times l$} and \mbox{$m \times l$} constant matrices, respectively. 
%If 
Assume that 
%we can diagonalize
the constant matrix \mbox{$\boldsymbol{Z} (\mu)$} can be diagonalized 
as 
\[
\boldsymbol{Z} X
%(\zeta) 
 = X \Lambda, \hspace{5mm} \Lambda 
	:= \mathrm{diag} (\lambda_1, \lambda_2, \ldots, \lambda_{l+m}), 
\]
using a nonsingular matrix $X$ whose 
$j$th column 
%vector 
is 
%the 
an eigenvector of  \mbox{$\boldsymbol{Z} (\mu)$} 
with an 
%the 
eigenvalue $\lambda_j$. 
Then, by setting 
\[
\left[
\begin{array}{c}
 {\mathscr C}_1 \\
 {\mathscr C}_2 \\
\end{array}
\right] = X \left[
\begin{array}{c}
 {\mathscr U}_1 \\
 {\mathscr U}_2 \\
\end{array}
\right], 
\]
we can compute 
%express 
the matrix exponential in (\ref{mat-ex1}) 
%in closed form. 
%compute 
%the right-hand side in a more form. 
explicitly. 
Thus, in view of (\ref{gauge1}), we obtain 
\begin{align}
\left[
\begin{array}{c}
 \Psi_1 
%(\zeta) 
\\
 \Psi_2 
%(\zeta) 
\\
\end{array}
\right] = 
%\mathscr{\Psi} 
%{\mathit \Psi}
\left[
\begin{array}{cc}
P_1 \mathrm{e}^{-2\mathrm{i} t AB}& O \\
O & \mathrm{i} P_2^{-1} \mathrm{e}^{-\mathrm{i} x \Gamma + \mathrm{i} t \Gamma^2} \\ 
%\mathscr{ABC}
%{\mathcal \Psi} {\mathbf \Psi} \boldsymbol{\mathit \Psi}
\end{array}
\right] X 
\mathrm{e}^{\mathrm{i} x (\Lambda + \mu I
%_{l+m} 
) + \mathrm{i}  t (-\Lambda^2 + 2 \mu^2 I
%_{l+m} 
)}
\left[
\begin{array}{c}
 {\mathscr U}_1 \\
 {\mathscr U}_2 \\
\end{array}
\right]. 
\label{}
\end{align}
In a similar 
%manner, 
way, 
%it is possible to 
%In the same way, 
we can also 
obtain 
%To obtain 
the linear eigenfunction of 
the adjoint Lax-pair representation 
(\ref{NLS-adUV}) 
with the potentials given by 
%for the case of 
%in the case where the potentials $Q$ and $R$ are given by 
%for the general 
%the plane-wave solution 
(\ref{plane2}). 
%can be 
%%we consider a gauge transformation, 
%expressed as 
%It turns out 
%Now, 
%It 
%is easy to see 
%becomes 
%is now clear that 
%Actually, 
The constant matrices $P_1$ and $P_2$ play no essential role 
in the 
%elementary 
%binary 
B\"acklund--Darboux transformations, 
%; 
%that is, 
%In other words, 
i.e., 
the order of applying a B\"acklund--Darboux transformation 
and 
%considering 
%the symmetry group 
a 
constant 
linear transformation using $P_1$ and $P_2$ 
%belonging to 
%that belongs to 
%from 
%the symmetry group 
%constant 
%linear transformations that leave the equations of motion invariant 
%commute with 
%does not manner. 
is irrelevant. 
%can be changed. 
Thus, 
%For brevity, 
in 
%In 
the following, we usually 
drop $P_1$ and $P_2$, 
%for brevity, 
%simplicity, 
keeping in mind 
%that 
%a group of linear transformations 
%any 
%a 
%constant 
%linear transformation 
%can be applied 
%to 
%any obtained solution 
%the 
%obtained 
%solutions 
%of the vector/matrix NLS equation
%if it 
the invariance 
%the symmetry group 
%which 
%that leaves  
%leaves 
%the multicomponent 
%vector/matrix 
%NLS system 
%equation
%that leave 
%the equation of motion 
%form-
%invariant.  
%can be considered 
%applied to 
%any obtained solution 
of the 
%vector/
matrix NLS system 
%equation. 
under the action of the symmetry group. 
%can be considered. 

%Thus, 
To express the matrix exponential explicitly, we need to diagonalize
%\footnote{In 
%this paper, we do not 
%%proactively 
%discuss the non-generic case where 
%%of a non-diagonalizable matrix 
%\mbox{$\boldsymbol{Z} (\zeta)$} is };  
the constant matrix \mbox{$\boldsymbol{Z} (\mu)$}; 
%in 
throughout this paper, 
we do not separately 
%proactively 
discuss the non-generic case where 
%of a non-diagonalizable matrix 
\mbox{$\boldsymbol{Z} (\mu)$} is 
%not 
non-diagonalizable (see, {\em e.g.},~\cite{Dega13}).   
%, which 
%reduces to 
For a 
%given set 
general choice 
of $\mu$, $\Gamma$, $A$ and $B$, 
this requires 
%us 
to solve the characteristic equation for \mbox{$\boldsymbol{Z} (\mu)$} 
of degree \mbox{$l+m$}. 
Thus, even for the 
%simplest nontrivial 
simple case of the two-component vector 
NLS equation, we generally need to resort to 
Cardano's formula~\cite{Park00,Wright}, 
%[Wright00-1], 
which is 
%apparently 
too complicated for practical purposes. 
%computations. 
%(also see~\cite{Dub88})
%In the following, we will 
%show that 
%consider 
However, as we will see below, 
%it is possible to introduce 
with a suitable re-parametrization of $\mu$, $\Gamma$, 
%$\Gamma$, 
$A$ and $B$, 
%and , 
%which can 
%to 
it is possible to 
%reduce the degree of the characteristic equation for \mbox{$\boldsymbol{Z} (\zeta)$} 
%by at least one. 
%and 
compute 
%this 
the 
matrix exponential 
explicitly 
without using Cardano's formula.
%, etc.
%for 
%%the most interesting cases such as 
%the two-component vector NLS equation, etc.   
%so 
%it is possible to  
%a simpler and more explicit solution 
%for \mbox{$\boldsymbol{Z} (\zeta)$} 

\section
%{Elementary B\"acklund--Darboux transformation: 
{Dark-soliton solutions}
\label{sec3}

In this section, we 
%formulate two 
apply the 
%an 
elementary B\"acklund--Darboux transformation 
defined in Proposition~\ref{prop2.1} 
%for 
%based on the Lax-pair representation for 
to the 
%vector/
matrix NLS 
%equation 
system 
(\ref{mNLS}), 
which 
includes 
%contains 
the vector NLS system as a special case. 
%described in subsection~\ref{sec2.1}. 
%Thus, to 
%To avoid confusion, 
%we denote 
%write 
%fix 
%write 
%real-valued. 
%The application of 
%each 
%either elementary B\"acklund--Darboux transformation 
%to 
%the 
%Starting with 
Taking a 
%general 
plane-wave solution as the seed solution, we 
%can 
obtain 
%provides 
%gives 
%a solution, which can be reduced 
%reduces 
%to a 
dark-soliton solutions  
of the multicomponent NLS equations 
with a self-defocusing or mixed focusing-defocusing nonlinearity.  
%by 
%a suitable choice of 
%%imposing suitable restrictions on 
%the parameters. 
The Hermitian conjugation reduction between the 
%two 
transformed potentials 
%dependent variables 
$\widetilde{Q}$ and $\widetilde{R}$ can be 
%is 
realized 
%by 
through a suitable choice 
of the parameters 
in the solution.  
%Note that 
In particular, 
the B\"acklund parameter 
%is 
%denoted as 
%by 
$\mu$, 
%$\zeta$ as $\mu$. 
%instead of 
which 
%comes from 
has its origin in 
the spectral parameter 
%not 
$\zeta$ in the Lax-pair representation, 
is 
%required to 
%should 
%be real-valued. 
restricted to be real. 

\subsection
%{Machinery}
%{Method}
%{General setup}
{Vector dark soliton without internal degrees of freedom}
\label{sec3.1}

We first 
%set \mbox{$l=1$} and 
consider 
the nonreduced 
%scalar/
vector 
%case of the nonreduced matrix 
NLS system, 
which can be obtained from (\ref{mNLS})
by setting \mbox{$l=1$}, \mbox{$Q=\vt{q}$} and \mbox{$R=\vt{r}^T$}, 
%in (\ref{mNLS}), 
i.e.\ 
%[(\ref{mNLS}) with \mbox{$Q=\vt{q}$} and \mbox{$R=\vt{r}^T$}], i.e.\ 
%: 
\begin{subnumcases}{\label{vNLS}}
{}
%\begin{equation} \left\{ \begin{split}
\label{vNLS1}
 \mathrm{i} \vt{q}_t + \vt{q}_{xx} - 2 \sca{\vt{q}}{\vt{r}}
	\vt{q} = \vt{0}, 
%\hspace{13mm}
\\[0.5mm]
%[1mm]
\label{vNLS2}
 \mathrm{i} \vt{r}_t - \vt{r}_{xx} + 2 \sca{\vt{r}}{\vt{q}}
	\vt{r} = \vt{0}.
%\hspace{13mm}
%\notag \\
%\end{split} \right. \end{equation}
\end{subnumcases}
%
%The $m$-component 
%vector NLS system 
%equation (\ref{rvNLS}) 
%(\ref{vNLS}) for row vectors 
Here, 
%$\vt{q}$ and $\vt{r}$ are $m$-component row vectors:  
\mbox{$\vt{q}:=(q_1, \ldots, q_m)$} and \mbox{$\vt{r}:=(r_1, \ldots, r_m)$} 
are $m$-component row vectors. 
%Obviously, 
Note that 
the simplest case of \mbox{$m=1$} 
%provides 
corresponds to 
%gives 
the scalar NLS system: 
\begin{subnumcases}{\label{sNLS}}
{}
%\begin{equation} \left\{ \begin{split}
\label{sNLS1}
 \mathrm{i} q_t + q_{xx} - 2 q^2 r = 0, 
\\[0.5mm]
%[1mm]
\label{sNLS2}
 \mathrm{i} r_t - r_{xx} + 2 r^2 q = 0.
\end{subnumcases}
A general plane-wave solution~\cite{Mak82} of 
the vector NLS system 
%equation 
(\ref{vNLS}) 
%admits the general plane-wave solution~\cite{Mak82} 
%(cf.~(\ref{plane2})):
%of 
%can be written explicitly as 
is 
%given in the form 
(cf.~(\ref{plane2}))
%, i.e.
%
\begin{equation}
%\vt{q} 
q_j (x,t) = 
%\mathrm{e}^{-2\mathrm{i} t \sum_{k=1}^m a_k b_k 
%%\sca{\vt{\scriptstyle a}}{\vt{b}}
%} 
%\vt{a} 
a_j \mathrm{e}^{\mathrm{i}  \gamma_j x - \mathrm{i} \left( \gamma_j^2 
 + 2 \sum_{k=1}^m a_k b_k \right) t}, 
\hspace{5mm}
 r_j (x,t) = b_j \mathrm{e}^{-\mathrm{i}  \gamma_j x + \mathrm{i} \left( \gamma_j^2 
 + 2 \sum_{k=1}^m a_k b_k \right) t}, 
\label{vector-plane}
\end{equation}
where \mbox{$j=1, 2, \ldots, m$}.  
%Here, 
%Note that 
%%Obviously, 
%(\ref{vNLS}) is invariant under 
%%any constant 
%any 
%%constant 
%linear transformation 
%%transformations 
%of the form 
%\mbox{$\vt{q} \mapsto \vt{q} P$}, \mbox{$\vt{r}^T \mapsto P^{-1} \vt{r}^T$} 
%with an invertible \mbox{$m \times m$} constant 
%matrix $P$; 
%%for any constant nonsingular \mbox{$m \times m$} matrix $P$, 
%%but 
%however, 
%for brevity
%%, 
%we suppress 
%%omit 
%%the 
%this 
%freedom 
%in the following 
%%discussion. 
%computation. 
%of applying constant 
%unitary 
%linear transformations. 
%
%To realize a Hermitian conjugation reduction between $\vt{q}$ and $\vt{r}$ 
%after the application of an elementary B\"acklund--Darboux transformation, 
%later on 
%(after the application of an elementary B\"acklund--Darboux transformation), 
We 
%mainly 
consider the generic case where 
%We assume that 
all $\gamma_j$ are pairwise distinct. 
%; the special cases where some $\gamma_j$ coincide with each other 
%can, in principle, be recovered 
%%restored
%by taking a coalescence limit. 
%real numbers and all $a_k b_k$ are nonzero; 
%these conditions are 
%to realize a Hermitian conjugation reduction between $\vt{q}$ and $\vt{r}$ 
%%after the application of an elementary B\"acklund--Darboux transformation, 
%later on 
%(after the application of an elementary B\"acklund--Darboux transformation), 
%%We 
%
%Thus, 
The 
%\mbox{$(m+1) \times (m+1)$} 
matrix 
%$\boldsymbol{Z} (\zeta)$ in (\ref{mat-ex1}) 
$\boldsymbol{Z} (\mu)$ (cf.~(\ref{mat-ex1})) 
in this case 
is given by
%as 
%by the \mbox{$(m+1) \times (m+1)$} matrix  
\begin{align}
%\det 
\boldsymbol{Z} (\mu
%\zeta
) = 
\left[
\begin{array}{ccccc}
-2 
%\zeta 
\mu & a_1 & \cdots & a_m \\
 -b_1 & \gamma_1 &  & \kern-6pt\raisebox{-6pt}{\mbox{{\Large $O$}}} \\
% -c_2 & 0 & \zeta a_2 & \vdots \\
 \vdots &  & \ddots & \\
 -b_m & \mbox{{\Large $O$}} &  & \gamma_m \\
\end{array}
\right].  
\label{Z-vector}
%\nonumber 
\end{align}
The 
%Its 
%the 
characteristic equation (up to an overall sign) 
%of which 
%of $\boldsymbol{Z} (\zeta)$ 
reads 
%is 
%
\begin{align}
\det \left( \lambda I - \boldsymbol{Z} 
%(\zeta) 
\right) = (\lambda + 2\mu 
%\zeta
) \prod_{j=1}^m (\lambda -\gamma_j) 
 + \sum_{k=1}^m a_k b_k \dprod{j=1}{j \neq k}^m (\lambda - \gamma_j) =0, 
\label{chara1}
\end{align}
%or equivalently 
which can be rewritten as 
(cf.~\cite{Dub88})
\begin{align} 
\lambda + \sum_{k=1}^m \frac{a_k b_k}{\lambda -\gamma_k} = -2\mu. 
%\zeta. 
\label{chara2}
\end{align}
For each eigenvalue $\lambda$ that solves 
%satisfies 
%the above characteristic equation, 
(\ref{chara2}), 
the corresponding eigenvector is given 
up to 
%a proportionality factor 
an overall constant 
as  
\begin{align}
%\det 
\boldsymbol{Z} (\mu
%\zeta
)
\left[
\begin{array}{c}
1 \\
-\frac{b_1}{\lambda -\gamma_1} \\
\vdots \\
-\frac{b_m}{\lambda -\gamma_m} \\
\end{array}
\right] = \lambda
\left[
\begin{array}{c}
1 \\
-\frac{b_1}{\lambda -\gamma_1} \\
\vdots \\
-\frac{b_m}{\lambda -\gamma_m} \\
\end{array}
\right]. 
\nonumber
\end{align}
To obtain 
%the genuine 
%a 
dark-soliton 
%solution 
solutions 
satisfying 
%realize 
a 
%Hermitian 
complex conjugation reduction between $\vt{q}$ and $\vt{r}$,  
we assume that 
%consider the case where 
%all 
the parameters 
%in (\ref{chara2}), i.e., 
$\gamma_k$, $a_k b_k$
%$\gamma_j$, $a_j b_j$ 
and $\mu$ 
%$\zeta$ 
are all real. 
Then, 
considering the behavior of the left-hand side of (\ref{chara2}) 
as a 
%real 
function of $\lambda$, we can identify 
%see that 
%the 
%under what 
%the 
conditions 
%under which 
for the algebraic equation (\ref{chara2}) 
%for 
in $\lambda$ 
%admits 
%has 
%can 
to have a pair of complex conjugate roots. 
%complex roots. 
%that appear in complex-conjugate pairs. 
%an imaginary solution. 
In particular, 
%it is clear that 
%we see that 
at least one of the 
$a_k b_k$ must be positive. 
Then, 
%Among 
among the \mbox{$m+1$} solutions of (\ref{chara2}), 
%assuming 
%by setting 
we 
%write 
express 
%set 
%requiring that 
%among the \mbox{$m+1$} solutions of  
%(\ref{chara2}) there is 
%exists  
%the 
%roots of 
%this algebraic equation for $\lambda$ 
%appear 
%admits 
%a pair of 
%as 
%has 
%contain 
%a 
the pair of complex conjugate 
%complex-conjugate 
roots 
%among the \mbox{$m+1$} solutions of (\ref{chara2}) 
as 
%pair, 
%of  
\mbox{$\lambda= \xi \pm \mathrm{i} \eta$} 
%and \mbox{$\lambda= \xi - \mathrm{i} \eta$} 
with \mbox{$\eta > 0$}
%, we obtain 
to obtain 
%have 
%
\begin{align} 
\xi \pm \mathrm{i} \eta + \sum_{k=1}^m 
	\frac{a_k b_k}{\xi \pm \mathrm{i} \eta -\gamma_k} = -2\mu
%\zeta
. 
\label{xi-eta-zeta}
%\nonumber 
\end{align}
%By 
%Separating the real and imaginary parts, 
%Thus, we obtain 
This can be rewritten as 
%implies that 
\begin{equation}
\xi + 
\sum_{k=1}^m \frac{(\xi -\gamma_k) a_k b_k}{(\xi -\gamma_k)^2 + \eta^2} = -2\mu
%\zeta
, 
\hspace{5mm} 
\sum_{k=1}^m \frac{a_k b_k}{(\xi -\gamma_k)^2 + \eta^2} = 1. 
\label{xi-eta}
\end{equation}
Using the second equation, 
it is in principle possible 
%using he second equation 
%can 
%enables us 
to determine
\mbox{$\eta \, (>0)$} from $\xi$, $\gamma_k$ and $a_k b_k$
%; 
%it requires 
%the main step is to solve 
by solving an algebraic equation 
of degree $m$ 
%$m$th degree equation 
for $\eta^2$. 
Subsequently, the first equation can be 
%is 
used to determine $\mu$ 
%$\zeta$ 
from $\xi$, $\gamma_k$ and $a_k b_k$. 
However, in practice, 
this procedure is 
%practical 
not effective 
%only for 
%\mbox{$m=1$} and 
%\mbox{$m \le 2$}, 
for \mbox{$m \ge 3$}, 
because we 
need to use 
Cardano's 
formula 
even for 
%the case 
\mbox{$m =3$}. 
%Thus, for \mbox{$m \ge 3$}, 
%it is better 
%%much more concies 
%to 
%%we 
%express  $a_k b_k$ and $\zeta$ in terms of 
%%a given set of free parameters 
%the \mbox{$m+1$} solutions of (\ref{chara2}) (delete?).  

\subsubsection{Scalar NLS equation}

It would be 
%is 
instructive to 
%first 
study 
%%investigate 
%the simplest case of 
the scalar NLS 
%case 
%equation 
system 
%(\ref{sNLS})
%(\mbox{$m=1$}),
by setting \mbox{$m=1$}, 
before considering 
%investigating 
%discussing 
the multicomponent case. 
From (\ref{xi-eta}), we obtain 
%have 
\begin{equation}
2\xi - \gamma_1 = -2 \mu
%\zeta
, \hspace{5mm} 
a_1 b_1 =  (\xi -\gamma_1)^2 + \eta^2. 
%\nonumber 
\label{a_1b_1}
\end{equation}
Thus, we have 
\mbox{$\eta = \sqrt{a_1 b_1 - (\xi -\gamma_1)^2}$} with the condition 
\mbox{$a_1 b_1 > (\xi -\gamma_1)^2$} and 
\[
\boldsymbol{Z} = 
\left[
\begin{array}{cc}
 2\xi - \gamma_1  & a_1 \\
 -b_1 & \gamma_1 \\
\end{array}
\right]. 
\]
We choose 
the 
%two 
eigenvectors corresponding to the 
%pair of 
eigenvalues 
\mbox{$\xi \pm \mathrm{i} \eta$} 
%of this \mbox{$2 \times 2$} matrix 
%can be given as
%are given 
as
\[
\boldsymbol{Z}
\left[
\begin{array}{c}
a_1 \\
 -\xi \pm \mathrm{i} \eta + \gamma_1 \\
\end{array}
\right] 
= (\xi \pm \mathrm{i}\eta) 
\left[
\begin{array}{c}
a_1 \\
 -\xi \pm \mathrm{i} \eta + \gamma_1 \\
\end{array}
\right]. 
\]
%
%Thus, 
%In the scalar NLS case, 
%Then, 
%In the scalar NLS case, 
The general 
%a generic 
solution (\ref{mat-ex1}) of the 
%\mbox{$2 \times 2$} 
linear problem 
(\ref{NLS-UV2}) 
%given by (\ref{mat-ex1}) 
in the scalar NLS case 
%in the simplest scalar case 
%in this case 
can be written explicitly 
as 
\begin{align}
\left[
\begin{array}{c}
 \boldsymbol{\mathit \Psi}_1 \\
 \boldsymbol{\mathit \Psi}_2 
%(\zeta) 
\\
\end{array}
\right] &= c_1 \mathrm{e}^{\mathrm{i} x \left( \frac{\gamma}{2} + \mathrm{i} \eta \right) 
+ \mathrm{i}  t \left( \xi^2 - 2 \mathrm{i} \xi \eta + \eta^2  
	-2\xi \gamma_1 + \frac{\gamma_1^2}{2} \right)}
\left[
\begin{array}{c}
a_1 \\
 -\xi + \mathrm{i} \eta + \gamma_1 \\
\end{array}
\right] 
\nonumber \\[1mm]
 & \hphantom{=}\; \mbox{}+ c_2 \mathrm{e}^{\mathrm{i} x \left( \frac{\gamma}{2} 
	- \mathrm{i} \eta \right) 
+ \mathrm{i}  t \left( \xi^2 + 2 \mathrm{i} \xi \eta + \eta^2  
	-2\xi \gamma_1 + \frac{\gamma_1^2}{2} \right)}
\left[
\begin{array}{c}
a_1 \\
 -\xi - \mathrm{i} \eta + \gamma_1 \\
\end{array}
\right], 
\nonumber
%\label{mat-ex2}
\end{align}
where $c_1$ and $c_2$ are arbitrary 
%nonzero 
constants; they 
%are assumed 
%required 
%to be 
%must
should  be nonzero 
%in order to provide 
so that a 
%new 
nontrivial solution 
%so that 
%by 
can be 
%obtained using 
%provided using 
generated by 
%obtained by 
applying 
%an 
the elementary 
%through a 
B\"acklund--Darboux transformation. 
%can provide a nontrivial solution.  
%
In view of (\ref{gauge1}) and (\ref{a_1b_1}), we obtain 
\begin{align}
\frac{\Psi_2}{\Psi_1}
& = \mathrm{i} \hspace{1pt}
%\frac{\mathrm{i}}{P_1 P_2} 
\mathrm{e}^{-\mathrm{i} \gamma_1  x
 + \mathrm{i}  (\gamma_1^2 + 2 a_1 b_1) t} 
\frac{\boldsymbol{\mathit \Psi}_2}{\boldsymbol{\mathit \Psi}_1} 
%\mathscr{ABC}
%{\mathcal \Psi} {\mathbf \Psi} \boldsymbol{\mathit \Psi}
%\label{gauge2}
\nonumber \\
&= 
%\frac{\mathrm{i}}{P_1 P_2} 
 \mathrm{i} \hspace{1pt} \mathrm{e}^{-\mathrm{i}  \gamma_1 x
 + \mathrm{i} \left[ \gamma_1^2 
%+ 2 a_1 b_1
+ 2(\xi -\gamma_1)^2 + 2 \eta^2 \right] t} 
 \frac{c_1 (-\xi + \mathrm{i} \eta + \gamma_1) 
 \mathrm{e}^{- \eta x + 2 \xi \eta t}
 + c_2 (-\xi - \mathrm{i} \eta + \gamma_1) \mathrm{e}^{ \eta x - 2 \xi \eta t}}
{a_1 \left( c_1 \mathrm{e}^{- \eta x + 2 \xi \eta t}
 + c_2 \mathrm{e}^{ \eta x - 2 \xi \eta t} \right) }.
\nonumber
\end{align}
%
%According to 
%Using 
%the formulas 
%%given 
%in 
%Thus, 
Then, 
%using 
formulas (\ref{eBD1})
%, 
provides 
%the following 
%a 
%we obtain 
the 
%corresponding 
solution of the 
%nonreduced 
scalar NLS system (\ref{sNLS}) 
%(cf.~(\ref{mNLS})) 
%is obtained 
%: 
as 
%reads 
%
\begin{align}
%\widetilde{R} 
 %R 
r &= 
%\left. \Psi_2 \Psi_1^{-1}\right|_{\zeta= \mu}, 
 \frac{\mathrm{i}}{a_1} \mathrm{e}^{-\mathrm{i}  \gamma_1 x
 + \mathrm{i} \left[ \gamma_1^2 
%+ 2 a_1 b_1
+ 2(\xi -\gamma_1)^2 + 2 \eta^2 \right] t} 
 \frac{c_1 (-\xi + \mathrm{i} \eta + \gamma_1) 
 \mathrm{e}^{- \eta x + 2 \xi \eta t}
 + c_2 (-\xi - \mathrm{i} \eta + \gamma_1) \mathrm{e}^{ \eta x - 2 \xi \eta t}}
	{c_1 \mathrm{e}^{- \eta x + 2 \xi \eta t}
 + c_2 \mathrm{e}^{ \eta x - 2 \xi \eta t}}, 
%\hspace{5mm} 
\nonumber \\[1mm]
%\widetilde{Q} 
%Q 
q &= \mathrm{i}
 a_1 \mathrm{e}^{\mathrm{i}  \gamma_1 x - \mathrm{i}  \left[ \gamma_1^2 
%+ 2 a_1 b_1
+ 2(\xi -\gamma_1)^2 + 2 \eta^2 \right] t} 
 \frac{c_1 (\xi + \mathrm{i} \eta - \gamma_1) 
 \mathrm{e}^{- \eta x + 2 \xi \eta t}
 + c_2 (\xi - \mathrm{i} \eta - \gamma_1) \mathrm{e}^{ \eta x - 2 \xi \eta t}}
{c_1 \mathrm{e}^{- \eta x + 2 \xi \eta t}
 + c_2 \mathrm{e}^{ \eta x - 2 \xi \eta t} },
%\label{eBD1}
\nonumber
\end{align}
%Here, 
where we omit the tilde 
%for 
%of 
%$r$ and $q$. 
to denote 
%denoting 
the application of the B\"acklund--Darboux transformation. 
%for brevity. 
%has been 
%%is 
%omitted. 
Thus, if 
%If 
we re-parametrize some 
%the old 
parameters as 
%set
\[
\mathrm{i} a_1 =: \mathrm{e}^{\mathrm{i}\theta}, 
%\hspace{5mm} -\mathrm{i} b_1 =: \mathrm{e}^{-\mathrm{i}\theta} d^2, 
\hspace{5mm} \frac{c_2}{c_1}=: \mathrm{e}^{2 \delta}
%, 
\]
with 
%real-valued 
%using 
%new 
real 
%parameters 
%real 
$\theta$
%, $d$ 
and $\delta$, 
%then 
we can realize the complex conjugation reduction \mbox{$r = q^\ast$} 
%\mbox{$R = Q^\ast$} 
and  
obtain the one-dark soliton solution 
of the self-defocusing 
%scalar 
NLS equation \mbox{$\mathrm{i} q_t + q_{xx} - 2 |q|^2 
%\left| q \right|^2 
q = 0$} 
%\mbox{$\mathrm{i} Q_t + Q_{xx} - 2 | Q |^2 Q = 0$} 
in the form~\cite{ZS2,Tsuzuki,Hase73-2,Hiro76}: 
\begin{align}
%Q
q(x,t) &= \mathrm{e}^{\mathrm{i} \gamma_1 x - \mathrm{i}  \left[ \gamma_1^2 
 + 2(\xi -\gamma_1)^2 + 2 \eta^2 \right] t
	+ \mathrm{i} \theta} 
 \frac{(\xi + \mathrm{i} \eta - \gamma_1) 
 \mathrm{e}^{- \eta ( x - 2 \xi t) -\delta}
 + (\xi - \mathrm{i} \eta - \gamma_1) \mathrm{e}^{ \eta (x - 2 \xi t) +\delta}}
{\mathrm{e}^{- \eta (x - 2 \xi t) -\delta}
 + \mathrm{e}^{ \eta (x - 2 \xi t) +\delta} }
\nonumber \\[1mm]
&= \mathrm{e}^{\mathrm{i} \gamma_1 x - \mathrm{i}  \left[ \gamma_1^2 
+ 2(\xi -\gamma_1)^2 + 2 \eta^2  \right] t
	+ \mathrm{i} \theta} \left\{ (\xi - \gamma_1) - \mathrm{i} \eta 
	\tanh \left[ \eta ( x - 2 \xi t) +\delta \right] \right\}. 
\nonumber
\end{align}
%
%Because \mbox{$\eta > 0$}, the asymptotic behaviors are 
%values 
This solution represents a simple dip in the background plane wave 
and the asymptotic value at 
%spatial 
infinity 
is 
given as 
%by 
\mbox{$\lim_{x \to \pm \infty} |q(x,t)| = \sqrt{(\xi - \gamma_1)^2 + \eta^2} $}. 
%Note that 
%Apparently, 
The elementary B\"acklund--Darboux transformation
defined in Proposition~\ref{prop2.1} 
%apparently 
does not
%, in general, 
%preserve 
%maintain 
retain the 
%Hermitian 
complex 
conjugation reduction 
%\mbox{$R = Q^\dagger$} 
%\mbox{$R = Q^\ast$} 
in general; 
however, 
if it is applied to a plane-wave solution in combination 
%we combine it 
with a simple 
%%scaling symmetry, 
rescaling of the 
%pair of 
%two 
dependent variables, 
we can 
%it is possible to 
realize the complex conjugation reduction 
%between the 
in both the old and new pair of variables. 
%potentials. 
%transformed 
%using a simple rescaling 

%\subsubsection{Two-component 
%{Vector 
%NLS}
%dark soliton}
%case}

%It can be shown using Proposition~\ref{prop2.1} that 
Using Proposition~\ref{prop2.1}, we can confirm 
%show 
that 
the one-dark soliton solution 
%indeed provides 
is associated with 
%produces 
a 
bound state 
%bound-state eigenfunction 
of the 
Lax-pair representation 
%Lax pair 
%linear problem 
(\ref{NLS-UV})
%, as well as 
%and 
%the 
%its adjoint 
%representation 
%(\ref{NLS-adUV}), 
at 
%the eigenvalue 
\mbox{$\zeta=\mu$}, i.e., when 
the spectral parameter 
%is equal to 
coincides with 
the parameter of the applied B\"acklund--Darboux  
%parameter. 
transformation. 
Indeed, 
%using 
the gauge transformation (\ref{eBT-g1})
%, 
%implies 
%shows that 
%gives 
provides 
a generic linear eigenfunction 
at \mbox{$\zeta=\mu$}
for 
%corresponding to 
the transformed potentials 
%at \mbox{$\zeta=\mu$}
%the elementary B\"acklund--Darboux transformation
%at \mbox{$\zeta=\mu$}
%can be expressed 
%obtained 
%using the gauge transformation (\ref{eBT-g1}) 
as 
%
%in terms of a linear eigenfunction 
%of the Lax-pair representation (\ref{NLS-UV}) at \mbox{$\zeta=\mu$} 
%as~\cite{Kono82,Calo84} 
%[Chinese87] 
%\begin{equation}
%\widetilde{R} = \left. \Psi_2 \Psi_1^{-1}\right|_{\zeta= \mu}, 
%%\label{eBD1}
%\end{equation}
%The corresponding gauge transformation~\cite{Adler94} [Date83, Chinese87], 
%given by~\cite{Adler94} [Chinese87], 
\begin{align}
& \left[
\begin{array}{c}
\widetilde{\Psi}_1 \\ 
\widetilde{\Psi}_2 \\
\end{array}
\right] 
%(\mu)
%:= g(\zeta,\mu) 
\propto
%= 
\frac{
\left|
\begin{array}{cc}
\Psi_1^{(1)} & \Psi_1 \\
\Psi_2^{(1)} & \Psi_2 \\
\end{array}
\right|
}{\Psi_1^{(1)}}
 \left[
\begin{array}{c}
 -q \\
 1 \\
\end{array}
\right], 
\nonumber
%\\ &= \nonumber 
%\label{eBT-g1}
\end{align}
up to an overall constant. 
%where 
Here, the superscript 
%${}^{(j)}
%\; (j=1, 2)
%$ with \mbox{$j=1$} or $2$ is used to 
${}^{(1)}$ 
means 
%denotes  
%designates 
%distinguish 
%a 
%two linearly independent eigenfunctions 
the linear eigenfunction used to define the elementary B\"acklund--Darboux transformation 
%as in 
(\ref{eBD1}). 
%at \mbox{$\zeta=\mu$}.  
%set of solutions 
%where $g(\zeta,\mu)$ is an arbitrary (but nonzero) 
%scalar function of $\zeta$ and $\mu$,  
%Note that 
In the above expression, 
the determinant in the numerator is
%, in general, 
%a 
the Wronskian of two 
linearly independent 
%in general 
eigenfunctions, 
which 
%so it 
is a nonzero constant; 
%In addition, 
%Moreover, 
$q(x,t)
$ in the seed solution is a plane wave and 
$1/\Psi_1^{(1)}$ is 
%a 
%nonzero everywhere 
%function 
%for \mbox{$c_2/c_1 \in \mathbb{R}$} 
%and \mbox{$| \Psi_1^{(1)} | \to \infty$} as \mbox{$x \to \pm \infty$}. 
nonsingular everywhere 
and 
%This indeed 
decays exponentially fast as \mbox{$x \to \pm \infty$}. 
Thus, 
%we have confirmed that 
%the above linear eigenfunction certainly gives 
the 
%applied 
application of the 
%an 
elementary 
B\"acklund--Darboux  
%parameter. 
transformation 
%(\ref{eBD1})
%certainly 
%produces 
%can 
indeed 
generates 
a 
%new 
bound state 
of 
%the Lax-pair representation. 
the associated linear 
%eigenvalue 
%linear 
problem. 
%at \mbox{$\zeta=\mu$}. 
%In fact, 
%In general, 
More generally, 
%this is also true for 
%a more general seed solution 
%that supports 
%%already has 
%bound states. 
%at the eigenvalue
%More generally, 
%it can be shown that an 
the elementary B\"acklund--Darboux  
%%parameter. 
transformation 
%(\ref{eBD1}) with \mbox{$\mu \in \mathbb{R}$} 
%preserves 
increases the number of bound states by one, 
by adding a new bound state 
%at \mbox{$\zeta=\mu$}, 
with a real eigenvalue $\mu$, 
when it is applied to 
%the general 
a 
%more 
%general 
seed solution
%; 
%however, if the seed solution 
%already has a bound state at the eigenvalue 
that does not 
%already 
have a bound state at
\mbox{$\zeta=\mu$}.
%, then 
%it is deleted by the elementary B\"acklund--Darboux  
%%parameter. 
%transformation (\ref{eBD1}). 
%More generally, 
%he gauge transformation (\ref{eBT-g1}) 
%maps an old 
%bound-state eigenfunction with \mbox{$\zeta$}
Thus, one can intuitively 
think of the elementary B\"acklund--Darboux  
%%parameter. 
transformation as a classical analog of the creation operator.

\subsubsection{Vector 
%soliton}
NLS equation} 

%The vector dark-soliton solutions for 
%the 
%%simplest 
%%nontrivial 
%two-component case 
%%\mbox{$m =2$}
%%of two components 
%(\mbox{$m =2$}) 
%have 
%already 
%been 
%%derived 
%obtained~\cite{Dub88}
%%; 
%%extensively 
%%studied 
%%in the literature~\cite{Dub88} 
%and 
%their properties 
%%behaviors 
%%were already studied 
%%have been 
%%investigated 
%were clarified, {\it e.g.}, in [Kivshar, Ohta]. 
%; however, 
%However, 
%Nonetheless, 
%it would be worthwhile
%%so we do not discuss them here. 
%%of some value 
%%interesting 
%meaningful 
%to 
%demonstrate 
%%show 
%that 
%derive 
%Thus, 
%In this subsection, 
%we 
%consider the general $m$-component case of the vector NLS equation (\ref{rvNLS}) 
%and 
Let us derive 
%We demonstrate that 
%the 
%vector 
a one-dark soliton solution  
of the vector NLS equation (\ref{rvNLS}) 
in the general $m$-component case. 
%(cf.~(\ref{rvNLS}))
%can be 
%easily 
%derived 
%in a simpler manner 
%using the 
%%only 
%%by applying 
%elementary B\"acklund--Darboux transformation. 
%transformations. 
%In the case \mbox{$m =2$}, 
%
%\mbox{$\boldsymbol{Z} (\zeta)$} in 
%(\ref{Z-vector}) 
%reads 
%can be diagonalized as 
%\begin{align}
%\boldsymbol{Z} (\zeta) = 
%\left[
%\begin{array}{ccc}
%-2 \zeta & a_1 & a_2 \\
% -b_1 & \gamma_1 & 0 \\
% -b_2 & 0 & \gamma_2 \\
%\end{array}
%\right], 
%\end{align}
%which implies that the the eigenvectors corresponding to the 
%%complex-conjugate 
%the pair of complex-conjugate eigenvalues can be written 
%%expressed 
%as 
%
%Following the 
%%above-mentioned 
%procedure described 
%%explained 
%above, 
%line 
Actually, 
using 
the elementary B\"acklund--Darboux transformation, 
we can 
%allows us to 
%it is possible to 
construct 
%interesting new 
%various 
%exact 
%various interesting 
%an 
%essentially 
more general 
%and interesting 
%class of 
solutions 
of 
%for 
%a more general system, 
the nonreduced vector NLS system (\ref{vNLS}), 
which 
%do not admit 
are not compatible with 
the complex conjugation reduction; 
%which appear to be interesting in their own right; 
%; 
%However,
%Here, 
%We 
%which 
%are essentially more general than 
%the one-dark soliton 
%cannot be reduced directly to 
%the solutions of 
%the reduced equation 
%(\ref{rvNLS}). 
however, 
such solutions apparently 
%do not 
have no physical significance, 
so 
%in this paper, 
%but 
%this 
we do not discuss 
%consider 
them 
%discuss it 
%but 
%this is beyond the scope of this paper. 
in this paper. 
% so 
%we do not 
%are only interested in the 
%discuss it in this paper. 
%
%only the 
%solutions that 
%%admits the complex conjugation reduction 
%can be directly 
%reduced 
%reduce 
%to solutions of 
%the vector NLS equation (\ref{rvNLS}) 
%through 
%by 
%imposing a suitable restriction on
%special 
%a suitable choice of 
%the parameters. 
Thus, 
%the general 
%a generic 
%so we chooce 
among the general solution (\ref{mat-ex1}) of 
%a special solution of 
the 
%\mbox{$(m+1) \times (m+1)$} 
linear problem (\ref{NLS-UV2}) in the vector NLS case, 
%at \mbox{$\zeta=\mu$}, 
%given by (\ref{mat-ex1}) 
%in the vector case 
%can be written explicitly 
we consider 
%choose 
a particular 
%special 
solution, 
%as 
\begin{align}
\left[
\begin{array}{c}
 \boldsymbol{\mathit \Psi}_1 \\
 \boldsymbol{\mathit \Psi}_2 
%(\zeta) 
\\
\end{array}
\right] &= c_1 \mathrm{e}^{\mathrm{i} x \left( \xi + \mathrm{i}\eta + \mu
%_{l+m} 
\right) + \mathrm{i}  t \left[ -(\xi + \mathrm{i}\eta )^2 + 2 \mu^2 
%_{l+m} 
\right]}
\left[
\begin{array}{c}
1 \\
-\frac{b_1}{\xi + \mathrm{i}\eta -\gamma_1} \\
\vdots \\
-\frac{b_m}{\xi + \mathrm{i}\eta -\gamma_m} \\
\end{array}
\right]
\nonumber \\[1mm]
 & \hphantom{=}\; \mbox{}
+ c_2 \mathrm{e}^{\mathrm{i} x \left( \xi - \mathrm{i}\eta + \mu
%_{l+m} 
\right) + \mathrm{i}  t \left[ -(\xi - \mathrm{i}\eta )^2 + 2 \mu^2 
%_{l+m} 
\right]}
\left[
\begin{array}{c}
1 \\
-\frac{b_1}{\xi - \mathrm{i}\eta -\gamma_1} \\
\vdots \\
-\frac{b_m}{\xi - \mathrm{i}\eta -\gamma_m} \\
\end{array}
\right]. 
\nonumber
%\label{mat-ex2}
\end{align}
%
%where 
Here, $c_1$ and $c_2$ are arbitrary 
%nonzero 
constants, 
which 
%; they are assumed 
%required 
%to 
should be nonzero 
%in order to 
so that the above solution can 
provide 
%result in 
a nontrivial solution 
of the vector NLS system. 
%Note 
Recall that the seed solution of the 
%nonreduced 
vector NLS system (\ref{vNLS})
is the 
%vector 
general 
plane-wave solution 
(\ref{vector-plane}) and 
%we have the conditions on the parameters 
the parameters satisfy the relations 
%conditions 
(\ref{xi-eta}). 
%(\ref{xi-eta-zeta}). 

%Using 
Then, 
by applying 
the elementary B\"acklund--Darboux  
%%parameter. 
transformation 
(\ref{eBD1}), 
%with \mbox{$\mu \in \mathbb{R}$} 
%defined in Proposition~\ref{prop2.1}, 
we obtain  
a new solution of the vector NLS system (\ref{vNLS}) as 
%can construct 
%the 
%%one-
%dark soliton solutions of the vector NLS equation (\ref{rvNLS}). 
%(\ref{eBD1})
%
\begin{align}
\widetilde{
%R
\vt{r}}^T &= 
%\left. 
%\Psi_2 \Psi_1^{-1}
%\right|_{\zeta= \mu} 
%= 
\mathrm{i} \hspace{1pt}
%P_2^{-1} 
\mathrm{e}^{-\mathrm{i} x \Gamma + \mathrm{i} t \Gamma^2} 
\boldsymbol{\mathit \Psi}_2 \boldsymbol{\mathit \Psi}_1^{-1} 
\mathrm{e}^{2\mathrm{i} t \sum_{k=1}^m a_k b_k 
%AB
} 
%P_1^{-1}
\nonumber \\
&= 
- \mathrm{i} 
%P_2^{-1} 
%\mathrm{e}^{-\mathrm{i} x \Gamma + \mathrm{i} t \Gamma^2} 
\frac{\mathrm{e}^{2\mathrm{i} t \sum_{k=1}^m a_k b_k}}
{c_1 \mathrm{e}^{ - \eta x + 2 \xi \eta t }
+ c_2 \mathrm{e}^{\eta x - 2 \xi \eta t }}
\left[
\begin{array}{c}
\mathrm{e}^{-\mathrm{i}  \gamma_1 x + \mathrm{i} \gamma_1^2 t} \left( 
c_1 \frac{b_1}{\xi + \mathrm{i}\eta -\gamma_1} \mathrm{e}^{ - \eta x + 2 \xi \eta t}
+c_2 \frac{b_1}{\xi - \mathrm{i}\eta -\gamma_1} \mathrm{e}^{\eta x - 2 \xi \eta t }
\right) \\
\vdots \\
\mathrm{e}^{-\mathrm{i} \gamma_m x + \mathrm{i} \gamma_m^2 t} \left( 
c_1 \frac{b_m}{\xi + \mathrm{i}\eta -\gamma_m}  \mathrm{e}^{ - \eta x + 2 \xi \eta t}
+c_2 \frac{b_m}{\xi - \mathrm{i}\eta -\gamma_m} \mathrm{e}^{\eta x - 2 \xi \eta t } 
\right) \\
\end{array}
\right], 
%P_1^{-1} 
\nonumber 
\end{align}
%
%Using 
%the relation
%(\ref{xi-eta-zeta}), 
%We also 
%can remove the parameter $\mu$ and 
%obtain with the aid of (\ref{xi-eta-zeta})
where 
\mbox{$\Gamma = \mathrm{diag} (\gamma_1, \gamma_2, \ldots, \gamma_m)$},
and 
\begin{align}
%\hspace{5mm} 
\widetilde{\vt{q}} &= -\vt{q}_x - 2 \mathrm{i} \mu \vt{q} 
 + \sca{\vt{q}}{\widetilde{\vt{r}}} \vt{q} 
\nonumber \\[1mm]
&= \left( \widetilde{q}_1, \widetilde{q}_2, \ldots, \widetilde{q}_m \right), 
%\left( q_1, q_2, \ldots, q_m \right)
%P_2, 
\nonumber
\end{align}
with 
\begin{align}
\widetilde{q}_j 
&= \mathrm{i} 
%\left( 
 a_j \mathrm{e}^{\mathrm{i}  \gamma_j x - \mathrm{i} \left( \gamma_j^2 
%+ 2 a_1 b_1
%+ 2(\xi -\gamma_1)^2 + 2 \eta^2 
 + 2 \sum_{k=1}^m a_k b_k \right) t} 
 \frac{c_1 (\xi + \mathrm{i} \eta - \gamma_j) 
 \mathrm{e}^{- \eta x + 2 \xi \eta t}
 + c_2 (\xi - \mathrm{i} \eta - \gamma_j) \mathrm{e}^{ \eta x - 2 \xi \eta t}}
{c_1 \mathrm{e}^{- \eta x + 2 \xi \eta t}
 + c_2 \mathrm{e}^{ \eta x - 2 \xi \eta t} }, 
 \hspace{5mm} j= 1, 2, \ldots, m. 
%, \ldots,  \right. \nonumber \\ & \hphantom{=} \;  \left. 
%\right) 
\nonumber 
\end{align}
%\[
%Q = \mathrm{e}^{-2\mathrm{i} t AB} A  
%\mathrm{e}^{\mathrm{i} x \Gamma - \mathrm{i} t \Gamma^2} P_2, 
%\]
Here, 
%We also 
%can remove 
the B\"acklund parameter $\mu$ 
%was 
has been 
eliminated 
%and 
%obtain 
with the aid of 
%using 
(\ref{xi-eta-zeta}).  
%(\ref{xi-eta}). 
The complex conjugation reduction 
%\mbox{$r_j = \sigma_j q_j^\ast$} 
\mbox{$\widetilde{r}_j = \sigma_j \widetilde{q}_j^{\hspace{2pt}\ast}$} 
can be realized by setting 
\begin{align}
& b_j = \sigma_j \left[ (\xi - \gamma_j)^2 + \eta^2 \right]  a_j^\ast, 
%\hspace{5mm} \sigma_j = \pm1, \hspace{5mm} j= 1, 2, \ldots, m, 
\nonumber \\[1mm] 
&  \frac{c_2}{c_1}= \mathrm{e}^{2 \delta}, 
\;\; \delta \in {\mathbb R}, 
\nonumber 
\end{align}
where 
%\mbox{$\sigma_j = 1$} for \mbox{$j= 1, 2, \ldots, n$} 
%and \mbox{$\sigma_j = -1$} for \mbox{$j= n+1, n+2, \ldots, m$}. 
\mbox{$\sigma_j = +1$} or  \mbox{$-1$}. 
%for \mbox{$j= n+1, n+2, \ldots, m$}. 
%with 
%real-valued 
%new real parameters 
%real 
%$\theta$
%, $d$ 
%and 
%$\delta$. 
%so 
Thus, omitting the tilde, 
we obtain 
%the 
a one-dark soliton solution of the 
vector NLS equation (\ref{rvNLS}) 
%with \mbox{$\Sigma = \mathrm{diag} (\sigma_1, \sigma_2, \ldots, \sigma_m)$} 
%\mbox{$\Sigma = \mathrm{diag} (\sigma_j)$} 
as 
%can construct 
%the 
%%one-
%dark soliton solutions of the vector NLS equation (\ref{rvNLS}). 
%
%with a mixed focusing-defocusing nonlinearity:
%as 
\begin{align}
%\vt{q} 
q_j (x,t) &= 
%\mathrm{i} 
%\left( 
 a_j \mathrm{e}^{\mathrm{i}  \gamma_j x - \mathrm{i} \left\{ \gamma_j^2 
%+ 2 a_1 b_1
%+ 2(\xi -\gamma_1)^2 + 2 \eta^2 
 + 2 \sum_{k=1}^m \sigma_k |a_k|^2 \left[ (\xi - \gamma_j)^2 
	+ \eta^2 \right]   \right\} t} 
	 \left\{ \mathrm{i} (\xi - \gamma_j) 
%- \mathrm{i} 
	+\eta \tanh \left[ \eta ( x - 2 \xi t) +\delta \right] \right\}, 
\nonumber \\
&
% \hspace{95mm} 
\hspace{75mm} 
j= 1, 2, \ldots, m. 
%\nonumber
\label{v-dark}
\end{align}
%Here, the tilde is omitted. 
Note that 
the second relation 
%equation 
in (\ref{xi-eta}) reduces to the 
%following constraint
normalization condition: 
\begin{equation}
\sum_{j=1}^m \sigma_j |a_j|^2 =1. 
\nonumber
\end{equation}
%
%Thus, 
This implies that 
%can be satisfied only when 
at least one of the 
%$m$ 
$\sigma_j$ must be positive, i.e., 
there must 
%exists 
exist one or more defocusing components 
%at least 
%one defocusing component
in the vector 
%dependent 
variable $\vt{q}(x,t)$. 
This is consistent with the previous observation 
that at least one of the 
$a_k b_k$ 
must be positive 
for (\ref{chara2}) 
%for 
%in $\lambda$ 
%admits 
%has 
%can 
to have a pair of complex conjugate roots. 

The vector 
%dark-soliton solution 
dark soliton (\ref{v-dark}) 
%is essentially 
%``static" 
has an essentially time-independent shape 
and 
%has 
contains no free parameters 
%representing 
%affording 
corresponding to 
%responsible for 
%the  
%does not admit 
the internal degrees of freedom; 
%[Ohta]; 
in the two-component case (\mbox{$m=2$}), 
this solution 
%its 
%the multisoliton 
%generalization 
%it 
%(\ref{v-dark}) 
%is well known
has 
%its properties have 
been extensively studied in the literature~\cite{Mak81,Mak82,Dub88,Kiv97,Ohta11}. 
%[Kivshar, OhtaYang]; 
%it has no 
%%does not admit 
%internal degrees of freedom [Ohta]. 
In section~\ref{sec5}, 
we will derive 
a 
more general 
%interesting 
vector 
dark-soliton 
%solutions 
solution 
that 
can exist for \mbox{$m \ge 3$} and 
%admits 
%can 
admits the 
internal degrees of freedom.
%for \mbox{$m \ge 3$}. 

%\subsubsection{Three-component vector NLS}
%dark soliton}

\subsection{Matrix dark soliton}

In this subsection, 
we construct new nontrivial 
dark-soliton solutions of 
the defocusing matrix NLS equation
%, 
%i.e., 
%namely, 
(\ref{rmNLS}) 
with \mbox{$\sigma= +1$}. 
Before 
focusing on 
%considering 
the square matrix case 
%of 
\mbox{$l=m$}, 
we 
%explain 
%briefly 
%%describe 
%discuss 
%and intuitively 
%describe the reason 
%in 
%for 
consider the more 
general case of \mbox{$l \le m$}
%why 
%when 
%under what 
%kind of 
and 
%how 
identify the 
conditions 
%discuss 
%can make 
under which 
the new potentials generated by 
%a single application of 
the 
elementary B\"acklund--Darboux transformation 
%make 
%the Hermitian conjugation reduction 
%can be imposed on 
%the new pair of potentials obtained 
%Using 
%using an 
%is 
%can become 
%be 
%becomes 
%%is 
%consistent with 
admit 
the Hermitian conjugation reduction. 
%an 
%the 
%elementary B\"acklund--Darboux transformation. 
 
%As 
By applying the elementary B\"acklund--Darboux transformation
%described in 
%From 
defined in Proposition~\ref{prop2.1} 
to the plane-wave solution (\ref{plane2})
%, 
with $P_1$ and $P_2$ 
%in (\ref{plane2}) are 
dropped, 
%and omitting 
%with the identity matrices 
%$P_1$ and $P_2$, 
%From Proposition~\ref{prop2.1}, 
we obtain the transformed potentials in the form: 
%have 
\begin{subequations}
\label{matrix-eBD1}
\begin{align}
\widetilde{R} &= 
%\left. 
\Psi_2 \Psi_1^{-1}
%\right|_{\zeta= \mu} 
\nonumber \\
 &= \mathrm{i} 
%P_2^{-1} 
\mathrm{e}^{-\mathrm{i} x \Gamma + \mathrm{i} t \Gamma^2} 
\boldsymbol{\mathit \Psi}_2 \boldsymbol{\mathit \Psi}_1^{-1} 
\mathrm{e}^{2\mathrm{i} t AB} 
%P_1^{-1}
, 
%\nonumber 
\\[2mm]
%\end{align}
%
%and 
%
%\begin{align}
\widetilde{Q} &= -Q_x - 2 \mathrm{i} 
\mu Q + Q \widetilde{R} Q 
\nonumber \\
&=  \mathrm{i} 
%P_1 
\mathrm{e}^{-2\mathrm{i} t AB} A 
\left( \boldsymbol{\mathit \Psi}_2 \boldsymbol{\mathit \Psi}_1^{-1} A 
 - \Gamma - 2 \mu I_m \right) 
\mathrm{e}^{\mathrm{i} x \Gamma - \mathrm{i} t \Gamma^2} 
%P_2
.
%\nonumber
\end{align}
\end{subequations}
Here, 
%$P_1$ and $P_2$ in (\ref{plane2}) 
%are dropped, 
%omitted, 
%and  
the \mbox{$l \times l$} matrix 
$\boldsymbol{\mathit \Psi}_1$ and 
the \mbox{$m \times l$} matrix 
$\boldsymbol{\mathit \Psi}_2$ 
%are the components of the linear eigenfunction satisfying 
satisfy the linear 
%the eigenvalue 
problem (\ref{NLS-UV2}). 
%at \mbox{$\zeta=\mu$}.
%
%(\ref{NLS-UV2}) 
From 
%Note that 
(\ref{NLS-U2}), we 
%can 
obtain 
%implies 
the matrix Riccati equation 
for $\boldsymbol{\mathit \Psi}_2 \boldsymbol{\mathit \Psi}_1^{-1}$: 
\begin{align}
-\mathrm{i} 
\left( \boldsymbol{\mathit \Psi}_2 \boldsymbol{\mathit \Psi}_1^{-1} \right)_x 
= -B + 2\mu \boldsymbol{\mathit \Psi}_2 \boldsymbol{\mathit \Psi}_1^{-1} 
 + \Gamma \boldsymbol{\mathit \Psi}_2 \boldsymbol{\mathit \Psi}_1^{-1} 
 - \boldsymbol{\mathit \Psi}_2 \boldsymbol{\mathit \Psi}_1^{-1} A 
  \boldsymbol{\mathit \Psi}_2 \boldsymbol{\mathit \Psi}_1^{-1}. 
\label{mRic}
\end{align}
%To realize 
%We 
%are interested in the case where 
%require that 
%would like 
%need to impose 
To realize 
the Hermitian conjugation reduction 
\mbox{$\widetilde{R}=\widetilde{Q}^{\hspace{1pt}\dagger}$}
%\mbox{$\widetilde{Q}=\widetilde{R}^{\hspace{1pt}\dagger}$}
%for 
%can be imposed 
%on 
%for $\widetilde{Q}$ and $\widetilde{R}$ 
in 
%as given in 
%given by 
(\ref{matrix-eBD1}), 
%Thus, 
%so we should 
we require 
%assume 
that 
\[
\Gamma^\dagger = \Gamma, \hspace{5mm} 
%P_1^{-1} = P_1^\dagger, \hspace{5mm} 
%P_2^{-1} = P_2^\dagger, \hspace{5mm} 
\left( AB \right)^\dagger =AB, 
\]
and 
\begin{align}
\left( \boldsymbol{\mathit \Psi}_2 \boldsymbol{\mathit \Psi}_1^{-1} \right)^\dagger 
= - A \left( \boldsymbol{\mathit \Psi}_2 \boldsymbol{\mathit \Psi}_1^{-1} A 
 - \Gamma - 2 \mu I \right).  
\label{Ric-red}
\end{align}
%
%In view of (\ref{mRic}), 
%the latter condition 
%this condition 
%we find that 
A direct 
%lengthy 
calculation 
%implies 
shows 
that the matrix Riccati equation 
(\ref{mRic}) 
%is consistent with the above 
%admits 
%this 
is consistent with 
the reduction (\ref{Ric-red})
if 
%can be realized by setting 
\begin{align}
A A^\dagger A = A, \hspace{5mm} 
B=\Gamma \left( I_m - A^\dagger A \right) \Gamma A^\dagger 
	+ 2 \mu \Gamma A^\dagger + A^\dagger A K A^\dagger, 
\hspace{5mm} 
\boldsymbol{\mathit \Psi}_2 \boldsymbol{\mathit \Psi}_1^{-1} = 
\Gamma A^\dagger + A^\dagger \boldsymbol{\mathit \Phi}, 
%\Phi
%\boldsymbol{\mathit \Phi}_2 \boldsymbol{\mathit \Phi}_1^{-1}, 
%\nonumber
\label{red-cond}
\end{align}
%and $\zeta$ is real, 
%and 
where $K$ is an \mbox{$m \times m$} constant Hermitian matrix, 
\mbox{$K^\dagger = K$}, $\boldsymbol{\mathit \Phi}$
%$\Phi$ is an 
%$\boldsymbol{\mathit \Phi}_1$ and $\boldsymbol{\mathit \Phi}_2$ 
%are 
is an \mbox{$l \times l$} \mbox{$(x,t)$}-dependent matrix, 
and the B\"acklund
%--Darboux  
parameter $\mu$ is 
%restricted to be 
real. 
%real-valued. 
%Thus, 
%If 
Note that for $B$ 
%is 
given 
%as 
above, 
%then 
the relation 
\mbox{$\left( AB \right)^\dagger =AB$} is 
automatically 
satisfied. 
%Using 
With the aid of the singular value decomposition, 
the first relation in (\ref{red-cond}) implies that 
the singular values of $A$ are either $0$ or $1$. 
%if $A A^\dagger$ and $H$ commute with each other, i.e.,  \mbox{$[AA^\dagger, H]=O$}.
%
%By 
To satisfy the last relation in (\ref{red-cond}),  
%motivates us to consider 
%implies that the ansatz can be 
%the linear change of 
%the 
%variables, 
we set 
\[
\boldsymbol{\mathit \Psi}_2  = 
\Gamma A^\dagger \boldsymbol{\mathit \Phi}_1 + A^\dagger 
%A  
\boldsymbol{\mathit \Phi}_2 
%\boldsymbol{\mathit \Phi}_1^{-1} A^\dagger\boldsymbol{\mathit \Psi}_1
, \hspace{5mm}  
\boldsymbol{\mathit \Psi}_1 = 
%A 
\boldsymbol{\mathit \Phi}_1, 
\]
%
%so that 
where \mbox{$\boldsymbol{\mathit \Phi}=\boldsymbol{\mathit \Phi}_2 
	\boldsymbol{\mathit \Phi}_1^{-1}$}. 
%which reduces 
Then, the linear 
%system 
problem 
%of PDEs <- undefined abbreviation 
(\ref{NLS-UV2})  
for $\boldsymbol{\mathit \Psi}_1$ and $\boldsymbol{\mathit \Psi}_2$ 
%at \mbox{$\zeta=\mu$}
is satisfied if the \mbox{$l \times l$} matrices 
$\boldsymbol{\mathit \Phi}_1$ and $\boldsymbol{\mathit \Phi}_2$ satisfy 
%the linear system 
%problem 
%pair of linear PDEs: 
%can be reduced 
%to a simpler form: 
%
\begin{subequations}
\label{Phi-lin}
%\begin{equation}
%\left\{ \begin{array}{l}
\begin{align}
& \left[
\begin{array}{c}
 \boldsymbol{\mathit \Phi}_1  \\
 \boldsymbol{\mathit \Phi}_2 \\
\end{array}
\right]_x 
= \mathrm{i} \left[
\begin{array}{cc}
-\mu I_l +  A \Gamma A^\dagger & A A^\dagger 
%I_l 
\\
-A K A^\dagger & \mu I_l  \\
\end{array}
\right] 
\left[
\begin{array}{c}
 \boldsymbol{\mathit \Phi}_1  \\
 \boldsymbol{\mathit \Phi}_2 \\
\end{array}
\right],
%\label{NLS-U2}
\\[1.5mm]
& \left[
\begin{array}{c}
 \boldsymbol{\mathit \Phi}_1  \\
 \boldsymbol{\mathit \Phi}_2 \\
\end{array}
\right]_t 
= \mathrm{i} \left( \partial_x^2 - 2 \mathrm{i} \mu \partial_x  + \mu^2 
%I 
\right) 
\left[
\begin{array}{c}
 \boldsymbol{\mathit \Phi}_1  \\
 \boldsymbol{\mathit \Phi}_2 \\
\end{array}
\right]. 
%\label{NLS-V2}
\end{align}
%\end{array}\right.
%\end{equation}
\end{subequations}
%Note that 
Thus, the problem of computing the 
%matrix 
exponential of 
an \mbox{$(l+m) \times (l+m)$} matrix is 
%has been 
reduced
to that of a \mbox{$2l \times 2l$} matrix, 
which is 
%a great simplification 
%very 
%quite 
a meaningful simplification 
for 
%if 
\mbox{$l<m$}. 
In particular, 
in the case of \mbox{$l=1$} 
%, the above system 
corresponding to the vector NLS equation
%, which was already 
considered 
%studied 
in the previous 
%sub-
subsection, we 
%only 
need only 
%to 
%consider the 
diagonalize a \mbox{$2 \times 2$} matrix 
to construct a 
%the 
one-dark soliton solution.  
%

%In the following, 
Now, we consider the 
%simple 
case of 
%where 
\mbox{$l=m$} corresponding to the square matrix NLS equation 
and assume that $A$ is a 
%assume a 
%an \mbox{$l \times l$} 
%square 
unitary matrix, 
%so that 
i.e., \mbox{$AA^\dagger=A^\dagger A=I_l$}. 
%so 
Thus, 
%Then, 
the first relation 
%condition 
in (\ref{red-cond}) is automatically satisfied and 
\mbox{$B= \left( 2 \mu \Gamma + K \right) A^\dagger$}, 
%We set \mbox{$B= S A^\dagger$}, 
where \mbox{$
%S:= 
2 \mu \Gamma + K$} 
is a constant Hermitian matrix. 
%identically. 
Actually, 
we can reduce  
%considering 
the seed plane-wave solution (\ref{plane2})
%we can reduce 
in this case 
to the simpler case of \mbox{$A=I_l$} 
%and \mbox{$B=B^\dagger$} 
by 
%a 
redefining 
%suitable 
%redefinition of 
$P_1$. 
%and $B$. 
Thus, in the following, 
we simply set 
\mbox{$A=I_l$} and \mbox{$B=B^\dagger$}. 
Then, it can be checked directly that 
(\ref{mRic}) is indeed 
%admits 
consistent with the reduction (\ref{Ric-red}). 
%In considering 
%Actually, 
%%Note that 
%without 
%%any 
%loss of generality, 
%%in the seed plane-wave solution (\ref{plane2}), 
%we can set \mbox{$A=A^\dagger=I_l$} by redefining $P_1$ 
%in the seed plane-wave solution (\ref{plane2}).    
%Then, 
%In this case, 
Because \mbox{$l=m$}, 
it is not so meaningful 
%effective 
to consider 
the linear problem 
%system 
(\ref{Phi-lin}) 
for $\boldsymbol{\mathit \Phi}_1$ and $\boldsymbol{\mathit \Phi}_2$. 
%can be rewritten as 
%does not provide an essential simplification of the problem, 
%so 
Thus, 
we 
%rather 
consider the original linear 
%system 
problem (\ref{NLS-UV2})
for $\boldsymbol{\mathit \Psi}_1$ and $\boldsymbol{\mathit \Psi}_2$ 
%at \mbox{$\zeta=\mu$} 
%original 
%problem 
%in a slightly rewritten form: 
in the form: 
\begin{subequations}
\label{NLS-UV3}
%\begin{equation}
%\left\{ \begin{array}{l}
\begin{align}
& \left[
\begin{array}{c}
 \mathrm{e}^{-\mathrm{i} \mu x- 2\mathrm{i} \mu^2 t} \boldsymbol{\mathit \Psi}_1  \\
 \mathrm{e}^{-\mathrm{i} \mu x- 2\mathrm{i} \mu^2 t} \boldsymbol{\mathit \Psi}_2 \\
\end{array}
\right]_x 
= \mathrm{i} \left[
\begin{array}{cc}
-2\mu I_l & I_l \\
-B & 
%2 \zeta I_l + 
\Gamma \\
\end{array}
\right] 
\left[
\begin{array}{c}
\mathrm{e}^{-\mathrm{i} \mu x- 2\mathrm{i} \mu^2 t} \boldsymbol{\mathit \Psi}_1  \\
\mathrm{e}^{-\mathrm{i} \mu x- 2\mathrm{i} \mu^2 t} \boldsymbol{\mathit \Psi}_2 \\
\end{array}
\right],
\label{NLS-mat1}
\\[1.5mm]
& \left[
\begin{array}{c}
 \mathrm{e}^{-\mathrm{i} \mu x - 2\mathrm{i} \mu^2 t} \boldsymbol{\mathit \Psi}_1  \\
 \mathrm{e}^{-\mathrm{i} \mu x - 2\mathrm{i} \mu^2 t} \boldsymbol{\mathit \Psi}_2 \\
\end{array}
\right]_t 
%= \mathrm{i} \left[
%\begin{array}{cc}
%-2\zeta^2 I_l + S & 2 \zeta I_l -\Gamma \\
% -2 \zeta S + \Gamma S & 2 \zeta^2 I_m + S - \Gamma^2 \\
%\end{array}
%\right]
%\left[
%\begin{array}{c}
%A^\dagger \boldsymbol{\mathit \Psi}_1  \\
% \boldsymbol{\mathit \Psi}_2 \\
%\end{array}
%\right]
%\nonumber \\ &
= \mathrm{i} 
%\left( 
\partial_x^2 
%- 2 \mathrm{i} \zeta \partial_x  
%+ 2 \zeta^2 
%%I 
%\right) 
\left[
\begin{array}{c}
 \mathrm{e}^{-\mathrm{i} \mu x - 2\mathrm{i} \mu^2 t} \boldsymbol{\mathit \Psi}_1  \\
 \mathrm{e}^{-\mathrm{i} \mu x - 2\mathrm{i} \mu^2 t} \boldsymbol{\mathit \Psi}_2 \\
\end{array}
\right], 
%\label{NLS-V2}
\end{align}
%\end{array}\right.
%\end{equation}
\end{subequations}
where $\Gamma$ is a real diagonal matrix: 
\mbox{$\Gamma = \mathrm{diag} (\gamma_1, \gamma_2, \ldots, \gamma_l)$}. 
%where $\gamma_j$ are real constants. 
Here, 
\begin{subequations}
\label{313}
\begin{align}
\widetilde{R} 
&= \mathrm{i} 
%P_2^{-1} 
\mathrm{e}^{-\mathrm{i} x \Gamma + \mathrm{i} t \Gamma^2} 
\left( \mathrm{e}^{-\mathrm{i} \mu x- 2\mathrm{i} \mu^2 t} \boldsymbol{\mathit \Psi}_2 
\right) \left( \mathrm{e}^{-\mathrm{i} \mu x- 2\mathrm{i} \mu^2 t} 
\boldsymbol{\mathit \Psi}_1 \right)^{-1} 
\mathrm{e}^{2\mathrm{i} t B} 
%P_1^{-1}
, 
%\nonumber 
\\[2mm]
%\end{align}
%
%and 
%
%\begin{align}
\widetilde{Q} &=  \mathrm{i} 
%P_1 
\mathrm{e}^{-2\mathrm{i} t B} 
\left[ \left( \mathrm{e}^{-\mathrm{i} \mu x- 2\mathrm{i} \mu^2 t}
\boldsymbol{\mathit \Psi}_2 \right) 
\left( \mathrm{e}^{-\mathrm{i} \mu x- 2\mathrm{i} \mu^2 t} 
 \boldsymbol{\mathit \Psi}_1 \right)^{-1} 
 - \Gamma - 2 \mu I_l \right]
\mathrm{e}^{\mathrm{i} x \Gamma - \mathrm{i} t \Gamma^2} 
%P_2
.
%\nonumber
\end{align}
\end{subequations}
%
%To be precise, $\zeta$ 
%in the above formulas
%%, 
%%$\zeta$ 
%should be replaced with $\mu$ as in (\ref{eBD1}), 
%but 
%%for 
%in the remainder 
%%remaining 
%of this section, 
%we omit this replacement for brevity. 
%the change \mbox{$\zeta=\mu$} as in . 
%replacement 
The characteristic polynomial of 
%equation for 
the 
%key 
square matrix in (\ref{NLS-mat1}) 
can be written (up to an overall constant) 
%sign) 
as 
%is
\begin{align}
%0 &= 
\det \left[
\begin{array}{cc}
(\lambda + 2\mu) I_l & -I_l \\
 B & 
%2 \zeta I_l + 
\lambda I_l - \Gamma \\
\end{array}
\right] 
%\nonumber \\[1mm]
&= 
\det \left[ ( \lambda + 2\mu ) \left( \lambda I_l - \Gamma \right) 
 + B
\right]. 
\label{quad-eigen}
\end{align}
Indeed, the \mbox{$2l \times 2l$} eigenvalue problem, 
\begin{equation}
\left[
\begin{array}{cc}
-2\mu I_l & I_l \\
-B & \Gamma \\
\end{array}
\right] 
\left[
\begin{array}{c}
\vt{f} \\
\vt{g} \\
\end{array}
\right] 
= \lambda 
\left[
\begin{array}{c}
\vt{f} \\
\vt{g} \\
\end{array}
\right], 
\label{fg-eigen}
\end{equation}
%can be reformulated as 
can be rewritten as an 
%is essentially equivalent to the 
\mbox{$l \times l$} nonstandard eigenvalue problem: 
\begin{equation}
\left[ (\lambda + 2 \mu) \left( \lambda I_l - 
%(\lambda + 2 \zeta) 
\Gamma \right)
+B  
\right] \vt{f} = \vt{0},
\label{nonst-eigen}
\end{equation}
where \mbox{$\vt{g}=(\lambda + 2 \mu) \vt{f}$}. 

Because of the complexity of the computation 
%and the solution, 
in the general case, 
%for
%when  
%%the general case of 
%for \mbox{$l \ge 3$}, 
we first 
%can 
consider 
%discuss only 
the 
%simple 
%simplest 
simpler special 
case 
where $B$ is a diagonal matrix. 
%, 
%By setting 
We set 
%\mbox{$
\[
B=\mathrm{diag}(b_1, b_2, \ldots, b_l), 
\hspace{5mm} b_j >0, 
\]
%$} 
%with \mbox{$b_j := (\xi_j -\gamma_j)^2 + \eta_j^2$} and 
%\mbox{$\zeta=$}, we obtain that
and assume that each diagonal element 
%entry 
%in 
of the matrix 
%in 
on the right-hand side of (\ref{quad-eigen}) 
can be factored as 
\[
 (\lambda + 2 \mu) (\lambda - \gamma_j) +b_j 
= \left[ \lambda - ( \xi_j +\mathrm{i} \eta_j ) \right]
	\left[ \lambda - ( \xi_j -\mathrm{i} \eta_j ) \right], 
\]
which is equivalent to 
%implies that 
%
\begin{align} 
\xi_j \pm \mathrm{i} \eta_j + \frac{b_j}{\xi_j \pm \mathrm{i} \eta_j-\gamma_j} 
	= -2\mu. 
%\label{chara4}
\nonumber
\end{align}
Thus, 
%From 
%(\ref{xi-eta}), 
%(\ref{chara4}), 
we obtain 
%have 
\begin{equation}
2\xi_j - \gamma_j = -2 \mu, \hspace{5mm} 
%\eta_j^2 = b_j - (\xi_j -\gamma_j)^2. 
 b_j = (\xi_j -\gamma_j)^2 + \eta_j^2.
\nonumber 
%\label{alpha_j}
\end{equation}
%\newpage

We consider that 
%for 
%real parameters 
$\gamma_j$, \mbox{$\eta_j(>0)$} and \mbox{$\mu
%\in \mathbb{R}
$} 
are free real parameters, which determine $\xi_j$ and $b_j$ as 
\begin{equation}
\xi_j =\frac{\gamma_j}{2} - \mu, \hspace{5mm} 
 b_j = \left( \frac{\gamma_j}{2} + \mu \right)^2 + \eta_j^2.
% (>0). 
\nonumber 
%\label{alpha_j2}
\end{equation}
Thus, the solution of the linear 
%system 
problem (\ref{NLS-UV3}) 
can be 
%given as 
written as 
\begin{align}
& \left( 
\mathrm{e}^{-\mathrm{i} \mu x- 2\mathrm{i} \mu^2 t} \boldsymbol{\mathit \Psi}_1 
\right)_{jk} = 
%\left[ 
 c_{jk} \mathrm{e}^{\mathrm{i} ( \xi_j +\mathrm{i} \eta_j )x 
	-\mathrm{i} ( \xi_j +\mathrm{i} \eta_j )^2 t}
+ d_{jk} \mathrm{e}^{\mathrm{i} ( \xi_j -\mathrm{i} \eta_j )x
	-\mathrm{i} ( \xi_j -\mathrm{i} \eta_j )^2 t} 
\nonumber \\
&= \mathrm{e}^{\mathrm{i} \left( \frac{\gamma_j}{2} - \mu  \right) x 
	-\mathrm{i} \left[ \left( \frac{\gamma_j}{2} - \mu  \right)^2 - \eta_j^2 \right] t}
\left\{ c_{jk} \mathrm{e}^{-\eta_j \left[ x 
	-\left( \gamma_j - 2 \mu  \right) t\right] }
+ d_{jk} \mathrm{e}^{\eta_j \left[ x 
	-\left( \gamma_j - 2 \mu \right) t\right]} \right\}, 
%\right]_{1 \le j,k \le l}, 
\nonumber \\[3mm]
& \left( \mathrm{e}^{-\mathrm{i} \mu x- 2\mathrm{i} \mu^2 t} \boldsymbol{\mathit \Psi}_2 
\right)_{jk} = 
( \xi_j +\mathrm{i} \eta_j + 2 \mu) c_{jk} \mathrm{e}^{\mathrm{i} 
( \xi_j +\mathrm{i} \eta_j )x -\mathrm{i} ( \xi_j +\mathrm{i} \eta_j )^2 t}
+ ( \xi_j -\mathrm{i} \eta_j + 2 \mu) d_{jk} \mathrm{e}^{\mathrm{i} 
( \xi_j -\mathrm{i} \eta_j )x -\mathrm{i} ( \xi_j -\mathrm{i} \eta_j )^2 t}
\nonumber \\ 
&= \mathrm{e}^{\mathrm{i} \left( \frac{\gamma_j}{2} - \mu  \right) x 
	-\mathrm{i} \left[ \left( \frac{\gamma_j}{2} - \mu \right)^2 - \eta_j^2 \right] t}
\left\{ \left(\frac{\gamma_j}{2} + \mu +\mathrm{i} \eta_j \right) c_{jk} 
	\mathrm{e}^{-\eta_j \left[ x 
	-\left( \gamma_j - 2 \mu \right) t\right]}
	+ \left( \frac{\gamma_j}{2} + \mu -\mathrm{i} \eta_j \right) 
	d_{jk} \mathrm{e}^{\eta_j \left[ x 
	-\left( \gamma_j - 2 \mu \right) t\right]} 
\right\}, 
\nonumber 
\end{align}
where $c_{jk}$ and $d_{jk}$ are arbitrary 
constants, 
\mbox{$1 \le j,k \le l$}. 
Using (\ref{313}) and omitting 
%$P_1$, $P_2$ and 
the tilde,  
%to denote the application of the elementary B\"acklund--Darboux transformation, 
we
%finally 
%obtain 
arrive at 
%the 
a one-dark soliton solution of the nonreduced 
matrix NLS system (\ref{mNLS})
in the form:
% as 
\begin{align}
%\widetilde{Q} 
Q 
%&=  \mathrm{i} P_1 \mathrm{e}^{-2\mathrm{i} t B} 
%\left[ \left( \mathrm{e}^{-\mathrm{i} \zeta x- 2\mathrm{i} \zeta^2 t}
%\boldsymbol{\mathit \Psi}_2 \right) 
%\left( \mathrm{e}^{-\mathrm{i} \zeta x- 2\mathrm{i} \zeta^2 t} 
% \boldsymbol{\mathit \Psi}_1 \right)^{-1} 
% - \Gamma - 2 \zeta I_l \right]
%\mathrm{e}^{\mathrm{i} x \Gamma - \mathrm{i} t \Gamma^2} P_2
%\nonumber \\
%&= -\mathrm{i} P_1 \mathrm{e}^{\mathrm{i} \frac{1}{2}x \Gamma 
%	- \mathrm{i} t \left(\frac{3}{4} \Gamma^2 + \zeta \Gamma + \eta^2 +\zeta^2 I_l \right)} 
%\left( \frac{1}{2}\Gamma + \zeta I_l + \mathrm{i} \frac{\partial Y}{\partial x} Y^{-1} \right)
%\mathrm{e}^{\mathrm{i} \frac{1}{2}x \Gamma 
%	- \mathrm{i} t \left(\frac{3}{4} \Gamma^2 + \zeta \Gamma + \eta^2 +\zeta^2 I_l 
%	\right)} P_2
%\nonumber \\
&= -\mathrm{i} \hspace{1pt}
%P_1 
\mathrm{e}^{\mathrm{i} \frac{1}{2}x \Gamma 
	- \mathrm{i} t \left(\frac{3}{4} \Gamma^2 + \mu \Gamma 
+ \mathcal{H}^2 +\mu^2 I_l \right)} 
\left\{ \frac{1}{2}\Gamma + \mu I_l -\mathrm{i} \mathcal{H} 
 \left[ \mathrm{e}^{-x\mathcal{H} + t \mathcal{H} 
 \left( \Gamma -2 \mu I_l \right) }
\mathcal{C} - \mathrm{e}^{x\mathcal{H} 
 -t \mathcal{H} \left( \Gamma -2 \mu I_l \right) }\mathcal{D} \right]
\right.
\nonumber \\ 
& \hphantom{=} \; \times \left. 
\left[ \mathrm{e}^{-x\mathcal{H} 
 + t \mathcal{H} \left( \Gamma -2 \mu I_l \right) }
\mathcal{C} + \mathrm{e}^{x\mathcal{H} 
 -t \mathcal{H} \left( \Gamma -2 \mu I_l \right) }\mathcal{D} 
 \right]^{-1}
\right\}
\mathrm{e}^{\mathrm{i} \frac{1}{2}x \Gamma 
	- \mathrm{i} t \left(\frac{3}{4} \Gamma^2 + \mu \Gamma + \mathcal{H}^2 +\mu^2 I_l 
	\right)} 
%P_2
,
%\nonumber
\label{m-one-dark} \\[2mm]
%\end{align}
%\begin{align}
%\widetilde{R} 
R 
%&= \mathrm{i} P_2^{-1} \mathrm{e}^{-\mathrm{i} x \Gamma + \mathrm{i} t \Gamma^2} 
%\left( \mathrm{e}^{-\mathrm{i} \zeta x- 2\mathrm{i} \zeta^2 t} \boldsymbol{\mathit \Psi}_2 
%\right) \left( \mathrm{e}^{-\mathrm{i} \zeta x- 2\mathrm{i} \zeta^2 t} 
%\boldsymbol{\mathit \Psi}_1 \right)^{-1} 
%\mathrm{e}^{2\mathrm{i} t B} P_1^{-1}
%\nonumber \\
%&= \mathrm{i} P_2^{-1} \mathrm{e}^{-\mathrm{i} \frac{1}{2}x \Gamma 
%	+ \mathrm{i} t \left(\frac{3}{4} \Gamma^2 + \zeta \Gamma + \eta^2 +\zeta^2 I_l \right)} 
%\left( \frac{1}{2}\Gamma + \zeta I_l - \mathrm{i} \frac{\partial Y}{\partial x} Y^{-1} \right)
%\mathrm{e}^{-\mathrm{i} \frac{1}{2}x \Gamma 
%	+ \mathrm{i} t \left(\frac{3}{4} \Gamma^2 + \zeta \Gamma + \eta^2 +\zeta^2 I_l 
%	\right)} P_1^{-1}
%\nonumber \\
&= \mathrm{i} \hspace{1pt}
%P_2^{-1} 
\mathrm{e}^{-\mathrm{i} \frac{1}{2}x \Gamma 
	+ \mathrm{i} t \left(\frac{3}{4} \Gamma^2 + \mu \Gamma + \mathcal{H}^2 +\mu^2 I_l \right)} 
\left\{ \frac{1}{2}\Gamma + \mu I_l 
 + \mathrm{i} \mathcal{H} \left[ \mathrm{e}^{-x\mathcal{H} 
 + t \mathcal{H} \left( \Gamma -2 \mu I_l \right) }
\mathcal{C} - \mathrm{e}^{x\mathcal{H} 
 -t \mathcal{H} \left( \Gamma -2 \mu I_l \right) }\mathcal{D} \right]
\right.
\nonumber \\ 
& \hphantom{=} \; \times \left. 
\left[ \mathrm{e}^{-x\mathcal{H} + t \mathcal{H} \left( \Gamma -2 \mu I_l \right) }
\mathcal{C} + \mathrm{e}^{x\mathcal{H} -t \mathcal{H} \left( \Gamma -2 \mu I_l 
\right) }\mathcal{D} 
 \right]^{-1} \right\}
\mathrm{e}^{-\mathrm{i} \frac{1}{2}x \Gamma 
	+ \mathrm{i} t \left(\frac{3}{4} \Gamma^2 + \mu \Gamma 
 + \mathcal{H}^2 +\mu^2 I_l 
	\right)} 
%P_1^{-1}
.
\nonumber 
\end{align}
%and
%
%where 
Here, \mbox{$\mathcal{H} 
%\mathscr{H}
:= \mathrm{diag} (\eta_1, \eta_2, \ldots, \eta_l)$}, 
\mbox{$\mathcal{C} := \left( c_{jk} \right)_{1 \le j,k \le l}$} 
and \mbox{$\mathcal{D} := \left( d_{jk} \right)_{1 \le j,k \le l}$}. 
The Hermitian conjugation reduction \mbox{$R=Q^\dagger$} 
%\mbox{$\widetilde{R}=\widetilde{Q}^{\hspace{1pt}\dagger}$}
can be realized by 
%requiring 
imposing 
the 
%additional 
%Hermiticity 
condition 
that the matrix $\mathcal{C}^\dagger \mathcal{H} \mathcal{D}$ is 
%should be 
Hermitian: 
\begin{align}
\left( \mathcal{C}^\dagger \mathcal{H} \mathcal{D} \right)^\dagger 
= \mathcal{C}^\dagger \mathcal{H} \mathcal{D}. 
\nonumber
\end{align}
%That is, the matrix $\mathcal{C}^\dagger \mathcal{D}$
%and 
%Because of the invariance of 
Because the above soliton solution is invariant under the 
change \mbox{$\mathcal{C} \to \mathcal{C}U$},  \mbox{$\mathcal{D} \to \mathcal{D}U$} 
%for any 
%with 
for any nonsingular 
%invertible 
\mbox{$l \times l$} matrix $U$, 
%so 
we 
%can 
assume without loss of generality that 
$\mathcal{C}^\dagger \mathcal{H} \mathcal{D}$ is a real diagonal matrix. 
%A simple sufficient condition is that 
%Typically, 
A simple way to satisfy this constraint 
%assumption 
%condition 
is to 
%choose 
%we may 
set 
the two matrices $\mathcal{C}$ and $\mathcal{D}$ as 
%are expressed as 
%in the form: 
%as 
\[
\mathcal{C} = V \left[
\begin{array}{cccc}
%\mu_1
\alpha_1 & & \\
& \alpha_2 & \\
& & \ddots & \\
& & & \alpha_l \\ 
\end{array}
\right], \hspace{5mm}
\mathcal{D} = \mathcal{H}^{-1} \left( V^{-1} \right)^\dagger \left[
\begin{array}{cccc}
%\nu_1 
\beta_1 & & \\
& \beta_2 & \\
& & \ddots \\ 
& & & \beta_l \\
\end{array}
\right],
\]
where 
%Here, 
$V$ is an 
%arbitrary 
\mbox{$l \times l$} 
nonsingular matrix
%, 
and \mbox{$\alpha_j, \hspace{1pt} \beta_j 
%\mu_j, \hspace{1pt} \nu_j 
%\ge 0, 
\in \mathbb{R} \;\, (j = 1, 2, \ldots, l)$}. 
%\mbox{$\alpha_j^\ast \beta_j \in \mathbb{R} \;\, (j = 1, 2, \ldots, l)$}. 
%
%
%The \mbox{$l \times l$} matrix $Y$ is defined as 
%\begin{align}
%\left( Y \right)_{jk} := 
%c_{jk} \mathrm{e}^{-\eta_j x +2 \xi_j \eta_j t}
%+ d_{jk} \mathrm{e}^{\eta_j x -2 \xi_j \eta_j t}, 
%\hspace{5mm} 1 \le j,k \le l.
%\nonumber
%\end{align}

To summarize, (\ref{m-one-dark}) provides a 
%non-generic 
special 
one-dark soliton solution 
of the matrix NLS equation (\ref{rmNLS}) 
in the self-defocusing case \mbox{$\sigma=+1$}; 
%the matrices $P_1$ and $P_2$ are unitary matrices, 
$\Gamma$ and $\mathcal{H}$ are real diagonal matrices, 
%$\zeta$ 
$\mu$ is a real parameter
%, 
and $\mathcal{C}^\dagger \mathcal{H} \mathcal{D}$ 
is a real diagonal matrix. 
In the more 
special case 
%wherein 
where $\Gamma$ is a scalar 
%diagonal 
matrix, i.e.\ 
\mbox{$\Gamma = \gamma I_l$}, (\ref{m-one-dark}) 
%can 
describes a 
%group of 
%traditional 
conventional 
dark 
%-type 
%pulse 
soliton 
moving with 
%a constant 
the velocity \mbox{$\gamma - 2 \mu$}; 
even this special one-soliton solution is 
more general than 
%that 
the 
%one-dark soliton 
%corresponding 
already known one-soliton 
solution derived 
%using the inverse scattering method 
%in [Ieda et al] 
using the inverse scattering method~\cite{Ieda07,Ieda06}. 
%-soliton solution
%can be considered as 
%, 
\\
\\
{\it Remark.} 
%The 
Another matrix NLS equation, 
\begin{equation}
%\label{redNLS}
 \mathrm{i} Q_t + Q_{xx} - 2Q Q^\ast Q = O,
\nonumber 
\end{equation} 
can be obtained 
%through 
by imposing the complex conjugation reduction 
\mbox{$R = Q^\ast$} 
%of the 
on the matrix NLS system (\ref{mNLS}). 
%This equation 
%admits a
A 
%special 
%the 
one-dark soliton solution 
%of the form 
of this equation is given by (\ref{m-one-dark}), 
where $\mathcal{C}$ and $\mathcal{D}$ are real matrices 
without any constraint on $\mathcal{C}^T \mathcal{H} \mathcal{D}$.
%$\mathcal{C}^\dagger \mathcal{D}$. 
\\

%Next, we 
%move on to 
%discuss 
%consider 
In the 
more 
general case 
where 
the matrix $B$ in the linear problem 
%system 
(\ref{NLS-UV3}) 
is a 
%(generally) 
non-diagonal Hermitian matrix, 
%In this case, 
%We can 
we rewrite (\ref{NLS-UV3}) as 
\begin{subequations}
\label{s-system}
%\label{NLS-UV3}
%\begin{equation}
%\left\{ \begin{array}{l}
\begin{align}
& \left[
\begin{array}{c}
%{l}
 \boldsymbol{\mathit S}_1 \\
 \boldsymbol{\mathit S}_2 \\
\end{array}
\right]_x 
%\nonumber \\ & 
= \mathrm{i} \left[
\begin{array}{cc}
-\mu I_l + \frac{1}{2} \Gamma & I_l \\
-B + \left( \mu I_l + \frac{1}{2}\Gamma \right)^2 & -\mu I_l + \frac{1}{2}\Gamma \\
\end{array}
\right] 
\left[
\begin{array}{c}
 \boldsymbol{\mathit S}_1 \\
 \boldsymbol{\mathit S}_2 \\
\end{array}
\right],
%\nonumber 
%\label{NLS-mat1}
\\[1.5mm]
& \left[
\begin{array}{c}
 \boldsymbol{\mathit S}_1 \\
 \boldsymbol{\mathit S}_2 \\
\end{array}
\right]_t 
= \mathrm{i} 
%\left( 
\partial_x^2 
\left[
\begin{array}{c}
 \boldsymbol{\mathit S}_1 \\
 \boldsymbol{\mathit S}_2 \\
\end{array}
\right]. 
%\nonumber 
%\label{NLS-V2}
\end{align}
%\end{array}\right.
%\end{equation}
\end{subequations}
%where 
Here, 
%the entries 
%\mbox{$l \times l$} matrices 
$\boldsymbol{\mathit S}_1$ and 
$\boldsymbol{\mathit S}_2$ are \mbox{$l \times l$} matrices 
defined as 
\[
\boldsymbol{\mathit S}_1 := \mathrm{e}^{-\mathrm{i} \mu x - 2\mathrm{i} \mu^2 t} 
\boldsymbol{\mathit \Psi}_1,
\hspace{5mm} \boldsymbol{\mathit S}_2 :=  \mathrm{e}^{-\mathrm{i} \mu x - 2\mathrm{i} \mu^2 t} \left( \boldsymbol{\mathit \Psi}_2 
 - \mu \boldsymbol{\mathit \Psi}_1
 - \frac{1}{2} \Gamma \boldsymbol{\mathit \Psi}_1 \right). 
\]
%; 
%however, 
%because of the complexity of the computation, 
%for practical reasons, 
%To 
In terms of these $\boldsymbol{\mathit S}_1$ and 
$\boldsymbol{\mathit S}_2$, 
%the matrix 
%one-dark soliton solution 
%obtained through 
formula 
(\ref{313}) can be 
%expressed 
rewritten 
in the symmetric form: 
%as  
%\begin{subequations}
%\label{313}
\begin{align}
\widetilde{R} 
&= \mathrm{i} 
%P_2^{-1} 
\mathrm{e}^{-\mathrm{i} x \Gamma + \mathrm{i} t \Gamma^2} 
\left( \mu I_l +\frac{1}{2}\Gamma + \boldsymbol{\mathit S}_2 \boldsymbol{\mathit S}_1^{-1} 
	 \right) \mathrm{e}^{2\mathrm{i} t B} 
%P_1^{-1}
, 
\nonumber 
\\[2mm]
%\end{align}
%
%
%\begin{align}
\widetilde{Q} &=  \mathrm{i} 
%P_1 
\mathrm{e}^{-2\mathrm{i} t B} 
\left(  - \mu I_l - \frac{1}{2} \Gamma + 
\boldsymbol{\mathit S}_2 \boldsymbol{\mathit S}_1^{-1} 
 \right)
\mathrm{e}^{\mathrm{i} x \Gamma - \mathrm{i} t \Gamma^2} 
%P_2
.
\nonumber
\end{align}
%\end{subequations}
Note that 
%the 
this solution is invariant under the change 
%transformation 
\mbox{$\boldsymbol{\mathit S}_j 
%\mapsto 
\to 
\boldsymbol{\mathit S}_j U \; (j=1,2)$} 
for any nonsingular \mbox{$l \times l$} 
constant matrix $U$.  
To realize 
the Hermitian conjugation reduction 
\mbox{$\widetilde{R}=\widetilde{Q}^{\hspace{1pt}\dagger}$}, 
%can be realized by choosing 
we need only choose the solution of the linear problem 
%system 
(\ref{s-system}) as 
\begin{align}
\left[
\begin{array}{c}
 \boldsymbol{\mathit S}_1 \\
 \boldsymbol{\mathit S}_2 \\
\end{array}
\right] &= \mathrm{exp}  \left\{  \mathrm{i} x \left[
\begin{array}{cc}
-\mu I_l + \frac{1}{2} \Gamma & I_l \\
-B + \left( \mu I_l + \frac{1}{2}\Gamma \right)^2 & -\mu I_l + \frac{1}{2}\Gamma \\
\end{array}
\right] \right.
\nonumber \\
& \hspace{12mm} 
\left. 
- \mathrm{i}t \left[
\begin{array}{cc}
-\mu I_l + \frac{1}{2} \Gamma & I_l \\
-B + \left( \mu I_l + \frac{1}{2}\Gamma \right)^2 & -\mu I_l + \frac{1}{2}\Gamma \\
\end{array}
\right]^2 
\right\}
\left[
\begin{array}{c}
 I_l \\
 \mathrm{i} J
%\boldsymbol{\mathit H} 
% \mathcal{K} \mathcal{A} 
%{\mathscr H}
\\
\end{array}
\right], 
\label{S-exp}
\end{align}
where $J$ 
%$\mathscr{H}$ 
is an \mbox{$l \times l$} constant Hermitian matrix. 
%as $B$. 
%is. 
%Note that $\zeta$ is a real parameter and 
Indeed, because 
\begin{align}
\left[
\begin{array}{c}
 \boldsymbol{\mathit S}_2 \\
 \boldsymbol{\mathit S}_1 \\
\end{array}
\right] &=
\left[
\begin{array}{cc}
 O & I_l \\
 I_l & O \\
\end{array}
\right] 
\left[
\begin{array}{c}
 \boldsymbol{\mathit S}_1 \\
 \boldsymbol{\mathit S}_2 \\
\end{array}
\right] 
\nonumber \\[1mm]
&= \mathrm{exp}  \left\{  \mathrm{i} x \left[
\begin{array}{cc}
-\mu I_l + \frac{1}{2} \Gamma & -B + \left( \mu I_l + \frac{1}{2}\Gamma \right)^2 \\
 I_l & -\mu I_l + \frac{1}{2}\Gamma \\
\end{array}
\right] \right.
\nonumber \\
& \hspace{12mm} 
\left. 
- \mathrm{i}t \left[
\begin{array}{cc}
-\mu I_l + \frac{1}{2} \Gamma & -B + \left( \mu I_l + \frac{1}{2}\Gamma \right)^2 \\
I_l & -\mu I_l + \frac{1}{2}\Gamma \\
\end{array}
\right]^2 
\right\}
\left[
\begin{array}{cc}
 O & I_l \\
 I_l & O \\
\end{array}
\right] 
\left[
\begin{array}{c}
 I_l \\
 \mathrm{i} J \\
\end{array}
\right], 
\nonumber 
\end{align}
we obtain 
%have 
the relation:
\[
\left[
\begin{array}{cc}
\! \boldsymbol{\mathit S}_2^\dagger \! &
\! \boldsymbol{\mathit S}_1^\dagger \! \\
\end{array}
\right]
%^\dagger
\left[
\begin{array}{c}
 \boldsymbol{\mathit S}_1 \\
 \boldsymbol{\mathit S}_2 \\
\end{array}
\right] 
= \left[
\begin{array}{cc}
\! - \mathrm{i} J \! &
\! I_l \! \\
\end{array}
\right]
\left[
\begin{array}{c}
 I_l \\
 \mathrm{i} J \\
\end{array}
\right] =O. 
\]
%Thus, we obtain 
This implies that 
\mbox{$\left( \boldsymbol{\mathit S}_2  \boldsymbol{\mathit S}_1^{-1} \right)^\dagger
= - \boldsymbol{\mathit S}_2  \boldsymbol{\mathit S}_1^{-1}$}, 
which indeed guarantees 
%confirms the Hermitian conjugation reduction 
the 
relation 
%reduction 
\mbox{$\widetilde{R}=\widetilde{Q}^{\hspace{1pt}\dagger}$}. 

To summarize, a general 
%the 
%matrix 
one-dark soliton solution of 
the matrix NLS equation 
(\ref{rmNLS}) in the self-defocusing case \mbox{$\sigma = +1$} 
%is given by 
can be expressed, up to unitary transformations,  
%using the matrix exponential function 
as 
\begin{align}
Q &=  
%\mathrm{i} P_1 
\mathrm{e}^{-2\mathrm{i} t B} 
\left(  \mu I_l + \frac{1}{2} \Gamma - 
\boldsymbol{\mathit S}_2 \boldsymbol{\mathit S}_1^{-1} 
 \right)
\mathrm{e}^{\mathrm{i} x \Gamma - \mathrm{i} t \Gamma^2}.  
%P_2.
%\nonumber
\label{gene-dark}
\end{align}
%with (\ref{S-exp}). 
Here, 
$\boldsymbol{\mathit S}_1$ and 
$\boldsymbol{\mathit S}_2$ 
%are defined 
%as 
%\mbox{$l \times l$} matrices 
%can be 
are given 
%determined 
%expressed 
using 
the matrix exponential 
%function 
as (\ref{S-exp}); $B$ and $J$ 
%$\mathscr{H}$ 
%is an \mbox{$l \times l$} constant 
are Hermitian matrices, $\Gamma$ is a real diagonal matrix 
and $\mu$ 
%$\zeta$ 
is a real parameter. 

%It is possible 
To express the above 
%dark-soliton 
solution 
more 
explicitly 
without using the matrix exponential as 
%introduced 
%used 
in (\ref{S-exp}), 
we need to 
%suitably 
re-parametrize the matrix $B$. 
%suitably.  
%in terms of the eigenvalues of the key matrix. 
%appropriately. 
%we 
%only 
%consider 
%discuss 
%only the simplest nontrivial 
%in 
We briefly consider 
the simplest nontrivial 
case of 
a \mbox{$2 \times 2$} Hermitian 
%non-diagonal 
matrix $B$, 
\begin{align}
B = \left[
\begin{array}{cc}
 b_{11}  & b_{12} \\
 b_{12}^\ast & b_{22} \\
\end{array}
\right], 
%\hspace{5mm} b_{12} \neq 0. 
%\nonumber 
\label{B2by2}
\end{align}
%Here, 
where \mbox{$b_{11}, \hspace{1pt} b_{22} \in \mathbb{R}$} 
and \mbox{$b_{12} \in \mathbb{C} 
%\backslash 
\setminus \{ 0 \}$}. 
%simplest 
Note that 
the eigenvalue problem (\ref{fg-eigen}) can be rewritten as 
\begin{equation}
 \left[
\begin{array}{cc}
-\mu I_l + \frac{1}{2} \Gamma & I_l \\
-B + \left( \mu I_l + \frac{1}{2}\Gamma \right)^2 & -\mu I_l + \frac{1}{2}\Gamma \\
\end{array}
\right] 
\left[
\begin{array}{c}
\vt{f} \\
\vt{g} - \mu \vt{f} - \frac{1}{2} \Gamma \vt{f} \\
\end{array}
\right] 
= \lambda 
\left[
\begin{array}{c}
\vt{f} \\
\vt{g} - \mu \vt{f} - \frac{1}{2} \Gamma \vt{f} \\
\end{array}
\right]. 
\nonumber 
\end{equation}
%which 
%can be reformulated as 
%can 
%also 
%be rewritten as an 
%is essentially equivalent to the 
%\mbox{$l \times l$} 
Thus, this 
also 
%This 
implies the relation \mbox{$\vt{g}=(\lambda + 2 \mu) \vt{f}$} 
and the nonstandard eigenvalue problem (\ref{nonst-eigen}). 
%\[
%\left[ \lambda (\lambda + 2 \zeta) I_l +B - (\lambda + 2 \zeta) \Gamma 
%\right] \vt{f} = \vt{0},
%\]
%where 
%and 
%as well as the relation \mbox{$\vt{g}=(\lambda + 2 \mu) \vt{f}$}. 
%Then, 
%Thus, 
We 
%require 
assume 
that  
the relevant characteristic polynomial 
%as given in 
(\ref{quad-eigen}) 
can be factored 
%factorized 
as 
\begin{align}
&
\det \left[
\begin{array}{cc}
(\lambda + 2\mu) (\lambda - \gamma_1) +b_{11}  & b_{12} \\
 b_{12}^\ast & (\lambda + 2\mu) (\lambda - \gamma_2) + b_{22} \\
\end{array}
\right] 
\nonumber \\[1mm]
& =  \left[ \lambda^2 + (2 \mu - \gamma_1) \lambda + b_{11} -2\mu \gamma_1 \right]
 \left[ \lambda^2 + (2 \mu - \gamma_2) \lambda + b_{22} -2\mu \gamma_2 \right] 
  - \left| b_{12} \right|^2 
%b_{12} b_{12}^\ast
\nonumber  \\[1mm] 
&= \left[ \lambda - ( \xi_1 +\mathrm{i} \eta_1 ) \right]
	\left[ \lambda - ( \xi_1 -\mathrm{i} \eta_1 ) \right]
 \left[ \lambda - ( \xi_2 +\mathrm{i} \eta_2 ) \right]
	\left[ \lambda - ( \xi_2 -\mathrm{i} \eta_2 ) \right]. 
%\nonumber
\label{fac1}
\end{align}
%where \mbox{$b_{11}, \hspace{1pt} b_{22} \in \mathbb{R}$}
%%$b_{11}$ and $b_{22}$ are real,  
%%parameters. 
%and \mbox{$b_{12} \in \mathbb{C}$}. 
%$b_{12}$ is complex. 
%Thus, 
%comparing the coefficients of the different powers of $\lambda$, 
%Because 
%\newpage \noindent
%We first 
%Next, 
We consider the 
generic 
%more 
%general 
case 
of 
\mbox{$\gamma_1 \neq \gamma_2$}, 
because the degenerate 
%special 
case 
%of 
\mbox{$\gamma_1 = \gamma_2$} 
does not 
lead to 
%provide 
an essentially 
%interesting 
%a 
new solution 
(cf.~(\ref{m-one-dark})). 
Noting that (\ref{fac1}) 
%this 
is an identity in $\lambda$, 
we obtain 
\begin{subequations}
\begin{align}
& 4 \mu - \gamma_1 - \gamma_2 = -2\xi_1 -2\xi_2, 
%\nonumber 
\label{coeff1}
\\[1mm]
& (2\mu - \gamma_1)(2\mu -\gamma_2) + b_{11} -2\mu \gamma_1 + b_{22} -2\mu \gamma_2
= \xi_1^2 + \eta_1^2 + \xi_2^2 + \eta_2^2 + 4 \xi_1 \xi_2, 
%- b_{11} - b_{22} 
%\nonumber 
\label{coeff2}
\\[1mm]
& (2 \mu - \gamma_2)(b_{11} -2\mu \gamma_1) 
 + (2 \mu - \gamma_1)(b_{22} -2\mu \gamma_2) 
 = -2 \xi_1 \left( \xi_2^2 + \eta_2^2 \right) - 2\xi_2 \left( \xi_1^2 + \eta_1^2 \right), 
%\nonumber 
\label{coeff3}
\\[1mm]
& (b_{11} -2\mu \gamma_1)(b_{22} -2\mu \gamma_2 ) - \left| b_{12} \right|^2 
 = \left( \xi_1^2 + \eta_1^2 \right)  \left( \xi_2^2 + \eta_2^2 \right).
%\nonumber 
\label{coeff4}
\end{align}
\end{subequations}
%Thus, 
%Using 
From (\ref{coeff1}), 
%the first relation, 
we can express $2\mu$ as 
\begin{equation}
2\mu = \frac{1}{2}(\gamma_1 + \gamma_2 ) - \xi_1 -\xi_2.
%\nonumber
\label{z-expr}
\end{equation}
%Substituting this into 
%Using 
%From 
Then, (\ref{coeff2}) 
%and (\ref{coeff3}) 
can be rewritten as  
%with the aid of 
%%this relation, 
%(\ref{z-expr}), 
%the second and third relations above, 
%we can 
%%write 
%express \mbox{$b_{11} -2\zeta \gamma_1$} and \mbox{$b_{22} -2\zeta \gamma_2$} as 
%obtain 
%have  
\begin{align}
&  (b_{11} -2\mu \gamma_1) + (b_{22} -2\mu \gamma_2)
= \eta_1^2 + \eta_2^2 + 2 \xi_1 \xi_2 
  + \frac{1}{4} (\gamma_1 -\gamma_2)^2. 
%\nonumber 
\label{coeff5}
%\\[1mm]
%& (2 \zeta - \gamma_2)(b_{11} -2\zeta \gamma_1) 
% + (2 \zeta - \gamma_1)(b_{22} -2\zeta \gamma_2) 
% = -2 \xi_1 \left( \xi_2^2 + \eta_2^2 \right) - 2\xi_2 \left( \xi_1^2 + \eta_1^2 \right), 
%\nonumber 
%\label{coeff6}
\end{align}
Thus, 
%using 
combining 
%this with 
%using 
(\ref{coeff3}) and (\ref{coeff5})  
and then using (\ref{coeff4}), 
we obtain
\begin{align}
& 
%(\gamma_1 - \gamma_2)(
b_{11} -2\mu \gamma_1
%) 
% = -2 \xi_1 \left( \xi_2^2 + \eta_2^2 \right) - 2\xi_2 \left( \xi_1^2 + \eta_1^2 \right)
%  - (2 \zeta - \gamma_1) \left[ \eta_1^2 + \eta_2^2 + 2 \xi_1 \xi_2 
%  + \frac{1}{4} (\gamma_1 -\gamma_2)^2 \right] 
\nonumber \\ 
&= \frac{\left( \xi_1 - \xi_2 \right)  \left( \eta_1^2 - \eta_2^2 \right)}
 {\gamma_1 - \gamma_2}
+\frac{1}{2} \left( \eta_1^2 + \eta_2^2 + 2 \xi_1 \xi_2 \right) 
  + \frac{1}{4} (\gamma_1 -\gamma_2) (\xi_1 + \xi_2)
  + \frac{1}{8} (\gamma_1 -\gamma_2)^2 , 
\nonumber \\[2mm]
& b_{22} -2\mu \gamma_2
\nonumber \\ 
&= -\frac{\left( \xi_1 - \xi_2 \right)  \left( \eta_1^2 - \eta_2^2 \right)}
 {\gamma_1 - \gamma_2}
+\frac{1}{2} \left( \eta_1^2 + \eta_2^2 + 2 \xi_1 \xi_2 \right) 
  - \frac{1}{4} (\gamma_1 -\gamma_2) (\xi_1 + \xi_2)
  + \frac{1}{8} (\gamma_1 -\gamma_2)^2 , 
\nonumber \\[2mm]
& \left| b_{12} \right|^2  
%= (b_{11} -2\zeta \gamma_1)(b_{22} -2\zeta \gamma_2 ) 
% - \left( \xi_1^2 + \eta_1^2 \right)  \left( \xi_2^2 + \eta_2^2 \right)
%\nonumber \\ &
= \left[ \frac{1}{2} \left( \eta_1^2 + \eta_2^2 + 2 \xi_1 \xi_2 \right) 
+ \frac{1}{8} (\gamma_1 -\gamma_2)^2 \right]^2 
- \left[ \frac{\left( \xi_1 - \xi_2 \right)  \left( \eta_1^2 - \eta_2^2 \right)}
 {\gamma_1 - \gamma_2} + \frac{1}{4} (\gamma_1 -\gamma_2) (\xi_1 + \xi_2) \right]^2
\nonumber \\ & \hphantom{\left| b_{12} \right|^2 =}\;
- \left( \xi_1^2 + \eta_1^2 \right)  \left( \xi_2^2 + \eta_2^2 \right). 
%\nonumber 
\label{b12}
%\label{coeff6}
\end{align}
%With the aid of 
Further using 
(\ref{z-expr}), 
%we can express the elements of 
the 
%entries in 
%elements of the \mbox{$2 \times 2$} 
matrix $B$ 
in 
(\ref{B2by2}) 
%can be expressed 
%is 
%can be 
%are 
%is 
can be 
determined from 
%in terms of 
$\xi_j$, $\eta_j$ and $\gamma_j$ \mbox{$(j=1,2)$}
%by 
through the relations 
%as 
%\begin{subequations}
%\label{}
\begin{align}
b_{11} &= \frac{\left( \xi_1 - \xi_2 \right)  \left( \eta_1^2 - \eta_2^2 \right)}
 {\gamma_1 - \gamma_2}
+\frac{1}{2} \left( \eta_1^2 + \eta_2^2 + 2 \xi_1 \xi_2 \right) 
  - \frac{1}{4} (3\gamma_1 +\gamma_2) (\xi_1 + \xi_2)
\nonumber \\ & \hphantom{=}\;
  + \frac{1}{8} \left( 5\gamma_1^2 +2 \gamma_1 \gamma_2 +\gamma_2^2 \right), 
%\nonumber 
\label{b11} \\[2mm] 
b_{22} 
%\nonumber 
& = -\frac{\left( \xi_1 - \xi_2 \right)  \left( \eta_1^2 - \eta_2^2 \right)}
 {\gamma_1 - \gamma_2}
 +\frac{1}{2} \left( \eta_1^2 + \eta_2^2 + 2 \xi_1 \xi_2 \right) 
  - \frac{1}{4} (\gamma_1 +3\gamma_2) (\xi_1 + \xi_2)
\nonumber \\ & \hphantom{=}\;
 + \frac{1}{8} \left( \gamma_1^2 + 2 \gamma_1 \gamma_2 +5\gamma_2^2 \right), 
\label{b22}
\end{align}
%\end{subequations}
and (\ref{b12}). 
Note that the right-hand side of (\ref{b12}) must be positive 
%nonnegative 
and we can freely choose \mbox{$\mathrm{arg} \, b_{12}$}. 

%Therefore, 
%Combining the above information, 
%Using 
%Combining 
From the above results, 
%these results, 
%Now, 
%we can express the 
%Now, 
the
%A 
%The 
%relevant 
solution (\ref{S-exp})
of the linear 
%system 
problem (\ref{s-system})
%, i.e., (\ref{S-exp}), 
%(\ref{NLS-UV3}) 
%is given by 
can be written 
explicitly as 
\begin{align}
%\left[
%\begin{array}{c}
% \mathrm{e}^{-\mathrm{i} \zeta x- 2\mathrm{i} \zeta^2 t} \boldsymbol{\mathit \Psi}_1  \\
% \mathrm{e}^{-\mathrm{i} \zeta x- 2\mathrm{i} \zeta^2 t} \boldsymbol{\mathit \Psi}_2 \\
%\end{array}
%\right] 
\left[
\begin{array}{c}
%{l}
 \boldsymbol{\mathit S}_1 \\
 \boldsymbol{\mathit S}_2 \\
\end{array}
\right]
&= W 
\left[
\begin{array}{cccc}
\! \mathrm{e}^{\mathrm{i}  ( \xi_1 +\mathrm{i} \eta_1 ) x 
	- \mathrm{i}  ( \xi_1 +\mathrm{i} \eta_1 )^2 t} \! \\
& \! \mathrm{e}^{\mathrm{i} ( \xi_1 -\mathrm{i} \eta_1 ) x 
 - \mathrm{i} ( \xi_1 -\mathrm{i} \eta_1 )^2 t} \! \\
& & \! \mathrm{e}^{\mathrm{i}  ( \xi_2 +\mathrm{i} \eta_2 ) x 
	- \mathrm{i}  ( \xi_2 +\mathrm{i} \eta_2 )^2 t} \! \\
& & & \! \mathrm{e}^{\mathrm{i} ( \xi_2 -\mathrm{i} \eta_2 ) x 
 - \mathrm{i} ( \xi_2 -\mathrm{i} \eta_2 )^2 t} \! \\
\end{array}
\right]
\nonumber \\ 
&\hphantom{=}\; 
\mbox{} \times W^{-1} 
\left[
\begin{array}{c}
 I_l \\
 \mathrm{i} J
%\boldsymbol{\mathit H} 
% \mathcal{K} \mathcal{A} 
%{\mathscr H}
\\
\end{array}
\right]. 
\nonumber 
\end{align}
Here, the \mbox{$4 \times 4$} constant matrix $W$ is defined as 
\begin{align}
W:= 
\left[ 
\begin{array}{cccc} 
\vt{v} (\xi_1 + \mathrm{i} \eta_1)  & \vt{v}  (\xi_1 - \mathrm{i} \eta_1) 
& \vt{v} (\xi_2 + \mathrm{i} \eta_2)  & \vt{v}  (\xi_2 - \mathrm{i} \eta_2) 
\end{array}
\right], 
\nonumber 
\end{align} 
%where 
using 
the four-component column vector
%s 
$\vt{v} (\lambda)$ 
%\mbox{$(j=1,2)$} 
%are defined 
%are 
given as 
\begin{align}
\vt{v} (\lambda) &:= 
\left[
\begin{array}{c}
- b_{12} \\
(\lambda + 2\mu) ( \lambda - \gamma_1) +b_{11}  \\
 - \left( \lambda+ \mu -\frac{1}{2} \gamma_1 \right)b_{12} \\
\left( \lambda + \mu - \frac{1}{2} \gamma_2 \right) \left[ 
 ( \lambda + 2\mu) ( \lambda - \gamma_1) + b_{11} \right] \\
\end{array}
\right]. 
\nonumber 
\end{align}
%
%Note 
%also 
%that 
%In addition, 
Incidentally, 
we can 
also 
%rewrite 
compute 
the matrix exponential 
\mbox{$\mathrm{e}^{-2\mathrm{i} t B} $} 
in (\ref{gene-dark}) 
%can be 
%expressed 
%computed 
explicitly, 
%wirtten 
%rewritten explicitly, 
%explicitly 
%in closed form, 
because $B$ is 
%just 
a \mbox{$2 \times 2$} Hermitian matrix. 

Thus, 
in contrast to the one-bright soliton solution of the self-focusing 
matrix NLS equation, 
the 
%obtained 
general one-dark soliton solution (\ref{gene-dark}) 
%of the self-defocusing matrix NLS equation 
in the self-defocusing case 
is hightly nontrivial 
and interesting. 
%itself. 
That is, it is far from the straightforward matrix generalization 
of the scalar dark-soliton solution as given by 
%
%Using (\ref{313}) and omitting the tilde,  
%%to denote the application of the elementary B\"acklund--Darboux transformation, 
%we finally 
%obtain 
%%the 
%a one-dark soliton solution of the matrix NLS system (\ref{mNLS}) as 
%
\begin{align}
Q(x,t) = U_1 
%Q_0  
\left[
\begin{array}{cccc}
f_1 (x,t) \\
& f_ 2 (x,t) \\
&& \ddots \\
&&& f_l (x,t) \\
\end{array}
\right]
U_2, 
\nonumber 
\end{align}
%
%\begin{align}
%Q= U_1 
%\mathrm{diag} \left(
%\mathrm{e}^{\mathrm{i} \gamma_j x - \mathrm{i}  \left[ \gamma_j^2 
%+ 2(\xi_j -\gamma_j)^2 + 2 \eta_j^2  \right] t
%	+ \mathrm{i} \theta_j} \left\{ (\xi_j - \gamma_j) - \mathrm{i} \eta_j
%	\tanh \left[ \eta_j ( x - 2 \xi_j t) +\delta_j \right] \right\} \right)_{
%%1 \le j \le l
%j=1,2,\ldots,l} U_2, 
%\nonumber 
%\end{align}
where $U_1$ and $U_2$ are 
%arbitrary 
constant unitary matrices and 
$f_j (x,t)$ are either a plane-wave solution or a scalar 
%one-dark 
dark-soliton solution
of the 
%scalar 
self-defocusing 
NLS equation. 
%
%(\ref{chara3})
%
%
%Then, 
%
%We set \mbox{$l=2$} and consider the general case 
%%\mbox{$2 \times 2$} matrix NLS equation.  
%wherein $B$ is 
%%, in general, 
%%not 
%a non-diagonal matrix in general. 
%
%
%We require 
%assume 
%that the eigenvalues appear in complex conjugate pairs as 

\section
%{Coalescence limit 
%%Limiting case 
%of a binary B\"acklund--Darboux transformation}
%
%\subsection{Taking the limit}
%\subsection
{Vector dark soliton with 
%a polarization}
internal degrees of freedom}
\label{sec5}

%\subsection{Matrix dark soliton with a polarization}

In this section, we apply the limiting case 
of 
%a 
the binary B\"acklund--Darboux transformation 
defined in Proposition~\ref{prop2.4} and 
construct a vector dark-soliton solution with 
internal degrees of freedom. 

%As the seed solution to apply 
%the B\"acklund--Darboux transformation, 
%we 
We 
consider 
the 
%seed 
plane-wave solution (\ref{vector-plane}) of 
the vector NLS system (\ref{vNLS}) 
in 
%and discuss 
the generic case where the 
wavenumbers 
%of the seed plane wave 
$\gamma_j$ 
%of 
%in 
%the seed plane-wave solution (\ref{vector-plane}) 
are pairwise distinct. 
Let us denote 
%In terms of 
the roots 
of the characteristic equation (\ref{chara1}) of the matrix $\boldsymbol{Z} (\mu)$ as 
\mbox{$\{ \lambda_1, \lambda_2, \ldots, \lambda_{m+1} \}$}; 
%\mbox{$\lambda_j \; (j=1, 2, \ldots, m)$},
%the left-hand side of 
%the 
%%Its 
%%the 
%characteristic equation (\ref{chara1})
%of which 
%then, 
thus, it 
%reads 
can be factored as 
%is 
%
\begin{align}
\det \left( \lambda I - \boldsymbol{Z} 
%(\zeta) 
\right) &= (\lambda + 2\mu 
%\zeta
) \prod_{j=1}^m (\lambda -\gamma_j) 
 + \sum_{k=1}^m a_k b_k \dprod{j=1}{j \neq k}^m (\lambda - \gamma_j) 
\nonumber \\
&= \prod_{j=1}^{m+1} (\lambda -\lambda_j) . 
%\label{chara1}
\nonumber 
\end{align}
%
%Thus, 
%Because this 
Noting that 
this is an identity in $\lambda$, 
%of degree \mbox{$m+1$}, 
%in the generic case where $\gamma_j$ are pairwise distinct, 
%by comparing the coefficients of the $\lambda^m$ terms and 
%setting \mbox{$\lambda = \gamma_k$}, 
we obtain 
%so it 
%which provides 
\mbox{$m+1$} relations: 
\begin{align}
& 2 \mu = \sum_{j=1}^m \gamma_j - \sum_{j=1}^{m+1} \lambda_j, \hspace{5mm}
a_k b_k = \frac{\displaystyle \prod_{j=1}^{m+1} (\lambda_j - \gamma_k)}
	{\displaystyle \dprod{j=1}{j \neq k}^m (\gamma_j - \gamma_k)} \hspace{1pt}, 
\hspace{5mm} k=1, 2, \ldots, m. 
%\nonumber 
\label{mu-ak-bk}
\end{align}

%In 
For simplicity, 
we 
%mainly 
%focus on 
consider 
%discuss 
the 
%three-
%\mbox{$(2n+1)$}
three-component case 
%(\mbox{$n \ge 1$}) 
(\mbox{$m=3$}) 
and 
%, we 
assume that 
\begin{subequations}
\label{gamma-ak-bk}
\begin{equation}
\gamma_1 < \gamma_2 < \gamma_3 
\;\;\, \mathrm{or} \;\;\, 
\gamma_1 > \gamma_2 > \gamma_3 
\end{equation}
%\hspace{5mm}
%\mathrm{and} 
and 
%\hspace{5mm}
\begin{equation}
a_1 b_1 > 0, \hspace{5mm} a_2 b_2 <0, \hspace{5mm} a_3 b_3 >0. 
%; 
%, 
\label{ab123}
\end{equation}
\end{subequations}
%In this case, 
%although 
%The 
%the generalization to 
%%the general $m$-component case 
%the 
The 
more general 
case 
%with 
of 
%more than three 
\mbox{$m \, (\ge 4)$} components 
%is 
%%almost 
%straightforward. 
%can be treated in 
%a similar manner. 
will be touched upon later. 
%essentially the same way. 
Then, 
%it is possible that the 
equation 
%for determining the eigenvalues of $\boldsymbol{Z} (\mu)$, 
(\ref{chara2}), 
which is equivalent to 
the characteristic equation (\ref{chara1}), 
%of 
%%the matrix 
%$\boldsymbol{Z} (\mu)$, 
%determines the eigenvalues of $\boldsymbol{Z} (\mu)$ 
%the number of real and imaginary roots depends on the value of $\mu$. 
can have two pairs of complex conjugate 
%complex-conjugate 
roots
%, if we choose the parameters appropriately. 
for 
%a suitable 
an appropriate 
choice of the parameters. 
%require that the roots of the characteristic equation 
Thus, we 
%set
%can 
parametrize 
%write 
%appear in 
the eigenvalues of the matrix $\boldsymbol{Z} (\mu)$ 
%appearing 
%that appear in complex-conjugate pairs 
as
%, i.e.\ 
\begin{align}
\lambda_1 = \xi_1 + \mathrm{i} \eta_1, \hspace{5mm}
\lambda_2 = \xi_1 - \mathrm{i} \eta_1, \hspace{5mm}
\lambda_3 = \xi_2 + \mathrm{i} \eta_2, \hspace{5mm}
\lambda_4 = \xi_2 - \mathrm{i} \eta_2, 
%\lambda_{2j-1} = \xi_j + \mathrm{i} \eta_j, \hspace{5mm}
%\lambda_{2j} = \xi_j - \mathrm{i} \eta_j, \hspace{5mm} 1 \le j \le n+1.
\nonumber 
\end{align}
where \mbox{$\xi_j \in \mathbb{R}$} and \mbox{$\eta_j > 0$}. 
From (\ref{mu-ak-bk}) 
with \mbox{$m=3$}, 
%Thus, 
we have 
\begin{align}
& 2 \mu = \gamma_1 + \gamma_2 + \gamma_3 - 2 \xi_1 - 2 \xi_2, 
%\hspace{5mm}
%\nonumber 
\label{3-mu}
\\[2mm]
& a_k b_k = \frac{\left[ (\xi_1 - \gamma_k)^2 + \eta_1^2 \right] 
  \left[ (\xi_2 - \gamma_k)^2 + \eta_2^2 \right]}
	{\displaystyle \dprod{j=1}{j \neq k}^3 (\gamma_j - \gamma_k)} 
\hspace{1pt}, 
\hspace{5mm} k=1, 2, 3. 
%, 
%\nonumber 
\label{3-akbk}
\end{align}
Note that (\ref{3-akbk}) 
%which 
is indeed consistent with (\ref{gamma-ak-bk}). 
%the conditions on $\gamma_j$ and $a_
%
%Note that \mbox{$\prod_{k=1}^3 a_k b_k <0$}, so at least one of $a_k b_k$ 
%must be negative. 

By setting 
\mbox{$m=3$} and 
\begin{equation}
b_1 = a_1^\ast, \hspace{5mm} b_2 = -a_2^\ast, \hspace{5mm} b_3 = a_3^\ast, 
\label{3-bk}
\end{equation}
%the inequality 
%(\ref{ab123}) is satisfied, and 
%we can realize the 
%obtain the seed
%complex conjugation reduction 
%\mbox{$ $}
%in 
%the inequality 
%(\ref{ab123}) is satisfied and 
the 
%seed 
plane-wave solution (\ref{vector-plane}) 
of the vector NLS system (\ref{vNLS}) 
%and obtain the plane 
reduces to the plane-wave solution of the vector NLS equation (\ref{rvNLS})
% \mathrm{i} \vt{q}_t + \vt{q}_{xx} - 2\sca{\vt{q}\Sigma}{\vt{q}^\ast} \vt{q} = \vt{0},
with 
%\mbox{$m=3$} and 
\mbox{$\Sigma = \mathrm{diag} (1, -1, 1)$}; 
%Note 
note that 
%the inequality 
(\ref{ab123}) is satisfied. 
%i.e.\ 
%That is, 
Thus, 
the 
%three-component 
vector NLS equation with one focusing and two defocusing components, 
\begin{equation}
 \mathrm{i} q_{j,t} + q_{j,xx} 
%\partial_t q_j + \partial_x^2 q_j
- 2 \left( |q_1|^2 - |q_2|^2 + |q_3|^2  \right) 
	q_j = 0, \hspace{5mm} j=1,2,3, 
\label{3vNLS}
\end{equation}
%admits 
has the plane-wave solution~\cite{Mak82}:
%given by  
\begin{equation}
q_j (x,t) = a_j \mathrm{e}^{\mathrm{i}  \gamma_j x - \mathrm{i} \left[ \gamma_j^2 
 + 2 \left( |a_1|^2 - |a_2|^2 + |a_3|^2 \right)  \right] t}, 
\hspace{5mm} j=1,2,3. 
\label{3plane}
%\nonumber 
\end{equation}
We employ 
%use 
%apply to 
this plane-wave solution as the seed solution 
and apply 
%the following 
a 
%slightly 
reduced version 
of Proposition~\ref{prop2.4}; 
%which 
%it 
this 
defines a B\"acklund--Darboux transformation
%that 
%preserves 
%preserving 
maintaining 
%a
the 
Hermitian 
%complex 
conjugation reduction 
%%\mbox{$\vt{r}= \vt{q}^\ast \Sigma $}
%\mbox{$\vt{r}^T = \Sigma \hspace{1pt} \vt{q}^\dagger$}
between 
the 
%two 
row and column 
vector 
%unknowns 
%dependent 
variables  
%$\vt{q}$ and $\vt{r}$. 
%appearing 
in the Lax pair. 
% as follows: 
%
\begin{proposition}
\label{prop5.1}
Consider 
the \mbox{$(m+1) \times (m+1)$} 
%following 
%The 
Lax-pair representation for the 
$m$-component vector NLS equation (\ref{rvNLS})
%system (\ref{mNLS}) is 
%given by~\cite{Zakh,Konop1} 
as given by 
\begin{subequations}
\label{vNLS-UV}
%\begin{equation}
%\left\{ \begin{array}{l}
\begin{align}
& \left[
\begin{array}{c}
% \psi_0  \\
%% \psi_2 \\
% \vdots \\
% \psi_{m} \\
\phi \\
\vt{\psi} \\
\end{array}
\right]_x 
= \left[
\begin{array}{cc}
-\mathrm{i}\mu  & \vt{q} \\
 \Sigma \hspace{1pt} \vt{q}^\dagger & \mathrm{i}\mu I_m\\
\end{array}
\right] 
\left[
\begin{array}{c}
%\psi_0 \\
%\vdots \\
%\psi_{m} \\
 \phi \\
 \vt{\psi} \\
\end{array}
\right],
\label{vNLS-U}
%\nonumber
\\[1.5mm]
& \left[
\begin{array}{c}
% \psi_0 \\
% \vdots \\
% \psi_{m} \\
\phi \\
\vt{\psi} \\
\end{array}
\right]_t 
= \left[
\begin{array}{cc}
-2\mathrm{i}\mu^2  -\mathrm{i} \vt{q} \Sigma \hspace{1pt} \vt{q}^\dagger 
	& 2 \mu \vt{q} + \mathrm{i} \vt{q}_x \\
 2 \mu \Sigma \hspace{1pt} \vt{q}^\dagger 
 - \mathrm{i} \Sigma \hspace{1pt} \vt{q}^\dagger_x 
	& 2\mathrm{i}\mu^2 I_m +\mathrm{i} \Sigma \hspace{1pt} \vt{q}^\dagger \vt{q} \\
\end{array}
\right]
\left[
\begin{array}{c}
% \psi_0 \\
% \vdots \\
% \psi_{m} \\
\phi \\
\vt{\psi} \\
\end{array}
\right].
\label{vNLS-V}
%\nonumber 
\end{align}
%\end{array}\right.
%\end{equation}
\end{subequations}
%where 
Here, 
the spectral parameter 
%$\zeta$ 
is set at a real value 
$\mu$, 
%is a real 
%%spectral 
%parameter, 
$\phi$ is a scalar 
function, 
$\vt{\psi}$ is an $m$-component column-vector 
function 
and \mbox{$\Sigma := \mathrm{diag} (\sigma_1, \ldots, \sigma_m)$} is a diagonal 
matrix with 
entries 
%each diagonal entry $\sigma_j$ equal to 
\mbox{$\sigma_j = +1$} or $-1$. 
%We require that at least one of $\sigma_j$ is negative
The quadratic form
%, 
%associated to a Hermitian form
%\begin{equation}
%\left| \psi_0 \right|^2 
% - \sum_{j=1}^{m} \sigma_j \left| \psi_j \right|^2 
\mbox{$\left| \phi \right|^2 
 - \vt{\psi}^\dagger\Sigma \vt{\psi}$}
%,
%\nonumber 
%\end{equation}
is 
%a constant
%\mbox{$(x,t)$}
independent of $x$ and $t$. 
%;
%, so we can 
Thus, 
if 
we consider 
%choose 
%the 
%vector 
a 
linear eigenfunction 
%in such a way that 
%we assume 
%%by requiring 
%%require 
%%by assuming 
%that 
%the linear eigenfunction 
%vanishes 
%it 
%decays 
%sufficiently 
decaying 
rapidly as \mbox{$x \to - \infty$}: 
\[
\lim_{x \to - \infty} 
\left[
\begin{array}{c}
\phi 
(x) 
\\
\vt{\psi} 
(x) 
\\
\end{array}
\right] 
= \vt{0}, 
\]
%we can 
%choose the 
%%vector 
%linear eigenfunction 
%in such a way that 
%so that 
this 
%the above 
quantity vanishes identically, i.e.\ 
\begin{equation}
%\left| \psi_0 \right|^2 
% - \sum_{j=1}^{m} \sigma_j \left| \psi_j \right|^2 
\left| \phi \right|^2 
 = \vt{\psi}^\dagger \Sigma \vt{\psi}. 
\nonumber 
\end{equation}
%
%Moreover, 
%To satisfy this condition, 
%we assume 
%This relation 
%%condition 
%is satisfied 
Then, 
the new potential 
$\vt{q}'$ 
%constructed 
defined 
by the formula
%,  
%as 
\begin{align}
%\skew{3}\widehat{\widetilde{Q}} 
%\breve{\vt{q}} 
\vt{q}' &:= 
 \vt{q} - \frac{\phi}{\displaystyle c_0 + \int^x_{-\infty} 
%\Phi_1 (\mu) 
\left| \phi (y) \right|^2 \mathrm{d} y} \hspace{1pt} 
 \vt{\psi}^\dagger \Sigma,
\nonumber
\end{align}
where $c_0$ is a real constant, 
provides 
%gives 
a new solution 
of the 
%$m$-component 
vector NLS equation (\ref{rvNLS}). 
A 
%bound-state 
%corresponding 
linear eigenfunction 
%for 
%corresponding to 
associated with 
the new potential $\vt{q}'$ 
at the spectral parameter equal to $\mu$ 
%(i.e., \mbox{$\zeta = \mu$})
is given by 
\begin{align}
 \frac{1}{\displaystyle c_0 + \int^x_{-\infty} \left| \phi (y) \right|^2 \mathrm{d} y}
\left[
\begin{array}{c}
\phi \\
\vt{\psi} \\
\end{array}
\right], 
%\nonumber 
%\label{vector-bound}
\nonumber 
\end{align}
which 
%can provide 
%gives 
defines 
%provides 
a 
%bound-state eigenfunction 
bound state for \mbox{$c_0>0$} and 
a proper choice of $\phi$ and $\vt{\psi}$. 
%Note that 
%By rescaling $\phi$ and $\vt{\psi}$, 
%we can 
%For 
Note that if \mbox{$c_0>0$}, 
we can set \mbox{$c_0 = 1$} by rescaling $\phi$ and $\vt{\psi}$ 
in the above formulas. 
%we can rescale the 
%set the value of $c_0$ as $1$. 
\end{proposition}

%In the following, 
%we 
%%set 
%%consider the case of 
%focus on 
In the three-component case 
%of 
(\mbox{$m=3$}), 
%case, 
%by 
%recalling the discussion 
%on the basis of the result 
%in subsections~\ref{sec2.6} and \ref{sec3.1}, 
%but 
%although 
%the generalization to 
%%the general $m$-component case 
%the case with more than three components 
%is 
%%almost 
%straightforward. 
%
we 
%can express 
choose 
a 
%particular 
%solution 
%%special 
%solution of 
linear eigenfunction 
%satisfying 
of the Lax-pair representation 
(\ref{vNLS-UV}) with 
$\vt{q}$ given by 
%the plane wave 
%plane-wave solution 
(\ref{3plane}) 
as (cf.~subsections~\ref{sec2.6} and \ref{sec3.1})
\begin{align}
& \left[
\begin{array}{c}
 \phi \\
 \vt{\psi} \\
\end{array}
\right]
=
\left[
\begin{array}{cccc}
 \mathrm{e}^{ - 2 \mathrm{i} 
\left( |a_1|^2 - |a_2|^2  + |a_3|^2 \right) 
t 
%\sum_{k=1}^3 a_k b_k
} \\
& \mathrm{i} \mathrm{e}^{-\mathrm{i}  \gamma_1 x + \mathrm{i} \gamma_1^2 t} \\
& & \mathrm{i} \mathrm{e}^{-\mathrm{i}  \gamma_2 x + \mathrm{i} \gamma_2^2 t} \\
& & & \mathrm{i} \mathrm{e}^{-\mathrm{i}  \gamma_3 x + \mathrm{i} \gamma_3^2 t} \\
\end{array}
\right] 
\nonumber \\[1mm]
& \times \mathrm{e}^{\mathrm{i} \mu x + 2 \mathrm{i} \mu^2 t}
\left\{ c_1 \mathrm{e}^{\mathrm{i} \left( \xi_1 - \mathrm{i}\eta_1 \right) x 
 - \mathrm{i}  \left( \xi_1 - \mathrm{i}\eta_1 \right)^2  t}
\left[
\begin{array}{c}
1 \\
-\frac{a_1^\ast
%b_1
}{\xi_1 - \mathrm{i}\eta_1 -\gamma_1} \\[1mm]
%\vdots \\
\frac{a_2^\ast
%-b_2
}{\xi_1 - \mathrm{i}\eta_1 -\gamma_2} \\[1mm]
-\frac{a_3^\ast
%b_3
}{\xi_1 - \mathrm{i}\eta_1 -\gamma_3} \\
\end{array}
\right]
%\nonumber \\[1mm] & \hphantom{=}\; \mbox{}
+ c_2 \mathrm{e}^{\mathrm{i} \left( \xi_2 - \mathrm{i}\eta_2 \right)  x
 - \mathrm{i}  \left( \xi_2 - \mathrm{i}\eta_2 \right)^2 t}
\left[
\begin{array}{c}
1 \\
-\frac{a_2^\ast
%b_1
}{\xi_2 - \mathrm{i}\eta_2 -\gamma_1} \\[1mm]
\frac{a_2^\ast
%-b_2
}{\xi_2 - \mathrm{i}\eta_2 -\gamma_2} \\[1mm]
%\vdots \\
-\frac{a_3^\ast
%b_3
}{\xi_2 - \mathrm{i}\eta_2 -\gamma_3} \\
\end{array}
\right] \right\}. 
\nonumber
%\label{mat-ex2}
\end{align}
%
%where 
Here, the parameters 
satisfy 
%are related through 
the 
%$a_j$ and $b_j$ are 
relations (\ref{3-mu})--(\ref{3-bk}); 
%among the parameters are assumed; 
$c_1$ and $c_2$ are 
%arbitrary 
%nonzero 
complex constants. 
Because \mbox{$\eta_j > 0$}, 
%the 
this linear eigenfunction 
%approaches zero 
decays exponentially 
%sufficiently rapidly 
as \mbox{$x \to - \infty$} and the 
%The 
quadratic form 
\mbox{$\left| \phi \right|^2 
 - \vt{\psi}^\dagger\Sigma \vt{\psi}$}
%does not contain 
%%involve 
%%any 
%a constant term and 
%%decays 
%approaches zero 
%%sufficiently rapidly 
%as \mbox{$x \to - \infty$}, so it 
%vanishes identically. 
%for all $x$ and $t$. 
is identically equal to zero. 
%identically. 
%as required in Proposition~\ref{prop5.1}. 
%which 
%%; they are assumed 
%%required 
%%to 
%should be nonzero in order to provide a nontrivial solution. 
%\begin{align}
%\Psi_1 = P_1 \mathrm{e}^{-2\mathrm{i} t AB} 
%	\boldsymbol{\mathit \Psi}_1, 
%\hspace{5mm} 
%\Psi_2 = \mathrm{i} P_2^{-1} \mathrm{e}^{-\mathrm{i} x \Gamma + \mathrm{i} t \Gamma^2} 
%\boldsymbol{\mathit \Psi}_2, 
%%\label{gauge1}
%\nonumber 
%\end{align}
%
%\begin{subequations}
%\label{}
%\begin{align}
%%\skew{3}\widehat{\widetilde{Q}} 
%\breve{Q} &= 
%  Q +  2 \Psi_1(\mu) F (\mu)^{-1} \Phi_2 (\mu)
%\nonumber \\
%&=   Q + 2 \hspace{1pt} \mathrm{i} \hspace{1pt} 
%	\mathrm{e}^{-2\mathrm{i} t \sca{\vt{a}}{\vt{b}}} \boldsymbol{\mathit \Psi}_1 
%	F^{-1} \boldsymbol{\mathit \Psi}_2^\dagger 
%	\mathrm{e}^{ \mathrm{i} x \Gamma - \mathrm{i} t \Gamma^2} \Sigma, 
%\\[0.5mm]
%%\skew{3}\widehat{\widetilde{R}}
%\breve{R} &= 
%  R - 2 \Psi_2 (\mu) F (\mu)^{-1} \Phi_1 (\mu). 
%\end{align}
%\end{subequations}
%%
%Note that 
%%In the above proposition, 
%we have introduced \mbox{$f := \frac{1}{2}F(\mu)$} as a 
%%the 
%solution of (\ref{F-xt}), 
%%instead of the limit (\ref{F-def}), 
%which can be informally written as 
%Thus, 
%Moreover, we obtain 
Then, we introduce a real function $f$ as 
\begin{align}
%F
f:= \;
%%(\mu) 
%&= 2C+ \int^x \left[ \Phi_1 (\mu) \Psi_1 (\mu) - \Phi_2 (\mu) \Psi_2 (\mu)
%	\right] \mathrm{d} x
%	\nonumber \\
& 
%\hphantom{=} \;
%= 
%2 \left[ 
c_0 + \int^x_{-\infty} 
%\Phi_1 (\mu) 
\left| \phi (y)
%\Psi_1 (\mu) 
\right|^2 
%\hspace{1pt} 
\mathrm{d} y
%\right]
	\nonumber \\[2mm]
=\; & 
%2 \left[ 
c_0 + \int^x_{-\infty}
\left| 
c_1 \mathrm{e}^{\mathrm{i} \left( \xi_1 - \mathrm{i}\eta_1 \right) y
	- \mathrm{i} \left( \xi_1 - \mathrm{i}\eta_1 \right)^2 t} 
	+ c_2 \mathrm{e}^{\mathrm{i}\left( \xi_2 - \mathrm{i}\eta_2 \right) y
	- \mathrm{i}  \left( \xi_2 - \mathrm{i}\eta_2 \right)^2 t} \right|^2 
\mathrm{d} y
%\right]
\nonumber \\[2mm]
=\; & 
%2 \left\{ 
 c_0 +\frac{|c_1|^2}{2 \eta_1} 
 \mathrm{e}^{2 \eta_1 ( x - 2 \xi_1 t)}
 +\frac{|c_2|^2}{2 \eta_2} 
 \mathrm{e}^{2 \eta_2 ( x - 2 \xi_2 t) }
% \right. 
 \nonumber \\[1mm]
 \; & \hphantom{c_0}
%\hphantom{=} \; \, 
%\hspace{10mm} 
%\left.  
 \mbox{} + \mathrm{e}^{\eta_1 ( x - 2 \xi_1 t) + \eta_2 ( x - 2 \xi_2 t)} 
 \left[ \frac{c_1 c_2^\ast}{\eta_1 + \eta_2 + \mathrm{i} (\xi_1 -\xi_2)} 
   \mathrm{e}^{\mathrm{i} (\xi_1 -\xi_2) x 
	- \mathrm{i} (\xi_1^2 - \xi_2^2 - \eta_1^2 + \eta_2^2) t} 
\right. 
%\right.
 \nonumber \\[1mm]
 & \hspace{43mm} 
%\left. 
\left. \mbox{}
	+ \frac{c_1^\ast c_2}{\eta_1 + \eta_2 - \mathrm{i} (\xi_1 -\xi_2)} 
   \mathrm{e}^{-\mathrm{i} (\xi_1 -\xi_2) x 
	+ \mathrm{i} (\xi_1^2 - \xi_2^2 - \eta_1^2 + \eta_2^2) t} 
 \right]. 
% \right\}. 
%\nonumber 
\label{f-def}
\end{align}
%where $C$ is 
%with a real constant 
%Here, we assume 
%that 
By assuming \mbox{$c_0 >0$}, 
%so 
%that 
%$F$ 
%Note that 
$f$ becomes positive definite. 
%if \mbox{$c_0 >0$}. 
%arbitrary 
%\mbox{$l \times l$} constant matrix $C$; 
%Using 
%this positive function 
%%the function 
%$f$ and 
We also introduce 
%define 
complex functions 
\mbox{$g_j \, (j=1, 2, 3)$} 
%defined 
as 
\begin{align}
g_j 
%&:= \left\{ c_1 \mathrm{e}^{\mathrm{i} x \left( \xi_1 
%	- \mathrm{i}\eta_1 \right) 
%	- \mathrm{i}  t \left( \xi_1 - \mathrm{i}\eta_1 \right)^2 } 
% + c_2 \mathrm{e}^{\mathrm{i} x \left( \xi_2 - \mathrm{i}\eta_2 
%\right) - \mathrm{i} t  \left( \xi_2 - \mathrm{i}\eta_2 \right)^2} \right\}
%\nonumber \\[1mm]
%& \hphantom{:=} \; 
%%\, 
%\mbox{} 
%\times  \left\{ \frac{c_1^\ast}{\xi_1 + \mathrm{i}\eta_1 -\gamma_j} 
% \mathrm{e}^{\mathrm{i} x \left( - \xi_1 - \mathrm{i}\eta_1 \right) 
%	+ \mathrm{i}  t \left( \xi_1 + \mathrm{i}\eta_1 \right)^2 } 
% + \frac{c_2^\ast}{\xi_2 + \mathrm{i}\eta_2 -\gamma_j}  
%\mathrm{e}^{\mathrm{i} x \left( - \xi_2 - \mathrm{i}\eta_2 
%\right) + \mathrm{i}  t \left( \xi_2 + \mathrm{i}\eta_2 \right)^2} \right\}
%\nonumber \\[2mm]
&:= \left\{ c_1 \mathrm{e}^{\eta_1 ( x - 2 \xi_1 t) + \mathrm{i} \xi_1 x 
	- \mathrm{i} (\xi_1^2 - \eta_1^2) t} 
 + c_2 \mathrm{e}^{ \eta_2 ( x - 2 \xi_2 t) + \mathrm{i} \xi_2 x 
	- \mathrm{i} (\xi_2^2 - \eta_2^2) t} \right\}
\nonumber \\[1mm]
& \hphantom{:=} \; 
%\, 
\mbox{} 
\times  \left\{ \frac{c_1^\ast}{\xi_1 + \mathrm{i}\eta_1 -\gamma_j} 
 \mathrm{e}^{\eta_1 ( x - 2 \xi_1 t) - \mathrm{i} \xi_1 x 
	+ \mathrm{i} (\xi_1^2 - \eta_1^2) t} 
 + \frac{c_2^\ast}{\xi_2 + \mathrm{i}\eta_2 -\gamma_j} 
 \mathrm{e}^{ \eta_2 ( x - 2 \xi_2 t) - \mathrm{i} \xi_2 x 
	+ \mathrm{i} (\xi_2^2 - \eta_2^2) t} \right\}
\nonumber \\[2mm]
& \hphantom{:} =
\frac{|c_1|^2}{\xi_1 + \mathrm{i}\eta_1 -\gamma_j} \mathrm{e}^{2 \eta_1 ( x - 2 \xi_1 t)}
 +\frac{|c_2|^2}{\xi_2 + \mathrm{i}\eta_2 -\gamma_j} 
 \mathrm{e}^{2 \eta_2 ( x - 2 \xi_2 t) }
  \nonumber \\[1mm]
 & \hphantom{:=} \; \, 
%\hspace{10mm} 
%\left.  
 \mbox{} + \mathrm{e}^{\eta_1 ( x - 2 \xi_1 t) + \eta_2 ( x - 2 \xi_2 t)} 
 \left[ \frac{c_1 c_2^\ast}{\xi_2 + \mathrm{i}\eta_2 -\gamma_j} 
   \mathrm{e}^{\mathrm{i} (\xi_1 -\xi_2) x 
	- \mathrm{i} (\xi_1^2 - \xi_2^2 - \eta_1^2 + \eta_2^2) t} 
\right. 
%\right.
 \nonumber \\[1mm]
 & \hphantom{:}\hspace{45mm} 
%\left. 
\left. \mbox{}
	+ \frac{c_1^\ast c_2}{\xi_1 + \mathrm{i}\eta_1 -\gamma_j} 
   \mathrm{e}^{-\mathrm{i} (\xi_1 -\xi_2) x 
	+ \mathrm{i} (\xi_1^2 - \xi_2^2 - \eta_1^2 + \eta_2^2) t} 
 \right]. 
\label{gj-def}
\end{align}
%
%Using 
%the positive function 
%%the function 
%$f$ and complex functions 
%\mbox{$g_j \, (j=1, 2, 3)$}, 
%defined above 
%Then, 
Thus, 
with the aid of 
Proposition~\ref{prop5.1}, 
we 
%can 
obtain 
%express 
%the 
a one-dark soliton solution 
of the three-component 
%vector 
NLS equation (\ref{3vNLS})
in the form: 
%as  
%
\begin{align}
%\vt{q} 
q_j (x,t) & = a_j\mathrm{e}^{\mathrm{i}  \gamma_j x - \mathrm{i} 
\left[ 
\gamma_j^2 
 + 2 \left( |a_1|^2 - |a_2|^2  + |a_3|^2 \right)
%\sum_{k=1}^m a_k b_k 
\right] 
t} 
 \left\{ 1 - \mathrm{i} \frac{g_j}{f}
 \right\}, \hspace{5mm} j=1, 2, 3.
\label{v-soliton1}
\end{align}
Here, the positive function 
%the function 
$f$ and the 
complex functions 
\mbox{$g_j 
%\, (j=1, 2, 3)
$} 
are given 
%as 
by (\ref{f-def}) and (\ref{gj-def}), respectively;  
%and 
the amplitudes of the background plane waves are 
%defined 
determined as (cf.~(\ref{3-akbk}) and (\ref{3-bk}))
%through the relations: 
\begin{align}
& |a_1|^2 = \frac{\left[ (\xi_1 - \gamma_1)^2 + \eta_1^2 \right] 
  \left[ (\xi_2 - \gamma_1)^2 + \eta_2^2 \right]}
	{(\gamma_2 - \gamma_1)(\gamma_3 - \gamma_1)}, 
\nonumber \\[2mm]
& |a_2|^2 = \frac{\left[ (\xi_1 - \gamma_2)^2 + \eta_1^2 \right] 
  \left[ (\xi_2 - \gamma_2)^2 + \eta_2^2 \right]}
	{(\gamma_2 - \gamma_1)(\gamma_3 - \gamma_2)}, 
\nonumber \\[2mm]
& |a_3|^2 = \frac{\left[ (\xi_1 - \gamma_3)^2 + \eta_1^2 \right] 
  \left[ (\xi_2 - \gamma_3)^2 + \eta_2^2 \right]}
	{(\gamma_3 - \gamma_1)(\gamma_3 - \gamma_2)}. 
\nonumber 
\end{align}
%Note that the following relation holds: 
%\[
%|a_1|^2 + |a_3|^2 = |a_2|^2. untrue 
%\]
%Each 
%This 
%For 
When 
\mbox{$c_1=0$} or \mbox{$c_2=0$}, 
%either 
%$c_1$ or $c_2$ vanishes, 
we have 
%a stationary 
%an essentially 
a conventional 
%``static" 
dark 
soliton, which has essentially 
a time-independent shape 
%, 
and velocity 
%and polarization 
(cf.~(\ref{v-dark})). 
%However, 
For nonzero values of $c_1$ and $c_2$, 
the vector dark soliton (\ref{v-soliton1}) 
%has 
%possesses 
entails 
the 
internal degrees of freedom 
%contains a complex free parameter $c_1/c_2$ 
and 
%is 
%%no longer a 
%%stantionary 
%an essentially non-static 
%solution; 
exhibits 
%a 
complicated behavior; 
%non-stationary; 
in fact, 
the 
soliton's shape
%, 
and velocity 
%and polarization
%it has different shape, 
can change in time, 
%; 
%,
%as time passes from $-\infty$ to $+\infty$, 
so 
%in particular, 
the final state 
%of the soliton 
%may be 
%can be totally 
is generally 
different from 
%its 
the 
initial state. 
%This can be called 
%We may call it 
This phenomenon 
can 
%could 
be called 
a 
%soliton mutation phenomenon. 
spontaneous soliton mutation. 
%metamorphosis. 

%Because of 
%the existence of 
%%the 
%two 
%%complex 
%parameters $c_1$ and $c_2$, 
%implies that 
%Note that the 
%the 
%this 
%soliton 
%solution 
%with 
%admits 
%has 
%internal degrees of freedom 
%%because of the existence of the two complex parameters $c_1$ and $c_2$ 
%and 
%has 
%a time-dependent shape. 
%its shape changes in time. 
%is time-dependent. 

In a similar manner, 
%way, 
%the same manner, 
%Note that 
%the above construction of 
we can 
%also 
obtain 
%construct 
a 
%the 
one-dark soliton solution 
%can be generalized to the case 
%with $m$ components
of the general $m$-component NLS equation (\ref{rvNLS}) 
%in a similar way. 
%as
in the same form as (\ref{v-soliton1}): 
\begin{align}
%\vt{q} 
q_j (x,t) & = a_j\mathrm{e}^{\mathrm{i}  \gamma_j x - \mathrm{i} \left( \gamma_j^2 
 + 2 \sum_{k=1}^m \sigma_k |a_k|^2 \right) t} 
 \left\{ 1 - \mathrm{i} \frac{g_j}{f}
 \right\}, 
\hspace{5mm} j=1, 2, 
%3
\ldots, m.
\label{v-soliton2}
\end{align}
%Here, 
In 
the general 
%%in 
%%For the 
$m$-component 
%this 
%general 
case, 
%(\mbox{$m \ge 3$}), 
$f$ and $g_j$ 
%for the $m$-component case 
are given 
%defined 
as 
\begin{align}
%F
f:= \;
%%(\mu) 
%&= 2C+ \int^x \left[ \Phi_1 (\mu) \Psi_1 (\mu) - \Phi_2 (\mu) \Psi_2 (\mu)
%	\right] \mathrm{d} x
%	\nonumber \\
& 
%\hphantom{=} \;
%= 
%%2 \left[ 
%c_0 + \int^x_{-\infty} 
%%\Phi_1 (\mu) 
%\left| \phi
%%\Psi_1 (\mu) 
%\right|^2 
%%\hspace{1pt} 
%\mathrm{d} x 
%%\right]
%	\nonumber \\[2mm]
%=\; & 
%2 \left[ 
c_0 + \int^x_{-\infty}
\left| \sum_{\alpha =1}^{m_1}
c_\alpha \mathrm{e}^{\mathrm{i} \left( \xi_\alpha - \mathrm{i}\eta_\alpha \right) y
	- \mathrm{i} \left( \xi_\alpha - \mathrm{i}\eta_\alpha \right)^2 t} \right|^2 
\mathrm{d} y
%\right]
\nonumber \\[2mm]
=\; & 
%2 \left\{ 
 c_0 +\sum_{\alpha=1}^{m_1} \frac{|c_\alpha|^2}{2 \eta_\alpha} 
 \mathrm{e}^{2 \eta_\alpha ( x - 2 \xi_\alpha t)}
% \right. 
 \nonumber \\[1mm]
 \; & 
%\hphantom{c_0}
%\hphantom{=} \; \, 
%\hspace{10mm} 
%\left.  
 \mbox{} + \sum_{1 \le \alpha < \beta \le m_1 } 
 \mathrm{e}^{\eta_\alpha ( x - 2 \xi_\alpha t) + \eta_\beta ( x - 2 \xi_\beta t)} 
 \left[ \frac{c_\alpha c_\beta^\ast}{\eta_\alpha + \eta_\beta 
	+ \mathrm{i} (\xi_\alpha -\xi_\beta)} 
   \mathrm{e}^{\mathrm{i} (\xi_\alpha -\xi_\beta) x 
	- \mathrm{i} (\xi_\alpha^2 - \xi_\beta^2 - \eta_\alpha^2 + \eta_\beta^2) t} 
\right. 
%\right.
 \nonumber \\[1mm]
 & \hspace{43mm} 
%\left. 
\left. \mbox{}
	+ \frac{c_\alpha^\ast c_\beta}{\eta_\alpha + \eta_\beta 
	- \mathrm{i} (\xi_\alpha -\xi_\beta)} 
   \mathrm{e}^{-\mathrm{i} (\xi_\alpha -\xi_\beta) x 
	+ \mathrm{i} (\xi_\alpha^2 - \xi_\beta^2 - \eta_\alpha^2 + \eta_\beta^2) t} 
 \right],
% \right\}. 
\nonumber 
%\label{f-def2}
\end{align}
%\mbox{$g_j \, (j=1, 2, 3)$} 
%defined 
%as 
\vspace{2mm}
\begin{align}
g_j &:= \left[ \sum_{\alpha=1}^{m_1} c_\alpha 
	\mathrm{e}^{\eta_\alpha ( x - 2 \xi_\alpha t) + \mathrm{i} \xi_\alpha x 
	- \mathrm{i} (\xi_\alpha^2 - \eta_\alpha^2) t} \right]
%\times 
\left[ \sum_{\beta=1}^{m_1} \frac{c_\beta^\ast}{\xi_\beta 
	+ \mathrm{i}\eta_\beta -\gamma_j} 
 \mathrm{e}^{\eta_\beta ( x - 2 \xi_\beta t) - \mathrm{i} \xi_\beta x 
	+ \mathrm{i} (\xi_\beta^2 - \eta_\beta^2) t} \right]
\nonumber \\[2mm]
& \hphantom{:} =
\sum_{\alpha=1}^{m_1}
\frac{|c_\alpha|^2}{\xi_\alpha + \mathrm{i}\eta_\alpha -\gamma_j} 
	\mathrm{e}^{2 \eta_\alpha ( x - 2 \xi_\alpha t)}
 \nonumber \\[1mm]
 & \hphantom{:=} \; \, 
%\hspace{10mm} 
%\left.  
 \mbox{} + \sum_{1 \le \alpha < \beta \le m_1 } 
 \mathrm{e}^{\eta_\alpha ( x - 2 \xi_\alpha t) + \eta_\beta ( x - 2 \xi_\beta t)} 
 \left[ \frac{c_\alpha c_\beta^\ast}{\xi_\beta + \mathrm{i}\eta_\beta -\gamma_j} 
   \mathrm{e}^{\mathrm{i} (\xi_\alpha -\xi_\beta) x 
	- \mathrm{i} (\xi_\alpha^2 - \xi_\beta^2 - \eta_\alpha^2 + \eta_\beta^2) t} 
\right. 
%\right.
 \nonumber \\[1mm]
 & \hphantom{:}\hspace{45mm} 
%\left. 
\left. \mbox{}
	+ \frac{c_\alpha^\ast c_\beta}{\xi_\alpha + \mathrm{i}\eta_\alpha -\gamma_j} 
   \mathrm{e}^{-\mathrm{i} (\xi_\alpha -\xi_\beta) x 
	+ \mathrm{i} (\xi_\alpha^2 - \xi_\beta^2 - \eta_\alpha^2 + \eta_\beta^2) t} 
 \right]. 
%\label{gj-def2}
\nonumber
\end{align}
%Then, 
%we can express 
The sign $\sigma_k$ 
%of 
in 
%the 
each nonlinear 
%terms 
term in (\ref{rvNLS}) 
%$\sigma_j$ 
and 
the amplitude $|a_k|$ 
%amplitudes 
of each background plane wave 
%the plane waves 
%$|a_j|^2$ 
are determined 
%as 
through the relations: 
%relation 
%
%\mbox{$\{ \lambda_1, \lambda_2, \ldots, \lambda_m \} 
%= \{ \xi_1 + \mathrm{i} \eta_1, \xi_1 - \mathrm{i} \eta_1,\ldots, 
%\xi_{m_1} + \mathrm{i} \eta_{m_1}, \xi_{m_1} - \mathrm{i} \eta_{m_1}, 
%\nu_1, \ldots, \nu_{m_2} \}$}
%
\begin{align}
%& 2 \mu = \sum_{j=1}^m \gamma_j - \sum_{j=1}^{m+1} \lambda_j, \hspace{5mm}
& \sigma_k |a_k|^2 = \frac{\displaystyle \prod_{i=1}^{m_1}
\left[ (\xi_i - \gamma_k)^2 + \eta_i^2 \right] \times 
\prod_{j=1}^{m_2} (\nu_j - \gamma_k)
%(\lambda_j - \gamma_k)
}
	{\displaystyle \dprod{j=1}{j \neq k}^m (\gamma_j - \gamma_k)} \hspace{1pt}, 
\hspace{5mm} k=1, 2, \ldots, m. 
\nonumber 
%\label{ak-bk2}
\end{align}
%
%We 
%%set 
%divide 
%express 
Here, 
%we set \mbox{$b_j = \sigma_j a_j^\ast \, (\sigma_j = 
%\pm 1
%%+1 \; \mathrm{or} -1
%)$} 
%in the seed plane-wave solution (\ref{vector-plane}); 
%the characteristic equation 
%(\ref{chara1})
%; 
%thus, 
%then ,
%and 
we assume that 
%we 
%%have 
%assumed that 
the 
%\mbox{$m+1
%%= 2 m_1 + m_2
%$} 
roots \mbox{$\{ \lambda_1, \lambda_2, \ldots, \lambda_{m+1} \}$} 
of 
%the characteristic 
equation (\ref{chara1}) 
%with real coefficients 
%with \mbox{$b_j = \sigma_j a_j^\ast$}
with \mbox{$b_k = \sigma_k a_k^\ast$}
%as 
%into 
%consist of 
comprise 
$m_1$ pairs of complex conjugate 
%complex-conjugate 
roots 
\mbox{$\{ \xi_1 \pm \mathrm{i} \eta_1, 
%\xi_1 - \mathrm{i} \eta_1,
\ldots, 
\xi_{m_1} \pm \mathrm{i} \eta_{m_1}
%, \xi_{m_1} - \mathrm{i} \eta_{m_1} 
\}$}
%, \mbox{$\eta_j > 0$}
%where 
%(\mbox{$\xi_j \in \mathbb{R}$}, 
%and 
(\mbox{$\eta_j > 0$})
and $m_2$ real roots 
\mbox{$\{ \nu_1, \ldots, \nu_{m_2} \}$}, 
%(\mbox{$\nu_j \in \mathbb{R}$}), 
where 
%whence 
\mbox{$m=2m_1 + m_2 -1
%m+1=2m_1 + m_2 
%=m+1
$}. 
%
%:
% as 
%\mbox{$\lambda_{2j-1} =  \xi_j + \mathrm{i} \eta_j, \; 
%\lambda_{2j} =  \xi_j - \mathrm{i} \eta_j \; (j=1, 2, \ldots, m_1)$}
%and 
%%$m_2$ real roots 
%\mbox{$\lambda_{j} =  \nu_{j-2 m_1+1} \; (j=2 m_1, 2 m_1 +1, 
%\ldots, 2m_1 + m_2 =m)$}. 

\section{Bright-soliton solutions on a plane-wave background}
%{Binary B\"acklund--Darboux transformation}
\label{sec4}

In this section, we apply 
%a 
the 
binary B\"acklund--Darboux transformation 
%defined in Proposition~\ref{prop2.3} 
to 
the 
%vector
%vector/matrix 
multicomponent NLS 
%equation (\ref{rvNLS}) 
equations 
and obtain their 
%construct 
bright-soliton 
%bright one-soliton 
solutions 
%its 
%one-bright soliton solutions 
on a general 
plane-wave background. 
%Specifically, 
To simplify the derivation, 
%computation, 
we 
%present 
use 
a 
%slightly 
reduced version of
Proposition~\ref{prop2.3}, which 
%; it 
defines the 
%a 
binary 
B\"acklund--Darboux transformation
%which 
%that preserves 
%preserving 
maintaining 
the 
%a 
Hermitian conjugation reduction 
between the two 
matrix 
dependent variables. 
%$Q$ and $R$.  
%appearing in the Lax pair. 
%Lax-pair representation.  
In contrast to 
the 
construction of the 
%preceding sections devoted to 
dark-soliton solutions, 
%Note that 
the B\"acklund parameter 
$\mu$ for generating bright-soliton solutions
%is complex-valued and 
%has to be 
is 
%should be 
complex-valued, i.e., 
%with 
it has a nonzero imaginary part. 
%and thus is complex-valued. 
%to generate bright-soliton solutions. 
%
\begin{proposition}
\label{prop4.1}
%The 
%%binary B\"acklund--Darboux 
%transformation 
%map 
%\mbox{$(Q,R) \mapsto (\widehat{\widetilde{Q}}, \widehat{\widetilde{R}})$} 
%can be expressed 
%explicitly 
%as an explicit mapping 
%rewritten 
%in terms of 
%Using 
%For 
%a pair of 
Consider 
%a 
%given 
%linear eigenfunction 
%satisfying 
%of 
the \mbox{$(l+m) \times (l+m)$} 
Lax-pair representation 
%(\ref{NLS-UV}) 
for the \mbox{$l \times m$} 
matrix NLS equation (\ref{redNLS}) 
%(\ref{rmNLS}) 
as given by (cf.~(\ref{NLS-UV}) with (\ref{Herm})) 
%with 
%the Hermitian conjugation reduction 
%(\ref{Herm})
%at \mbox{$\zeta=\mu \in \mathbb{C}$}, i.e.\ 
%\begin{subequations}
%\label{NLS-UV3}
%\begin{equation}
%\left\{ \begin{array}{l}
\begin{align}
& \left[
\begin{array}{c}
 \Psi_1  \\
 \Psi_2 \\
\end{array}
\right]_x 
= \left[
\begin{array}{cc}
-\mathrm{i}\mu I_l & Q \\
\Sigma \hspace{1pt} Q^\dagger \hspace{1pt} \Omega & \mathrm{i} \mu I_m\\
\end{array}
\right] 
\left[
\begin{array}{c}
 \Psi_1  \\
 \Psi_2 \\
\end{array}
\right],
%\label{NLS-U3}
\\[1.5mm]
& \left[
\begin{array}{c}
 \Psi_1  \\
 \Psi_2 \\
\end{array}
\right]_t 
= \left[
\begin{array}{cc}
-2\mathrm{i}\mu^2 I_l -\mathrm{i} Q \hspace{1pt} \Sigma 
 \hspace{1pt} Q^\dagger \hspace{1pt} \Omega 
	& 2 \mu Q + \mathrm{i} Q_x \\
 2 \mu \Sigma \hspace{1pt} Q^\dagger \hspace{1pt} \Omega 
	- \mathrm{i} \Sigma \hspace{1pt} Q^\dagger_x \hspace{1pt} \Omega 
 & 2\mathrm{i}\mu^2 I_m 
 +\mathrm{i} \Sigma \hspace{1pt} Q^\dagger \hspace{1pt} \Omega \hspace{1pt}Q \\
\end{array}
\right]
\left[
\begin{array}{c}
 \Psi_1  \\
 \Psi_2 \\
\end{array}
\right]. 
%\label{NLS-V3}
\end{align}
%\end{array}\right.
%\end{equation}
%\end{subequations}
%
%(\ref{Herm})
%R = \Sigma \hspace{1pt} Q^\dagger \hspace{1pt} \Omega
Here, the 
%original 
spectral parameter 
$\zeta$ 
is 
%set 
fixed 
at a 
%specific 
complex 
value 
\mbox{$\mu
%$} 
%\in \mathbb{C}
$}, 
%is a complex parameter, 
$\Psi_1$ is an \mbox{$l \times l$} matrix and  
$\Psi_2$ is an \mbox{$m \times l$} matrix;  
$\Sigma$ and $\Omega$ are  \mbox{$m \times m$} and \mbox{$l \times l$}
diagonal matrices 
with 
%their 
%each 
diagonal 
entries 
%entry 
equal to 
%either 
$+1$ or $-1$. 
%as~\cite{Kono82,Calo84} 
%
%For a linear eigenfunction 
%of the Lax-pair representation (\ref{NLS-UV}) 
%and 
%its adjoint 
%the 
%an adjoint linear eigenfunction 
%satisfying 
%of the Lax-pair representation 
%(\ref{NLS-adUV}) at \mbox{$\zeta=\nu$}, 
% as 
Then, 
%the binary B\"acklund--Darboux transformation
%%an elementary auto-B\"acklund transformation 
%%\mbox{$(Q,R) \mapsto (\widetilde{Q}, \widetilde{R})$}
%%of
%%\mbox{$(Q,R) \mapsto (\skew{3}\widehat{\widetilde{Q}}, \skew{3}\widehat{\widetilde{R}})$}
%\mbox{$Q \mapsto Q' 
%%\skew{3}\widehat{\widetilde{Q}}
%$}
%for 
%%between two solutions, $(Q,R)$ and $(\widetilde{Q}, \widetilde{R})$, of 
%the matrix NLS equation (\ref{redNLS}) 
%(\ref{rmNLS}) 
%system (\ref{mNLS}) 
%, which 
%connects 
%connecting 
%%relates 
%two solutions $(Q,R)$ and $(\widehat{\widetilde{Q}}, \widehat{\widetilde{R}})$, 
%corresponds to the composition (\ref{path1}), 
%can be expressed as 
%explicitly 
%as
the new potential defined by the formula 
(see~\cite{Chen1,
%Lamb74,
KW75} for the scalar case and 
\cite{Steu88,Park00,Park02,Wright,Forest} 
%[Wright00-01,02] 
for the vector case), 
%and [DegaLom] for the matrix case), 
%\begin{subequations}
%\label{}
\begin{align}
%\skew{3}\widehat{\widetilde{Q}} 
Q' &:= 
%Q +  2 \mathrm{i} (\nu -\mu) 
%\Psi_1 (\mu)  \left[ \Phi_1 (\nu) \Psi_1 (\mu)
% + \Phi_2 (\nu) \Psi_2 (\mu)  \right]^{-1} \Phi_2 (\nu)
%\nonumber \\ &= 
Q +  2 \mathrm{i} (\mu - \mu^\ast) 
%P_{12} (\mu)
\Psi_1
%(\mu) 
\left( \Psi_1^\dagger \hspace{1pt} \Omega \hspace{1pt} \Psi_1
 - \Psi_2^\dagger \hspace{1pt} \Sigma \hspace{1pt} \Psi_2 \right)^{-1} 
\Psi_2^\dagger \hspace{1pt} \Sigma, 
%\\
%\skew{3}\widehat{\widetilde{R}} &= 
% R - 2 \mathrm{i} (\nu-\mu) \Psi_2 (\mu) \left[ \Phi_1 (\nu) \Psi_1 (\mu)
% + \Phi_2 (\nu) \Psi_2 (\mu)  \right]^{-1} \Phi_1 (\nu)
%\nonumber \\ &=
%  R - 2 \mathrm{i} (\nu-\mu) P_{21} (\mu,\nu). 
\nonumber 
\end{align}
%\end{subequations}
provides a new solution of the matrix NLS equation (\ref{redNLS}). 
%Here, $P_{12}$ and $P_{21}$ denote 
%the upper-right and lower-left block
%%blocks 
%of 
%the 
%%%\mbox{$2 \times 2$} block 
%matrix 
%$P$, 
%respectively. 
%The 
%corresponding 
%gauge transformation 
%of the 
A linear eigenfunction 
%satisfying the Lax-pair representation 
%for 
associated with 
the new 
%transformed 
potential $Q'$ 
at the spectral parameter 
$\zeta$ 
set 
equal to 
%fixed at 
$\mu^\ast$ is given by 
%that preserves the form of the Lax-pair representation 
%is 
%given 
%defined 
%can be written 
%is given 
%in 
%using 
%in 
%as 
%
\begin{align}
%\left[
%\begin{array}{c}
%%\widehat{\widetilde{\Psi}}_1 \\ 
%\Psi'_1 (\mu^\ast) \\ 
%%\widehat{\widetilde{\Psi}}_2 \\
%\Psi'_2 ( \mu^\ast) \\
%\end{array}
%\right] 
%\propto & 
%\hspace{3pt} 
%h(\zeta, \nu) g(\zeta,\mu) 
\left[
\begin{array}{c}
 \Psi_1
%(\mu) 
\\
 \Psi_2 
%(\mu) 
\\
\end{array}
\right] \left( \Psi_1^\dagger \hspace{1pt} \Omega \hspace{1pt} \Psi_1 
 - \Psi_2^\dagger \hspace{1pt} \Sigma \hspace{1pt} \Psi_2 \right)^{-1}, 
%\nonumber 
\label{bound-matrix}
\end{align}
%where 
%Here, 
%we omit 
%have omitted 
%omit the unessential 
%the 
%up to an overall constant. 
%Here, 
which 
%gives 
%With a 
%A suitable 
%%re-combination 
%combination of the column vectors, 
%this 
defines a bound state 
%provides a bound-state eigenfunction 
for a proper choice of $\Psi_1(\mu)$ and $\Psi_2(\mu)$.\footnote{In 
%It 
%In 
some cases, 
%It 
%is possible that not all of the column vectors in (\ref{bound-matrix}) 
%provides 
%correspond to independent 
%it is necessary to 
we need to consider a 
suitable 
linear combination 
of the columns 
%column vectors 
in (\ref{bound-matrix}). 
%are linearly independent and decay as 
%In addition, 
Note 
%also 
that 
%a square-matrix solution $\Psi$ of 
%to 
%the linear system 
%eigenvalue problem 
the spatial part of the Lax-pair representation 
\mbox{$\Psi_x = U \Psi
%,\; \Psi_t = V \Psi
$} 
admits the gauge transformation 
\mbox{$\Psi \to \mathrm{e}^{\mathrm{i} kx 
%+\omega t
} \Psi,\; U \to U + \mathrm{i} k I
%, \; V \to V + \omega I
$}, 
where $k$ is 
%with 
an arbitrary 
%{\em complex} 
complex constant; 
%$k$; 
%which 
%This 
this transformation is sometimes 
%useful 
necessary 
%useful 
to 
%construct 
%necessary to 
obtain 
a 
square-integrable 
%bound-state 
eigenfunction. 
%decaying rapidly as \mbox{$x \to \pm \infty$}. 
%bound state.
}
%Then, 
%Similarly, 
An adjoint linear eigenfunction, 
%at the spectral parameter 
%$\zeta$ 
%set 
%equal to 
%fixed at 
%$\mu$, 
\begin{align}
\left( \Psi_1^\dagger \hspace{1pt} \Omega \hspace{1pt} \Psi_1 
 - \Psi_2^\dagger \hspace{1pt} \Sigma \hspace{1pt} \Psi_2 \right)^{-1}
\left[
\begin{array}{cc}
%\Psi_{1,n}^{\mathrm{ad}} \! & \! \Psi_{2,n}^{\mathrm{ad}} 
\! \Psi_1^\dagger \Omega \! & \! -\Psi_2^\dagger \Sigma \!
\end{array}
\right], 
\nonumber 
\end{align}
%provides 
can provide 
%one more
%gives 
another 
%the other 
bound state at the spectral parameter 
$\zeta$ 
%set 
equal to 
%fixed at 
$\mu$.\footnote{Thus, 
%for the multicomponent case, 
%it is more convenient to consider 
a multicomponent bright soliton is associated with 
a pair of bound states:\ 
%, i.e., 
a 
%bound-state 
linear eigenfunction at \mbox{$\zeta=\mu^\ast$} 
and 
an 
%a bound-state 
adjoint linear eigenfunction at \mbox{$\zeta=\mu$}. 
%Note that 
For 
%In the case of 
the scalar NLS equation, 
%it is easy to rewrite 
%this 
the adjoint linear eigenfunction 
can 
%directly 
%give 
be 
%directly 
immediately 
%easily 
rewritten 
as the 
%conventional 
usual 
%a 
%the linear 
(i.e., column-vector) 
linear eigenfunction at \mbox{$\zeta=\mu$}.} 
%Then, 
%we 
%define a 
%%the 
%projection matrix $P(\mu
%%,\mu^\ast
%)$ 
%in the \mbox{$2 \times 2$} block-matrix 
%form: 
%%matrix $P$, respectively. 
%%as 
%%matrix 
%\begin{align}
%P(\mu
%%,\mu^\ast
%) &:= 
%\left[
%\begin{array}{c}
% \Psi_1
%%(\mu) 
%\\
% \Psi_2 
%%(\mu) 
%\\
%\end{array}
%\right] \left( \Psi_1^\dagger \hspace{1pt} \Omega \hspace{1pt} \Psi_1 
% + \Psi_2^\dagger \hspace{1pt} \Sigma \hspace{1pt} \Psi_2 \right)^{-1} 
%\left[
%\begin{array}{cc}
%\! \Psi_1^\dagger \hspace{1pt} \Omega 
%\! 
% & 
%\! 
%\Psi_2^\dagger \hspace{1pt} \Sigma \!
%\end{array}
%\right].
%\nonumber 
%%\\[1mm]
%%&=: 
%%\left[
%\begin{array}{cc}
%P_{11} & P_{12} \\
%P_{21} & P_{22} \\
%\end{array}
%\right] 
%\end{align}
%
%which satisfies the 
%Note that 
%Indeed, 
%it indeed satisfies 
%\mbox{$P^2 = P$}. 
%\mbox{$\left[ P(\mu,\nu) \right]^2 = P(\mu,\nu)$}.
%Similarly, 
%In 
%the same way, 
%a similar way, 
%manner, 
\end{proposition}

%Proposition~\ref{prop4.1} 
This proposition 
enables us to obtain 
%We 
%Let us 
%first 
%consider 
%construct 
%can obtain 
a formal expression for the 
%the 
%square matrix case \mbox{$l=m$} 
%matrix 
%case 
%The 
%bright-soliton 
%The 
bright 
one-soliton 
solution on a plane-wave background 
%can be expressed formally 
%formally, 
%i.e., 
using the matrix exponential 
%function 
of a constant 
non-diagonal matrix. 
%and 
%construct 
%
\subsection{Matrix bright soliton} 
%Bright-soliton solutions of matrix NLS 
%on a plane-wave background}
%The seed 
%Note that 
%a general plane-wave solution of 
%Let us 
We first consider 
the matrix NLS equation (\ref{rmNLS}) 
for 
%in the slightly generalized case of 
an 
%general 
\mbox{$l \times m$} matrix $Q$, 
%is 
which admits the 
%a 
general plane-wave solution  
%given by 
(cf.~(\ref{plane2})): 
\begin{align}
Q = 
%P_1 
\mathrm{e}^{-2\mathrm{i} t \sigma A A^\dagger} A  
\mathrm{e}^{\mathrm{i} x \Gamma - \mathrm{i} t \Gamma^2} 
%P_2
. 
%\hspace{5mm} 
%R= P_2^{-1} \mathrm{e}^{-\mathrm{i} x \Gamma + \mathrm{i} t \Gamma^2}
%	 B \mathrm{e}^{2\mathrm{i} t AB} P_1^{-1}, 
%%\label{plane2}
\nonumber 
\end{align}
%where 
Here, \mbox{$\sigma=+1$} and \mbox{$\sigma=-1$} correspond 
to the self-defocusing case 
and 
%\mbox{$\sigma=-1$} corresponds to 
%the 
self-focusing case, respectively; 
$\Gamma$ is an \mbox{$m \times m$} real diagonal matrix and 
$A$ is an \mbox{$l \times m$} matrix 
such that 
%where 
$A A^\dagger$ is a real diagonal matrix. 
%For simplicity, in the following computation, 
%we omit the unitary matrices $P_1$ and $P_2$ 
%%in the following computation
%%, {\it e.g.}, 
%by setting them as the identity matrix. 
%discussion. 
%
%By employing 
We employ this plane-wave 
solution as the seed solution and 
apply the binary 
B\"acklund--Darboux transformation defined in Proposition~\ref{prop4.1}. 
%By recalling 
Following the procedure 
%discussion given 
%outlined 
described in subsection~\ref{sec2.6}, 
%Then, 
we obtain a new solution of the matrix NLS equation (\ref{rmNLS}) 
in the form: 
\begin{align}
Q &= \mathrm{e}^{-2\mathrm{i} t \sigma A A^\dagger} A  
\mathrm{e}^{\mathrm{i} x \Gamma - \mathrm{i} t \Gamma^2} 
+  2 \mathrm{i}  (\mu - \mu^\ast) 
%P_{12} (\mu)
\Psi_1
%(\mu) 
\left( \sigma \Psi_1^\dagger \Psi_1 -  \Psi_2^\dagger \Psi_2 \right)^{-1} \Psi_2^\dagger 
\nonumber \\
&= \mathrm{e}^{-2\mathrm{i} t \sigma A A^\dagger} \left[ A 
+ 2 
%\mathrm{i} 
% \sigma 
 (\mu - \mu^\ast) 
  \boldsymbol{\mathit \Psi}_1
 \left( \sigma \boldsymbol{\mathit \Psi}_1^\dagger  \boldsymbol{\mathit \Psi}_1 
 - \boldsymbol{\mathit \Psi}_2^\dagger \boldsymbol{\mathit \Psi}_2 \right)^{-1}
 \boldsymbol{\mathit \Psi}_2^\dagger \right]  
\mathrm{e}^{\mathrm{i} x \Gamma - \mathrm{i} t \Gamma^2}. 
%\nonumber 
\label{matrix-bright}
\end{align}
%where 
Here, we omit the prime to distinguish the new solution from the 
old 
one.
%seed 
%solution. 
In the self-focusing case \mbox{$\sigma=-1$}, 
the new solution is 
%generally 
regular for 
%finite 
%real 
%values of 
all 
$x$ and $t$. 
%; 
%The 
The 
%components of 
gauge-transformed linear eigenfunction 
%appearing in 
used in (\ref{matrix-bright}) 
%quantities 
is given as 
(cf.~(\ref{mat-ex1})), 
\begin{align}
\left[
\begin{array}{c}
 \boldsymbol{\mathit \Psi}_1 \\
 \boldsymbol{\mathit \Psi}_2 
%(\zeta) 
\\
\end{array}
\right] =  \mathrm{e}^{\mathrm{i} \mu x + 2 \mathrm{i} \mu^2 t}
\mathrm{e}^{\mathrm{i}x
%\mathrm{i} x \boldsymbol{Z} 
%+ \mathrm{i}  t (-\boldsymbol{Z}^2 + 2 \zeta \boldsymbol{Z} + \zeta^2 I)
% (
\boldsymbol{Z} 
%+ \mu I
%_{l+m} 
%) 
- \mathrm{i}  t \boldsymbol{Z}^2 
%+ 2 \mu^2 I
%_{l+m} 
%)
}
\left[
\begin{array}{c}
 {\mathscr C}_1 \\
 {\mathscr C}_2 \\
\end{array}
\right], 
%{Z x + 2 \zeta Z^2 t},
%\exp \left( Z x \right)
\hspace{5mm}
\boldsymbol{Z} 
%(\mu) 
%Z
:= \left[
\begin{array}{cc}
-2 \mu I_l & A \\
 -\sigma A^\dagger & \Gamma \\
\end{array}
\right], 
\nonumber
%\label{mat-ex3}
\end{align}
where ${\mathscr C}_1$ and ${\mathscr C}_2$ are 
\mbox{$l \times l$} and \mbox{$m \times l$} constant matrices, respectively. 
Note that the 
%scalar 
%common 
prefactor 
\mbox{$\mathrm{e}^{\mathrm{i} \mu x + 2 \mathrm{i} \mu^2 t}$} plays no role 
in formula (\ref{matrix-bright}). 
%solution. 
This expression is rather formal and 
appears to be not so 
%very 
useful in the present form 
for 
practical applications. 
%the 
%asymptotic analysis, 
%so 
Thus, 
we 
next 
consider the 
%simplest 
%nontrivial 
simpler 
case of 
the 
%two-component 
vector NLS equation 
%case 
(\mbox{$l=1$}), 
%and then mainly 
particularly targeting 
%focusing 
on 
%at 
the 
two-component case (\mbox{$m=2$}), 
and 
%compute the matrix exponential explicitly. 
obtain a more explicit expression 
%solution 
%without using the matrix exponential. 
for 
the 
%bright one-soliton 
%this
solution. 
%solutions. 

%\subsection{Composition of two elementary B\"acklund--Darboux transformations}

\subsection{Vector bright soliton}
%Bright-soliton solutions of vector NLS 
%on a plane-wave background}
%Thus, 
%We 
%Let us 
%next consider the 
%simplest 
%nontrivial 
%simpler case of 
%the 
%two-component 
%vector NLS equation 
%case 
%(\mbox{$l=1$}), 
%and then mainly 
%particularly focusing on the 
%two-component case (\mbox{$m=2$}). 
%
%In this subsection, we 
%%follow the same procedure as in 
%%re-use 
%use some formulas 
%in section~\ref{sec5}; 
%the main difference from is that 
%note, however, that 
%%unlike in section~\ref{sec5}, 
%%in contrast to dark-soliton solutions, 
%the B\"acklund parameter $\mu$ 
%corresponding to 
%a 
%the 
%bright-soliton solution 
%solutions 
%here 
%%now 
%is 
%a complex 
%number. 
%parameter. 
%with a nonzero-imaginary part. 
%not real but 
%complex-valued. 
%takes complex values.  
%a complex parameter. 
%with a nonzero imaginary part. 
A general plane-wave solution of the vector 
%general $m$-component 
NLS equation (\ref{rvNLS}) 
is given by setting \mbox{$b_j =\sigma_j a_j^\ast$} in (\ref{vector-plane}), i.e.\ 
%:  
%(cf.~(\ref{vector-plane}))
%a general plane-wave solution 
%%can be generalized to the case 
%%with $m$ components
%of the general $m$-component NLS equation (\ref{rvNLS}) 
\begin{align}
%\vt{q} 
q_j (x,t) & = a_j\mathrm{e}^{\mathrm{i}  \gamma_j x - \mathrm{i} \left( \gamma_j^2 
 + 2 \sum_{k=1}^m \sigma_k |a_k|^2 \right) t} , 
\hspace{5mm} j=1, 2, 
%3
\ldots, m. 
\end{align}
%where 
Here, we assume that 
$\gamma_j \; (j=1, 2, \ldots, m)$ are 
%assumed to be 
pairwise distinct real constants. 
%numbers. 
%
%From (\ref{mu-ak-bk}), we obtain 
%Let us 
We recall the 
%The 
relations 
(\ref{mu-ak-bk}): 
% give 
%: 
\begin{align}
& 2 \mu = \sum_{j=1}^m \gamma_j - \sum_{j=1}^{m+1} \lambda_j
%, 
%\hspace{5mm}
\nonumber 
\end{align}
and 
\begin{align}
\sigma_k |a_k|^2 = \frac{\displaystyle \prod_{j=1}^{m+1} (\lambda_j - \gamma_k)}
	{\displaystyle \dprod{j=1}{j \neq k}^m (\gamma_j - \gamma_k)} \hspace{1pt}, 
\hspace{5mm} k=1, 2, \ldots, m, 
%\nonumber 
\label{mu-ak-bk2}
\end{align}
where 
%Note that 
%the B\"acklund parameter 
$\mu$ 
%here 
is a complex parameter. 
%complex-valued. 
Thus, 
%the characteristic equation (\ref{chara1}) of the matrix $\boldsymbol{Z} (\mu)$ as 
the eigenvalues \mbox{$\{ \lambda_1, \lambda_2, \ldots, \lambda_{m+1} \}$} 
of the matrix $\boldsymbol{Z} (\mu)$
must satisfy the condition that 
\mbox{$\prod_{j=1}^{m+1} (\lambda_j - \gamma_k)$} is 
%a real number 
%real 
real-valued 
for 
all $k$. 
%\mbox{$k=1, 2, \ldots, m$}. 
%every $k$. 
%\mbox{$1 \le k \le m$}. 
That is, if we set 
\[
\lambda_j = \xi_j + \mathrm{i} \eta_j, \hspace{5mm} j=1, 2, \ldots, m+1
%, 
\]
with real $\xi_j$ and $\eta_j$, then we have 
\[
\prod_{j=1}^{m+1} \left[ (\xi_j - \gamma_k) + \mathrm{i} \eta_j \right] \in \mathbb{R}, 
\hspace{5mm} k=1, 2, \ldots, m. 
\]
%This implies that 
This means that 
%Thus, 
%if we fix 
%when 
if the 
%\mbox{$m+1$} 
eigenvalues 
\mbox{$\lambda_j = \xi_j + \mathrm{i} \eta_j \; (j=1, 2, \ldots, m+1)$} 
are given, 
%then 
the 
%$m$ 
wavenumbers \mbox{$\gamma_k 
\; (k=1,2, \ldots, m)$} 
are determined as the 
solutions of an
%the 
algebraic equation 
%of degree \mbox{$m$} 
for $\gamma$, 
\begin{equation}
%\mathrm{Imaginary \;
%%\hphantom{a} 
%part \;
%%\hphantom{a} 
%of \;
%%\hphantom{a}
%} 
\mathrm{Im} \left\{ 
\prod_{j=1}^{m+1} \left[ (\xi_j - \gamma) + \mathrm{i} \eta_j \right] \right\} = 0. 
\label{Im0}
\end{equation}
%
%Because 
%In addition, 
This 
%equation for $\gamma$ 
%is degree $m$ and 
%has 
is 
%an equation 
a real-coefficient equation of degree \mbox{$m$}, 
%; 
%of degree less than \mbox{$m+1$} 
%with real coefficients, 
%it 
which is much easier to solve 
%deal with 
than the characteristic equation for determining 
the eigenvalues 
\mbox{$\lambda_j = \xi_j + \mathrm{i} \eta_j$} 
%\; 
\mbox{$(j=1, 2, \ldots, m+1)$} 
%\mbox{$\{ \lambda_1, \lambda_2, \ldots, \lambda_{m+1} \}$} 
%of the matrix $\boldsymbol{Z} (\mu)$ 
(cf.~(\ref{chara1})). 
%, which 
%Note that 
Indeed, the latter is a complex-coefficient equation 
of degree \mbox{$m+1$}. 
This is why we consider 
%in the following 
that 
the eigenvalues 
%\mbox{$\lambda_j = \xi_j + \mathrm{i} \eta_j$}
% \; (j=1, 2, \ldots, m+1)$} 
\mbox{$\{ \lambda_1, \lambda_2, \ldots, \lambda_{m+1} \}$} 
are free complex parameters, 
%and 
from which the wavenumbers 
\mbox{$\{ \gamma_1, \gamma_2, \ldots, \gamma_m \}$} 
of the background plane waves are determined. 
%from them. 

Then, 
%computing 
by expressing 
the right-hand side of (\ref{mu-ak-bk2}) 
only in terms of 
%using only 
\mbox{$
%\lambda_j 
\xi_j$} and 
\mbox{$\eta_j 
%= \xi_j + \mathrm{i} \eta_j 
\; (j=1, 2, \ldots, m+1)
$}, 
%only, 
%determines 
%the sign 
we can determine 
the sign 
$\sigma_k$
%, 
%of 
%which defines 
%%determines 
%the sign of 
in each cubic term in the vector NLS equation (\ref{rvNLS})
%, as well as 
and 
the amplitude $|a_k|$ of each 
%the 
background 
plane wave. 
%plane-wave. 
%modulus square 
%is determined 
%through 
%by 
%the sign of 
%the right-hand side of (\ref{mu-ak-bk2}). 
%For fixed eigenvalues 
It is very useful to consider geometrically 
%in a geometric manner 
how the argument of 
the complex function 
%value of 
\mbox{$
%\mathrm{arg} 
\prod_{j=1}^{m+1} (\lambda_j - \gamma)$}
%changing 
changes 
%if 
as $\gamma$ moves on the real axis 
from $-\infty$ to $+\infty$. 
For example, 
%if 
%the case where 
%if 
when all the eigenvalues 
\mbox{$\{ \lambda_1, \lambda_2, \ldots, \lambda_{m+1} \}$} lie 
in the upper-half (or lower-half) complex plane, 
%, i.e., \mbox{$\mathrm{Im} \, \lambda_j > 0$} for all $j$ 
%corresponds to 
we have \mbox{$\sigma_k=-1$} for all $k$, which corresponds to 
the self-focusing case. 
%
%Because $\mu$ is a complex parameter, 
%the characteristic equation for determining 
%the eigenvalues \mbox{$\{ \lambda_1, \lambda_2, \ldots, \lambda_{m+1} \}$} 
%of the matrix $\boldsymbol{Z} (\mu)$ is 
%an algebraic equation of degree \mbox{$m+1$} 
%with complex coefficients (cf.~(\ref{chara1})).

%If we consider that 
In the scalar 
NLS 
case (\mbox{$m=1$}), (\ref{Im0}) 
%gives 
is 
%provides 
%reduces to 
%becomes 
%we have 
a linear equation 
for 
determining 
$\gamma_1$. 
%; 
%, which provides the solution 
%its 
%when 
%if 
If \mbox{$\eta_1 +\eta_2 \neq 0$}, 
%the 
%its solution is 
%given by 
we have 
\begin{equation}
\gamma_1 = \frac{\xi_1 \eta_2 + \xi_2 \eta_1}{\eta_1 +\eta_2}, 
%\nonumber 
\label{gamma1}
\end{equation}
which gives 
%implies that 
\begin{align}
\sigma_1 |a_1|^2 
&= (\lambda_1 - \gamma_1) (\lambda_2 - \gamma_1)
\nonumber \\[1mm]
&= - \eta_1 \eta_2 \left[ 1+ \left( \frac{\xi_1 - \xi_2}{\eta_1 +\eta_2} \right)^2 \right]. 
%\frac{\eta_1 \eta_2}{(\eta_1 +\eta_2)^2} 
%\left[ (\xi_1 - \xi_2)^2 +(\eta_1 +\eta_2)^2 \right]. 
%\nonumber 
%\label{mu-a1}
\nonumber 
\end{align}
Thus, \mbox{$\sigma_1 = -\mathrm{sgn} \left( \eta_1 \eta_2 \right)$}. 
To obtain a 
%non-singular 
regular 
%one-
%soliton 
solution of the NLS equation, 
we assume 
%that 
\mbox{$\eta_1 \eta_2 >0$} 
and 
%focus on 
consider 
the self-focusing case (\mbox{$\sigma_1 = -1$}). 
%In the self-focusing case (\mbox{$\sigma_1 = -1$}), 
%we have \mbox{$\eta_1 \eta_2 >0$}; 
%%$\eta_1$ and $\eta_2$ have the same sign, 
%%while 
%in the self-defocusing case (\mbox{$\sigma_1 = +1$}), 
%we have \mbox{$\eta_1 \eta_2 <0$}. 
%Because \mbox{$2 \mu = \gamma_1 - \lambda_1 - \lambda_2$}, 
%In this 
%%the scalar NLS 
%case, 
Then, 
the matrix $\boldsymbol{Z}(\mu)
$ 
%in the scalar NLS case
%that appear in 
%is given by the \mbox{$2 \times 2$} matrix 
%now 
takes the 
%\mbox{$2 \times 2$} 
%matrix 
form 
%(cf.~(\ref{mat-ex1}) and 
(cf.~(\ref{Z-vector})):  
%For convenience, we re-parametrize $a_1$ as 
%\[
%a_1 =: \frac{d
%%\eta_1 \eta_2
%}{\eta_1 +\eta_2} 
%\left[ (\xi_1 - \xi_2) + \mathrm{i} (\eta_1 +\eta_2) \right],
%%d, 
%\hspace{5mm} d \in \mathbb{C}.   
%\]
\[
\boldsymbol{Z} = 
\left[
\begin{array}{cc}
% -2\mu  
\lambda_1 + \lambda_2 -\gamma_1 & a_1 \\
% -\sigma_1 
a_1^\ast & \gamma_1 \\
\end{array}
\right]. 
%\hspace{5mm} \lambda_j = \xi_j + \mathrm{i} \eta_j. 
\]
We choose 
the 
%two 
eigenvectors corresponding to the 
%pair of 
eigenvalues 
\mbox{$\lambda_j = \xi_j + \mathrm{i} \eta_j$} 
%of this \mbox{$2 \times 2$} matrix 
%can be given as
%are given 
as
\begin{align}
\boldsymbol{Z}
\left[
\begin{array}{c}
a_1 \\
%2 \mu + \lambda_1 
%\gamma_1 - \lambda_2 
\eta_ 2 \left( \frac{\xi_1 -\xi_2 }{\eta_1 +\eta_2} - \mathrm{i} \right) \\
\end{array}
\right] 
&= \lambda_1 
%(\xi_1 + \mathrm{i}\eta_1) 
\left[
\begin{array}{c}
a_1 \\
\eta_ 2 \left( \frac{\xi_1 -\xi_2 }{\eta_1 +\eta_2} - \mathrm{i} \right) \\
% -\xi \pm \mathrm{i} \eta + \gamma_1 \\
\end{array}
\right],
\nonumber \\[2mm] 
\boldsymbol{Z}
\left[
\begin{array}{c}
a_1 \\
%2 \mu + \lambda_1 
%\gamma_1 - \lambda_2 
- \eta_ 1 \left( \frac{\xi_1 -\xi_2 }{\eta_1 +\eta_2} + \mathrm{i} \right) \\
\end{array}
\right] 
& = \lambda_2 
%(\xi_2 + \mathrm{i}\eta_2) 
\left[
\begin{array}{c}
a_1 \\
-\eta_ 1 \left( \frac{\xi_1 -\xi_2 }{\eta_1 +\eta_2} + \mathrm{i} \right) \\
% -\xi \pm \mathrm{i} \eta + \gamma_1 \\
\end{array}
\right]. 
\nonumber
\end{align}
Then, the matrix exponential in (\ref{mat-ex1})  
%of a 
%%linear 
%simple function of 
%$\boldsymbol{Z}
%%(\mu)
%$ 
can be computed explicitly to provide 
%a 
the gauge-transformed 
linear eigenfunction as 
%
%Thus, 
%In the scalar NLS case, 
%Then, 
%In the scalar NLS case, 
%The general 
%a generic 
%solution of the 
%\mbox{$2 \times 2$} 
%linear problem 
%(\ref{NLS-UV2}) given by (\ref{mat-ex1}) 
%in the simplest scalar case 
%in this case 
%can be written explicitly as 
\begin{align}
\left[
\begin{array}{c}
 \boldsymbol{\mathit \Psi}_1 \\
 \boldsymbol{\mathit \Psi}_2 
%(\zeta) 
\\
\end{array}
\right] &= c_1 \mathrm{e}^{\mathrm{i} x 
 \frac{1}{2} (\gamma_1 + \lambda_1 - \lambda_2) 
+ \mathrm{i}  t \frac{1}{2} \left[ \gamma_1^2 - 2 \gamma_1 (\lambda_1 + \lambda_2) 
-\lambda_1^2 + \lambda_2^2 +2 \lambda_1 \lambda_2
\right] }
\left[
\begin{array}{c}
a_1 \\
\eta_ 2 \left( \frac{\xi_1 -\xi_2 }{\eta_1 +\eta_2} - \mathrm{i} \right) \\
\end{array}
\right] 
\nonumber \\[1mm]
 & \hphantom{=}\; \mbox{}+ c_2 \mathrm{e}^{\mathrm{i} x 
  \frac{1}{2} (\gamma_1 - \lambda_1 + \lambda_2 ) 
+ \mathrm{i}  t 
\frac{1}{2} \left[ \gamma_1^2 - 2 \gamma_1 (\lambda_1 + \lambda_2) 
+\lambda_1^2 - \lambda_2^2 +2 \lambda_1 \lambda_2 \right]}
\left[
\begin{array}{c}
a_1 \\
-\eta_ 1 \left( \frac{\xi_1 -\xi_2 }{\eta_1 +\eta_2} + \mathrm{i} \right) \\
\end{array}
\right], 
\nonumber
%\label{mat-ex2}
\end{align}
where $c_1$ and $c_2$ are arbitrary complex 
%nonzero 
constants. 
%; they 
Thus, 
%using 
by adapting formula (\ref{matrix-bright}) 
to 
%the case of the 
%in 
the scalar 
%self-focusing 
NLS equation \mbox{$\mathrm{i} q_t + q_{xx} 
%- 2 \sigma 
+2|q|^2 
%\left| q \right|^2 
q = 0$}, 
%case, 
we obtain 
%a 
the bright one-soliton solution on the plane-wave background 
as (cf.~\cite{Kuz77,Kawa78,Ma79,Ste86}) 
%[Y-C.Ma])
%given by 
%\begin{subequations}
%\label{scalar-1-plane}
\begin{align}
\label{scalar-1-plane}
%q_j 
q (x,t) & = a_1 \mathrm{e}^{\mathrm{i}  \gamma_1 x - \mathrm{i} \left( \gamma_1^2 
 %+ 2 \sigma 
-2 |a_1|^2 \right) t} 
%\left\{ 1 + 2\mathrm{i} ( \eta_1 + \eta_2 ) 
\frac{ 
%\mathrm{e}^{}
g}{f} \hspace{1pt}.
%\right\}, 
\end{align}
Here, the positive 
%definite 
%real 
function $f(x,t)$ and the complex function 
$g(x,t)$ are 
%given by 
%as 
%given by
\begin{align}
f &:= |c_1|^2 \eta_2 \left[ (\xi_1 - \xi_2)^2 + ( \eta_1 +\eta_2 )^2  
\right] 
\mathrm{e}^{-(\eta_1 -\eta_2)x +2 (\xi_1 \eta_1 - \xi_2 \eta_2)t}
\nonumber \\ & \hphantom{:=}\;
 + |c_2|^2 \eta_1 \left[ (\xi_1 - \xi_2)^2 + ( \eta_1 +\eta_2 )^2  
\right]  
\mathrm{e}^{(\eta_1 -\eta_2)x -2 (\xi_1 \eta_1 - \xi_2 \eta_2)t}
\nonumber \\ & \hphantom{:=}\;
 - 2\mathrm{i} c_1^\ast c_2 \eta_1 \eta_2 \left[ (\xi_1 - \xi_2) 
	+ \mathrm{i} ( \eta_1 +\eta_2 ) \right] \mathrm{e}^{-\mathrm{i}(\xi_1 -\xi_2)x 
	+ \mathrm{i} \left( \xi_1^2 - \xi_2^2 - \eta_1^2 + \eta_2^2 \right)t}
\nonumber \\ & \hphantom{:=}\;
 + 2\mathrm{i} c_1 c_2^\ast \eta_1 \eta_2 \left[ (\xi_1 - \xi_2) 
	- \mathrm{i} ( \eta_1 +\eta_2 ) \right] \mathrm{e}^{\mathrm{i}(\xi_1 -\xi_2)x 
	- \mathrm{i} \left( \xi_1^2 - \xi_2^2 - \eta_1^2 + \eta_2^2 \right)t}
%\label{58b}
\nonumber 
%, 
%\\[2mm]
\end{align}
and
\begin{align}
g &:= |c_1|^2 \eta_2 \left[ (\xi_1 - \xi_2) + \mathrm{i} ( \eta_1 +\eta_2 ) \right]^2
\mathrm{e}^{-(\eta_1 -\eta_2)x +2 (\xi_1 \eta_1 - \xi_2 \eta_2)t}
\nonumber \\ & \hphantom{:=}\;
 + |c_2|^2 \eta_1 \left[ (\xi_1 - \xi_2) - \mathrm{i} ( \eta_1 +\eta_2 ) \right]^2
\mathrm{e}^{(\eta_1 -\eta_2)x -2 (\xi_1 \eta_1 - \xi_2 \eta_2)t}
\nonumber \\ & \hphantom{:=}\;
 + 2\mathrm{i} c_1^\ast c_2 \eta_2^2 \left[ (\xi_1 - \xi_2) 
	+ \mathrm{i} ( \eta_1 +\eta_2 ) \right] \mathrm{e}^{-\mathrm{i}(\xi_1 -\xi_2)x 
	+ \mathrm{i} \left( \xi_1^2 - \xi_2^2 - \eta_1^2 + \eta_2^2 \right)t}
\nonumber \\ & \hphantom{:=}\;
 - 2\mathrm{i} c_1 c_2^\ast \eta_1^2 \left[ (\xi_1 - \xi_2) 
	- \mathrm{i} ( \eta_1 +\eta_2 ) \right] \mathrm{e}^{\mathrm{i}(\xi_1 -\xi_2)x 
	- \mathrm{i} \left( \xi_1^2 - \xi_2^2 - \eta_1^2 + \eta_2^2 \right)t}, 
%\label{58c}
\nonumber 
\end{align}
%\end{subequations}
respectively. 
%and 
The amplitude and the wavenumber
of the background plane wave are 
%is 
determined 
%as 
by 
\begin{align}
|a_1|^2 
&= \eta_1 \eta_2 \left[ 1+ \left( \frac{\xi_1 - \xi_2}{\eta_1 +\eta_2} \right)^2 \right]
%, 
\nonumber
\end{align}
%\end{subequations}
and (\ref{gamma1})
%, 
respectively, 
%
%\mbox{$\eta_1 \eta_2 >0$}
%
while the 
%argument 
complex phase 
of $a_1$ is arbitrary, which reflects 
%corresponds to 
%represents 
the $U(1)$ symmetry of the scalar NLS equation. 
%Note that we have supressed 
%%have omitted 
%the subscript $1$ 
%%${}_1$ 
%%from 
%of the dependent variable and some 
%%the 
%parameters. 
%in (\ref{scalar-1-plane}). 
%In addition, 
Note that 
%the expression 
%(\ref{scalar-1-plane}) 
%(\ref{58b}) 
%The denominator 
$f$
and 
%the numerator 
$g$ 
%(\ref{58c}) 
%in (\ref{scalar-1-plane}) 
can be 
%slightly 
%further 
simplified 
%using 
by setting 
%a 
%new 
%complex parameter 
\mbox{$c_2/c_1 =: \mathrm{e}^{\delta + \mathrm{i}\theta}$} 
and using real $\delta$ and $\theta$ as new 
%real 
parameters. 
%, 
%%as a new parameter 
%instead of the 
%%original 
%complex parameters $c_1$ and $c_2$. 
%using $c_1$ and $c_2$ directly. 
%Alternatively, 
Thinking 
more intuitively, 
we can set 
%intuitively consider that 
\mbox{$c_1=c_2=1$} by 
%shifting 
constant shifts of 
$x$ and $t$. 
%by 
%appropriate 
%%arbitrary 
%constants. 
%For 
%In the generic case of 
%We need to assume 
The additional 
condition 
\mbox{$\eta_1 \neq \eta_2$} 
is necessary for 
%to 
%so that 
%ensures that 
(\ref{scalar-1-plane}) 
%is indeed a 
%can 
%describes 
to 
%describe 
provide 
a genuine 
%proper 
soliton 
solution 
associated with a pair of 
bound states of the Lax pair. 
%Incidentally, 
%the solution 
In fact, 
in the special case of \mbox{$\eta_1=\eta_2$} 
and \mbox{$\xi_1 \neq \xi_2$},  
(\ref{scalar-1-plane}) describes an 
oscillating 
solution 
%localized in time $t$ 
on the plane-wave background, 
%that 
which is localized in time $t$ 
%on the plane-wave background 
(cf.~\cite{Akh85,Ste86}). 

%Thus, 
%Let us focus on 
%From now on, 
Next, 
%in the remaining part, 
%of this subsection,  
we 
consider 
%focus on 
%In 
the 
%case of the 
two-component 
%vector NLS equation 
case 
%of \mbox{$m=2$}; 
(\mbox{$m=2$}). 
%; 
%In this case, 
Then, 
(\ref{Im0}) 
%results in 
%provides 
%we have 
%the two solutions of 
%the 
%is 
becomes 
a quadratic equation for $\gamma$,  
\begin{equation}
(\xi_1 -\gamma)(\xi_2 -\gamma) \eta_3 + 
(\xi_2 -\gamma)(\xi_3 -\gamma) \eta_1 + 
(\xi_3 -\gamma)(\xi_1 -\gamma) \eta_2 -\eta_1 \eta_2 \eta_3 =0,
%; 
%\hspace{5mm} k =1,2, \hspace{5mm} \gamma_1 \neq \gamma_2.  
\nonumber 
\end{equation}
%for determining 
%the two solutions of which determine 
%its 
which 
%so it 
%must have 
has 
two real 
%distinct 
solutions 
%provide 
%determine 
%give 
$\gamma_1$ and $\gamma_2$ (\mbox{$\gamma_1 \neq \gamma_2$}). 
We consider only the generic case of \mbox{$\eta_1 + \eta_2 + \eta_3 \neq 0$}
%; 
%so that 
when the above equation is indeed quadratic. 
% in $\gamma$.
Thus, the discriminant denoted by $\Delta$ 
must 
%should 
be positive: 
\begin{align}
\Delta := \,&
\left[ (\xi_1 + \xi_2) \eta_3 + (\xi_2 + \xi_3) \eta_1 + (\xi_3 + \xi_1) \eta_2 \right]^2 
\nonumber \\
\, & - 4 (\eta_1 + \eta_2 + \eta_3) (\xi_1 \xi_2 \eta_3 + \xi_2 \xi_3 \eta_1 
	+ \xi_3 \xi_1 \eta_2 - \eta_1 \eta_2 \eta_3) >0, 
%\label{D-def} 
\nonumber 
\end{align}
%
%The 
and the wavenumbers $\gamma_1$ and $\gamma_2$ 
are given by 
\begin{subequations}
\label{gamma1-2}
\begin{align}
& \gamma_1 = \frac{(\xi_1+\xi_2)\eta_3 + (\xi_2 + \xi_3)\eta_1 + (\xi_3 + \xi_1)\eta_2 
	- \sqrt{\Delta}}{2 (\eta_1 + \eta_2 + \eta_3)}, 
\\[1mm] 
& \gamma_2 = \frac{(\xi_1+\xi_2)\eta_3 + (\xi_2 + \xi_3)\eta_1 + (\xi_3 + \xi_1)\eta_2 
	+ \sqrt{\Delta}}{2 (\eta_1 + \eta_2 + \eta_3)}. 
\end{align} 
\end{subequations}
%where 
%with $D$ 
%is defined as 
%by
%as in 
%We do not consider the 
%%less interesting 
%exceptional 
%case of \mbox{$\eta_1 + \eta_2 + \eta_3=0$}. 
%given by (\ref{D-def}). 
From (\ref{mu-ak-bk2}), we have the relations: 
%also obtain 
%the amplitudes of the background planewave as 
\begin{subequations}
\label{mu-ak-bk3}
\begin{align}
& \sigma_1 |a_1|^2 = \frac{(\xi_1 - \gamma_1) (\xi_2 - \gamma_1) (\xi_3 - \gamma_1) 
-\eta_1 \eta_2 (\xi_3-\gamma_1) -\eta_2 \eta_3 (\xi_1-\gamma_1) 
	-\eta_1 \eta_3 (\xi_2-\gamma_1)
%\prod_{j=1}^{3} (\xi_j - \gamma_1 + \mathrm{i} \eta_j)
}{\gamma_2 - \gamma_1} 
%\hspace{1pt}
, 
%\nonumber \\[1mm]
\\[1mm]
& \sigma_2 |a_2|^2 = \frac{(\xi_1 - \gamma_2) (\xi_2 - \gamma_2) (\xi_3 - \gamma_2) 
-\eta_1 \eta_2 (\xi_3-\gamma_2) -\eta_2 \eta_3 (\xi_1-\gamma_2) 
	-\eta_1 \eta_3 (\xi_2-\gamma_2)}
	{\gamma_1 - \gamma_2} 
%\hspace{1pt}
. 
%\nonumber 
%\label{mu-ak-bk3}
\end{align}
\end{subequations}
%
%It is very useful to consider in a geometric manner 
%how the argument of 
%the complex function 
%%value of 
%\mbox{$
%%\mathrm{arg} 
%\prod_{j=1}^{m+1} (\lambda_j - \gamma)$}
%%changing 
%changes 
%%if 
%as $\gamma$ moves on the real axis 
%from $-\infty$ to $+\infty$. 
%For example, 
%if 
%the case where 
%if 
%when 
%We first consider 
%According to 
%On the basis of the geometric consideration 
%mentioned 
As mentioned above, 
%previously, 
%before, 
%using 
considering 
%how 
%the quantity 
%how the complex 
the range of 
the complex 
argument of 
%the complex function 
%value of 
\mbox{$
%\mathrm{arg} 
\prod_{j=1}^{3} (\lambda_j - \gamma)$} 
%for 
%varies 
as a function of 
%for 
real $\gamma$, 
we can 
%classify 
establish a correspondence between 
the configuration of \mbox{$\{ \lambda_1, \lambda_2, \lambda_3 \}$} 
%$\lambda_j$ 
and 
the combination of the signs 
%$\sigma_k$ 
%of nonlinearity 
\mbox{$\{ \sigma_1, \sigma_2 \}$}. 
%each cubic term. 
%the nonlinearity $\sigma_k$. 
%changing 
%changes 
%if 
%as $\gamma$ moves on the real axis 
%from $-\infty$ to $+\infty$. 
%We assume that 
That is, 
(i) 
when 
%if 
\mbox{$\eta_j >0 \; (j=1,2,3)$},
%(or \mbox{$\eta_j <0 \; (j=1,2,3)$}), 
%\mbox{$\eta_1, \eta_2, \eta_3 >0$}, 
i.e., 
all 
%of 
the eigenvalues 
\mbox{$\{ \lambda_1, \lambda_2, \lambda_3 \}$} lie 
in the upper-half 
%(or lower-half)
%complex 
plane,  
%, i.e., \mbox{$\mathrm{Im} \, \lambda_j > 0$} for all $j$ 
%corresponds to 
%we have \mbox{$\sigma_k=-1$} for all $k$, 
%which corresponds to 
%then 
we have a 
%the 
self-focusing nonlinearity 
%case 
\mbox{$\sigma_1=\sigma_2=-1$}; 
(ii) when 
\mbox{$\eta_1,
%>0$}, \mbox{$
\eta_2 >0$}, \mbox{$-(\eta_1 + \eta_2) < \eta_3 < 0$} and 
%\mbox{$\eta_1 + \eta_2 + \eta_3 <0$}
\mbox{$\Delta>0$}, we have a 
%the 
mixed focusing-defocusing 
%case 
nonlinearity \mbox{$\{ \sigma_1, \sigma_2 \}= \{ +1, -1 \}$}; 
%\mbox{$\sigma_1 \sigma_2=-1$}; 
(iii) when 
%if 
\mbox{$\eta_1, 
%>0$}, \mbox{$
\eta_2 >0$}
%, 
%\mbox{$\eta_3 < 0$} 
and \mbox{$\eta_3 < -(\eta_1 + \eta_2)$}, 
%\mbox{$\eta_1 + \eta_2 + \eta_3 <0$}, 
we have a 
%the 
self-defocusing nonlinearity 
%case 
\mbox{$\sigma_1=\sigma_2=+1$};
%, 
and 
so on. 
%etc.
%; and 
%similar for the other cases. 

%We first consider 
%In the case where 
When 
%If 
at least one of \mbox{$\{ \sigma_1, \sigma_2 \}$} is $-1$,
%the two components 
i.e., there exist 
%exists 
%is 
%a 
focusing component(s) in the vector dependent variable, 
%In this case, 
the associated 
%underlying 
eigenvalue 
%spectral 
problem 
%in the Lax-pair representation 
is not self-adjoint 
%(see 
(cf.~subsection~\ref{subs2.2}); thus, 
%Thus, 
%so 
%the 
%B\"acklund 
%parameter 
%value 
%of 
%$\mu$ (or 
%$\mu^\ast$
%)  
%of the spectral parameter 
%at which  
%\mbox{$2\mu = \gamma_1 + \gamma_2 - (\lambda_1 + \lambda_2 + \lambda_3)$} 
%corresponding to 
%associated with 
%a bound state 
%a 
the bound-state eigenvalues 
%is produced 
can 
%indeed 
%be complex-valued. 
take complex values. 
%, so \mbox{$\mu \in \mathbb{C}$}. 
%for a square-integrable 
%for 
%to admit a bound-state eigenfunction. 
%In contrast, 
%Contrastingly, 
%when 
In the self-defocusing case 
%where 
%of 
\mbox{$\sigma_1 = \sigma_2 = +1$}, 
% is $-1$
%in the case where 
%th two components 
the eigenvalue 
%spectral 
problem is self-adjoint, 
so 
%the parameter $\mu$ 
%every 
%a 
the bound-state eigenvalues 
%corresponding to a bound state 
%eigenfunction  
%must be 
%is 
%should be 
%real-valued; 
are restricted to be real; 
%always real; 
%, i.e., \mbox{$\mu-\mu^\ast=0$}; 
thus, 
we cannot construct 
%so 
%Proposition 
%the 
%proper bright-soliton 
%true 
a genuine bright-soliton 
%solutions 
solution 
%cannot be 
%obtained 
%constructed 
%using 
by applying the binary B\"acklund--Darboux transformation  
%method 
%considered 
%as 
used in this section. 
%, but 
However, we can 
%alternatively 
%we can 
%still 
obtain a ``soliton-like" 
%solutions 
solution 
in the self-defocusing case 
using a complex-valued 
B\"acklund
%non-real B\"acklund 
parameter $\mu$, 
%which are 
not associated with 
%square-integrable bound-state eigenfunctions. 
%any 
a bound state. 

%of the Lax pair. 
%In this 
%the scalar NLS 
%case, 
The matrix $\boldsymbol{Z}(\mu)$ 
with \mbox{$2\mu = \gamma_1 + \gamma_2 - (\lambda_1 + \lambda_2 + \lambda_3)$} 
%in the scalar NLS case
%that appear in 
%is given by the \mbox{$2 \times 2$} matrix 
%now 
takes the 
%\mbox{$2 \times 2$} 
%matrix 
form (cf.~(\ref{Z-vector})): 
%(\ref{mat-ex1})):  
%For convenience, we re-parametrize $a_1$ as 
%\[
%a_1 =: \frac{d
%%\eta_1 \eta_2
%}{\eta_1 +\eta_2} 
%\left[ (\xi_1 - \xi_2) + \mathrm{i} (\eta_1 +\eta_2) \right],
%%d, 
%\hspace{5mm} d \in \mathbb{C}.   
%\]
\[
\boldsymbol{Z} = 
\left[
\begin{array}{ccc}
% -2\mu  
\lambda_1 + \lambda_2 + \lambda_3 -\gamma_1 -\gamma_2 & a_1 & a_2 \\
% -\sigma_1 
- \sigma_1 a_1^\ast & \gamma_1 & 0 \\
- \sigma_2 a_2^\ast & 0 & \gamma_2  \\
\end{array}
\right]. 
%\hspace{5mm} \lambda_j = \xi_j + \mathrm{i} \eta_j. 
\]
We choose 
the 
%two 
eigenvectors corresponding to the 
%pair of 
eigenvalues 
\mbox{$\lambda_j = \xi_j + \mathrm{i} \eta_j$} 
%of this \mbox{$2 \times 2$} matrix 
%can be given as
%are given 
as
\begin{align}
\boldsymbol{Z}
\left[
\begin{array}{c}
1 \\
-\frac{\sigma_1 a_1^\ast}{\lambda_j -\gamma_1} \\[1mm]
-\frac{\sigma_2 a_2^\ast}{\lambda_j -\gamma_2} \\
\end{array}
\right] 
&= \lambda_j 
%(\xi_1 + \mathrm{i}\eta_1) 
\left[
\begin{array}{c}
1 \\
-\frac{\sigma_1 a_1^\ast}{\lambda_j -\gamma_1} \\[1mm]
-\frac{\sigma_2 a_2^\ast}{\lambda_j -\gamma_2} \\
\end{array}
\right], \hspace{5mm} j=1,2,3. 
\nonumber 
\end{align}
Then, the matrix exponential 
%of 
in (\ref{mat-ex1})  
%a 
%%linear 
%simple function of 
%$\boldsymbol{Z}
%%(\mu)
%$ 
can be computed explicitly to provide 
%a 
the gauge-transformed 
linear eigenfunction as 
%\newpage
\begin{align}
\left[
\begin{array}{c}
 \boldsymbol{\mathit \Psi}_1 \\
 \boldsymbol{\mathit \Psi}_2 
%(\zeta) 
\\
\end{array}
\right] &= \mathrm{e}^{\mathrm{i} \mu x + 2 \mathrm{i} \mu^2 t}
\sum_{j=1}^3 c_j \mathrm{e}^{\mathrm{i} \lambda_j x 
- \mathrm{i}  \lambda_j^2 t}
\left[
\begin{array}{c}
1 \\
-\frac{\sigma_1 a_1^\ast}{\lambda_j -\gamma_1} \\[1mm]
-\frac{\sigma_2 a_2^\ast}{\lambda_j -\gamma_2} \\
\end{array}
\right],
\nonumber 
%\label{mat-ex2}
\end{align}
where $\boldsymbol{\mathit \Psi}_1$ is a scalar 
%quantity 
and 
$\boldsymbol{\mathit \Psi}_2$ is a two-component column vector; 
$c_j$ are arbitrary complex 
%nonzero 
constants. 
%Thus, 
%by slightly modifying 
%using 
%With the aid of 
%Using 
Substituting 
%these quantities 
this expression 
into 
%Proposition~\ref{prop4.1} or 
a slightly 
%amended 
modified version of 
%a vector analog of 
formula 
%by adapting formula 
(\ref{matrix-bright}),  
%so that it can be applied to 
%to 
%the case of the 
%in 
we obtain a 
desired 
%new 
solution of 
the
two-component 
vector 
%scalar 
%self-focusing 
NLS equation [(\ref{rvNLS}) with \mbox{$m=2$}] 
as 
%in the form: 
%, 
%
%%((\ref{rvNLS}) with \mbox{$m=2$}), 
%%we obtain 
%\begin{align}
%%\skew{3}\widehat{\widetilde{Q}} 
%Q' & = Q +  2 \mathrm{i} (\mu - \mu^\ast) \Psi_1
%\left( 
%\Psi_1^\dagger \Psi_1 - \Psi_2^\dagger \hspace{1pt} \Sigma \hspace{1pt} \Psi_2
%\right)^{-1} 
%\Psi_2^\dagger \hspace{1pt} \Sigma, 
%\nonumber \\
%&= \mathrm{e}^{-2\mathrm{i} t \sum_{k=1}^2 \sigma_k |a_k|^2} \left[ A 
%+ 2 
%%\mathrm{i} 
%% \sigma 
% (\mu - \mu^\ast) 
%%\frac{
%\boldsymbol{\mathit \Psi}_1
%\left( 
%\boldsymbol{\mathit \Psi}_1^\ast
%%\dagger  
%\boldsymbol{\mathit \Psi}_1 
% - \boldsymbol{\mathit \Psi}_2^\dagger \hspace{1pt} \Sigma \hspace{1pt} 
% \boldsymbol{\mathit \Psi}_2 \right)^{-1}
% \boldsymbol{\mathit \Psi}_2^\dagger \hspace{1pt} \Sigma \right]  
%\mathrm{e}^{\mathrm{i} x \Gamma - \mathrm{i} t \Gamma^2}. 
%\nonumber 
%\end{align}
%
%\begin{subequations}
%\label{}
\begin{align}
q_j = a_j \mathrm{e}^{\mathrm{i}  \gamma_j x - \mathrm{i} 
\left( \gamma_j^2 
%+2 \sigma_1 |a_1|^2 +2 \sigma_2 |a_2|^2 
+2 \sum_{k=1}^2 \sigma_k |a_k|^2 
\right) t} 
\left\{ 1 + 2\mathrm{i} \left( \eta_1 + \eta_2 +\eta_3 \right) 
\frac{ 
%\mathrm{e}^{}
g_j}{f} \right\}, 
%\hspace{1pt}.
\hspace{5mm} j=1,2.
\label{one-vector-bright}
\end{align}
Here, the real function $f(x,t)$ and the complex functions $g_j (x,t)$ are 
%defined as 
\begin{align}
f &:= 
%\mathrm{e}^{\mathrm{i} (\mu^\ast -\mu) x 
%+ 2 \mathrm{i} \left( \mu^{\ast\, 2}-\mu^2 \right) t} \left( 
%\boldsymbol{\mathit \Psi}_1^\ast
%%\dagger  
%\boldsymbol{\mathit \Psi}_1 
% - \boldsymbol{\mathit \Psi}_2^\dagger \hspace{1pt} \Sigma \hspace{1pt} 
% \boldsymbol{\mathit \Psi}_2 \right)
% \nonumber \\
%&\hphantom{:}= 
\sum_{\alpha=1}^3 \left| c_\alpha \right|^2
 \mathrm{e}^{ -2 \eta_\alpha (x - 2 \xi_\alpha t)} \left[ 1 
 - \frac{\sigma_1 | a_1 |^2}{\left| \lambda_\alpha -\gamma_1 \right|^2 
%\left( \xi_\alpha -\gamma_1 \right)^2 + \eta_\alpha^2
%\left( \lambda_j -\gamma_1 \right) \left( \lambda_j^\ast -\gamma_1 \right)
}  - \frac{\sigma_2 | a_2 |^2}
 { \left| \lambda_\alpha -\gamma_2 \right|^2 
%\left( \xi_\alpha -\gamma_2 \right)^2 + \eta_\alpha^2
%\left( \lambda_j -\gamma_2 \right) \left( \lambda_j^\ast -\gamma_2 \right) 
} \right]
\nonumber \\[1mm] & \hphantom{:=}\; 
+ 
%\sum_{\alpha=1}^3  \sum_{\beta=1}^3 
\sum_{1 \le \alpha < \beta \le 3}
 \mathrm{e}^{ - \eta_\alpha (x - 2 \xi_\alpha t)- \eta_\beta (x - 2 \xi_\beta t)} 
%\times
% \mathrm{e}^{\mathrm{i} \lambda_\alpha x - \mathrm{i}  \lambda_\alpha^2 t}
% \mathrm{e}^{-\mathrm{i} \lambda_\beta^\ast x + \mathrm{i}
%\lambda_\beta^{\ast \hspace{1pt}2} t}
\nonumber \\[1mm] & \hphantom{:= \sum} \times \left\{
  c_\alpha c_\beta^\ast \hspace{1pt} \mathrm{e}^{\mathrm{i} (\xi_\alpha -\xi_\beta) x 
	- \mathrm{i} (\xi_\alpha^2 - \xi_\beta^2 - \eta_\alpha^2 + \eta_\beta^2) t} 
\left[ 1 - \frac{\sigma_1 | a_1 |^2}{\left( \lambda_\alpha -\gamma_1 \right) 
 \left( \lambda_\beta^\ast -\gamma_1 \right)}  - \frac{\sigma_2 | a_2 |^2}
 {\left( \lambda_\alpha -\gamma_2 \right) \left( \lambda_\beta^\ast -\gamma_2 \right) 
} \right] \right.
\nonumber \\[1mm] & \hphantom{:= \; \sum +} + \left.
  c_\alpha^\ast c_\beta \hspace{1pt} \mathrm{e}^{-\mathrm{i} (\xi_\alpha -\xi_\beta) x 
	+ \mathrm{i} (\xi_\alpha^2 - \xi_\beta^2 - \eta_\alpha^2 + \eta_\beta^2) t} 
\left[ 1 - \frac{\sigma_1 | a_1 |^2}{\left( \lambda_\alpha^\ast -\gamma_1 \right) 
 \left( \lambda_\beta -\gamma_1 \right)}  - \frac{\sigma_2 | a_2 |^2}
 {\left( \lambda_\alpha^\ast -\gamma_2 \right) \left( \lambda_\beta -\gamma_2 \right) 
} \right] \right\}
\nonumber 
\end{align} 
and 
\begin{align}
g_j &:= 
%-2\mathrm{i} \left( \eta_1 +\eta_2 + \eta_3 \right) 
 \sum_{\alpha=1}^3 \left| c_\alpha \right|^2
 \mathrm{e}^{ -2 \eta_\alpha (x - 2 \xi_\alpha t)} 
%\left[ 
 \frac{1}{\lambda_\alpha^\ast -\gamma_j} 
%\right]
%\nonumber \\[1mm] & \hphantom{:=}\; 
+ 
%\sum_{\alpha=1}^3  \sum_{\beta=1}^3 
\sum_{1 \le \alpha < \beta \le 3}
 \mathrm{e}^{ - \eta_\alpha (x - 2 \xi_\alpha t)- \eta_\beta (x - 2 \xi_\beta t)} 
%\times
% \mathrm{e}^{\mathrm{i} \lambda_\alpha x - \mathrm{i}  \lambda_\alpha^2 t}
% \mathrm{e}^{-\mathrm{i} \lambda_\beta^\ast x + \mathrm{i}
%\lambda_\beta^{\ast \hspace{1pt}2} t}
\nonumber \\[1mm] & \hphantom{:= }
%\sum} 
\times \left[
  c_\alpha c_\beta^\ast \hspace{1pt} \mathrm{e}^{\mathrm{i} (\xi_\alpha -\xi_\beta) x 
	- \mathrm{i} (\xi_\alpha^2 - \xi_\beta^2 - \eta_\alpha^2 + \eta_\beta^2) t} 
 \frac{1}{\lambda_\beta^\ast -\gamma_j} 
%\right.
%\nonumber \\[1mm] & \hphantom{:= \; \sum +} + \left.
+  c_\alpha^\ast c_\beta \hspace{1pt} \mathrm{e}^{-\mathrm{i} (\xi_\alpha -\xi_\beta) x 
	+ \mathrm{i} (\xi_\alpha^2 - \xi_\beta^2 - \eta_\alpha^2 + \eta_\beta^2) t} 
	 \frac{1}{\lambda_\alpha^\ast -\gamma_j}
\right], 
\nonumber 
\end{align} 
%\end{subequations}
respectively. 
The 
%wavenumber 
wavenumbers 
$\gamma_j$ 
%, the signs $\sigma_j$ 
and the 
%amplitude 
amplitudes 
$|a_j|$ of 
the 
%each 
%a 
background 
plane 
%wave 
waves 
as well as the signs 
$\sigma_j$ 
%in (\ref{rvNLS}) 
%in the background 
are determined 
%by
%as
through  
(\ref{gamma1-2}) and (\ref{mu-ak-bk3}). 
%On 
Concerning 
%the signs 
$\sigma_j$, 
%of the cubic 
%nonlinear 
%terms, 
we briefly 
comment on 
%discuss 
the 
%above-mentioned 
three cases (i)--(iii) mentioned above (cf.~\cite{Dub88}).
\begin{itemize}
\item[(i)] 
%when 
%if 
\mbox{$\eta_j >0 \; (j=1,2,3)$}: 
%(or \mbox{$\eta_j <0 \; (j=1,2,3)$}), 
%\mbox{$\eta_1, \eta_2, \eta_3 >0$}, 
%i.e., 
%all 
%%of 
%the eigenvalues 
%\mbox{$\{ \lambda_1, \lambda_2, \lambda_3 \}$} lie 
%in the upper-half 
%%(or lower-half)
%%complex 
%plane,  
%, i.e., \mbox{$\mathrm{Im} \, \lambda_j > 0$} for all $j$ 
%corresponds to 
%we have \mbox{$\sigma_k=-1$} for all $k$, 
%which corresponds to 
%then 
%we have 
%which corresponds 
%corresponding 
%to 
%This is 
the self-focusing 
%nonlinearity 
case 
%of 
\mbox{$\sigma_1=\sigma_2=-1$}. 
%
%In the self-focusing 
%%nonlinearity 
%case of \mbox{$\sigma_1=\sigma_2=-1$}, 
%In this case, 
The function 
$f(x,t)$ is 
%always 
positive and the solution (\ref{one-vector-bright}) 
is regular for all $x$ and $t$. 
%However, for 
For this 
%the 
solution 
to be 
%obtain 
a genuine 
%bright-
soliton solution, 
the
%The 
special 
case \mbox{$\eta_1 = \eta_2 = \eta_3$} should be excluded. 
%For 
%is prohibited. 
%the requirement 
%condition 
%that 
If 
%. This 
%requirement 
%implies that 
%implies 
%that 
%imposes a condition that 
%the imaginary part of 
%imposes 
the additional condition 
\mbox{$\eta_1 > \eta_2 + \eta_3$} 
is satisfied
%, 
up to a re-numbering of \mbox{$\{ \lambda_1, \lambda_2, \lambda_3 \}$}, 
the linear eigenfunction 
given by (cf.~Proposition~\ref{prop4.1} and (\ref{gauge1}))
%, 
%; 
\begin{align}
\left[
\begin{array}{c}
 \mathrm{e}^{-2\mathrm{i} t \sum_{k=1}^2 \sigma_k |a_k|^2} \boldsymbol{\mathit \Psi}_1
%(\mu) 
\\
 \mathrm{i} \hspace{1pt}
 \mathrm{e}^{-\mathrm{i} x \Gamma + \mathrm{i} t \Gamma^2} \boldsymbol{\mathit \Psi}_2
%(\mu) 
\\
\end{array}
\right] \left( 
\boldsymbol{\mathit \Psi}_1^\ast
%\dagger  
\boldsymbol{\mathit \Psi}_1 
% -
+ \boldsymbol{\mathit \Psi}_2^\dagger 
%\hspace{1pt} \Sigma \hspace{1pt} 
 \boldsymbol{\mathit \Psi}_2 \right)^{-1}
%, 
\nonumber 
\end{align}
%indeed 
%should 
%define 
%provides 
%gives 
%a bound state 
%decaying 
decays rapidly 
%exponentially 
as \mbox{$x \to \pm \infty$} and 
%defines 
provides 
a bound state; otherwise, we need to multiply it by 
\mbox{$\mathrm{e}^{
\mathrm{i} k
%\alpha 
x}$} with a 
%suitably chosen 
suitable 
complex 
parameter 
%\mbox{$\alpha \in \mathbb{R}$}. 
%a suitable exponential factor
\mbox{$k \in \mathbb{C}$}. 

\item[(ii)] 
%when 
\mbox{$\eta_1,
%>0$}, \mbox{$
\eta_2 >0$}, \mbox{$-(\eta_1 + \eta_2) < \eta_3 < 0$} and 
%\mbox{$\eta_1 + \eta_2 + \eta_3 <0$}
\mbox{$\Delta>0$}: 
%, we have 
%This is 
the 
mixed 
focusing-defocusing case 
%case 
%nonlinearity 
\mbox{$\{ \sigma_1, \sigma_2 \}= \{ +1, -1 \}$}.
%; 
%\mbox{$\sigma_1 \sigma_2=-1$}; 
We need to choose 
the parameters appropriately so that \mbox{$f(x,t) 
%\neq
> 0$} or \mbox{$<0$} 
for all $x$ and $t$.  

\item[(iii)]  
%when 
%if 
\mbox{$\eta_1, 
%>0$}, \mbox{$
\eta_2 >0$}
%, 
%\mbox{$\eta_3 < 0$} 
and \mbox{$\eta_3 < -(\eta_1 + \eta_2)$}: 
%\mbox{$\eta_1 + \eta_2 + \eta_3 <0$}, 
%we have 
%This is 
the self-defocusing case 
%nonlinearity 
%case 
\mbox{$\sigma_1=\sigma_2=+1$}. 
%, and so on. 
%etc.
%; and 
%similar for the other cases
%but 
%For To obtain a 
We 
%need to 
set \mbox{$c_3=0$} in the definitions of 
$f(x,t)$ and $g_j(x,t)$ 
%Then, 
so that 
%the solution 
(\ref{one-vector-bright}) 
can be 
%may 
%provides 
a regular solution without singularities. 
%Then, 
Indeed, 
for an appropriate
choice of the parameters, 
%this solution 
%can 
%define 
%be 
%gives 
it 
%exhibits 
%provides 
corresponds to the 
%a 
``soliton-like" 
%behavior, 
solution
%, which was 
obtained 
%first 
%observed 
%uncovered 
by 
%in 
Park and Shin
%derived 
%considered 
in a more implicit form 
%and was depicted 
%by 
%%in 
%Park and Shin
using Cardano's formula~\cite{Park00} (also see~\cite{Park02}). 
%[Park--Shin02?].
% (also see~\cite{Dub88}). 
\end{itemize}

%Note that 
For (ii) and (iii), 
the existence of the conserved 
%quantity 
%density 
densities 
\mbox{$|q_j|^2
%|q_1|^2 + |q_2|^2 
%- |a_1|^2 - |a_2|^2 
- \mathrm{const.}$} \mbox{$(j=1, \hspace{1pt} 2)$}~\cite{Mak81,Mak82,MMP81}
%(up to a constant shift) 
%\mbox{$\int_{-\infty}^{+\infty} \left( |q_1|^2 + |q_2|^2 - |a_1|^2 - |a_2|^2 
%\right) \mathrm{d}x$} 
implies that the regularity 
%well-posed-ness 
of the solution is generally preserved under the time evolution. 
%
%in the other cases, we need to choose 
%the parameters appropriately so that \mbox{$f(x,t) 
%%\neq
%> 0$} or \mbox{$<0$} 
%for all $x$ and $t$.  
%\end{itemize}
%
%\mbox{$2\mu = \gamma_1 + \gamma_2 - (\lambda_1 + \lambda_2 + \lambda_3)$} 

\section{Concluding remarks}

In this paper, we 
have 
%developed 
%presented 
%a new 
%both 
%the 
%general 
%formulation of 
%formulated 
%and the practical application of 
%formulated 
%considered 
%the 
%the application of 
applied 
%the procedure of 
%have elucidated how to apply 
three 
%kinds 
types 
of B\"acklund--Darboux transformations 
%and 
%developed 
%of 
%have 
%applied them 
%described 
%the practical application of 
%three 
%different 
%types 
%kinds of 
%B\"acklund--Darboux transformations 
%that 
%They can be applied 
%them 
%and have applied them 
%applicable 
to the 
%vector/matrix 
multicomponent 
NLS equations 
with a 
%The 
self-focusing, 
%nonlinearity, 
%the 
%a 
self-defocusing 
%nonlinearity 
or 
%and 
%the 
%a 
mixed focusing-defocusing nonlinearity. 
%areconsidered.  
%wherein the dependent variable is 
%with a vector or 
%a 
%square 
%matrix dependent variable. 
%and have constructed their exact 
%under 
Using 
%By employing 
%the 
a general plane-wave solution 
%boundary conditions 
as the seed solution, 
%and 
%noting , 
we 
%obtained 
%can 
constructed  
%new 
various 
%kinds of 
interesting 
%exact 
solutions 
%under 
%satisfying the general plane-wave boundary conditions, 
such as the vector/matrix 
dark-soliton solutions 
%dark solitons 
and 
%the 
%vector/matrix 
%bright solitons 
bright-soliton solutions 
on the 
%a 
%general 
plane-wave background. 
%In contrast to 
%%the 
%common belief,  
%other approaches 
%the existing literature
%that
%the simplest one among the three transformations, i.e., 
%some dark-soliton solutions 
%can be constructed using 
%an elementary B\"acklund--Darboux transformation, 
%can be used to construct 
%their 
%dark-soliton solutions, 
%although the limiting case of the binary B\"acklund--Darboux transformation 
%is 
%turns out to be 
%necessary in some cases 
%more useful and efficient 
%efficient to 
%in some cases. 
These 
%multicomponent 
%solitons 
%soliton 
solutions 
generally admit the internal degrees of freedom 
and 
%indeed 
provide 
highly 
nontrivial 
%vector/matrix analogs 
%multicomponent 
generalizations 
of the corresponding scalar NLS 
solitons. 
%; that is, the solutions contain free parameters besides soliton's width and velocity. 
%which cannot be constructed. 
%by guesswork. 
The main 
%contribution 
new feature 
%point 
%key 
of 
our approach 
%this work 
is 
%the introduction of 
%lies in 
%By employing 
%a proper 
an appropriate 
%suitable 
re-parametrization of the 
parameters 
%appearing in 
%applying 
that appear in the application of 
the 
B\"acklund--Darboux transformations; 
%which 
%enables 
this 
allows us 
to express  
%we obtained 
%The obtained 
%results in 
%expressions 
%for the 
the soliton solutions 
%are 
%more 
%to be written 
in more 
%expressed 
%explicitly 
explicit and 
%clearly 
%and 
tractable forms 
than those reported 
in the literature (cf.~\cite{Dub88,Park00}). 
%the known ones 
%as opposed to those 
%More specifically, 
%Specifically, 
%instead of 
%it is important to give up  
%the point is that 
%instead of treating 
The price to pay is that
%, 
%in contrast to 
unlike 
%other na\"ive 
%%known 
%%usual 
%approaches,  
the usual na\"ive 
parametrization, 
%already 
%known 
%reported 
%in the 
%%existing 
%literature, 
%both 
not all of 
%of 
the amplitudes 
and the wavenumbers 
of the background plane waves 
%may not be 
%cannot be 
%always as 
%become 
%are, in general, 
%may 
%become 
%are 
%can be taken as 
remain 
%as 
%not 
%%necessarily 
%not all 
%fully 
independent 
free parameters; 
%in general;
%free 
%parameters
%; 
rather, 
%and 
%then, 
%some of them 
%may become 
%are
%we consider 
%we consider 
%either or both of them as 
%express them as 
%functions of 
%they 
%are expressed 
%in terms of 
%may 
they can depend on  
%become 
%are 
%functions of 
other 
%soliton 
parameters 
characterizing 
%each soliton, 
the soliton 
%solitons 
such as 
%which characterize 
%, loosely speaking, 
%like 
%the soliton 
soliton's 
%its 
%the 
%its 
width and 
%their widths and 
%soliton 
%velocities. 
velocity. 
More precisely, 
%To be precise, 
%the 
a multicomponent soliton 
%solitons 
%with internal degrees of freedom 
%usually 
generally 
%each 
%have 
has a 
%strongly 
time-dependent 
and 
nonrecurrent 
shape 
%because of 
%by 
%owing to 
%because they suffer from 
%oscillated by 
%the beating effects 
%that 
%because of 
%exhibits 
%by 
due 
in part 
to the 
beating 
%oscillating 
%effects 
effect with the background plane waves;  
%so 
thus, 
it is 
%uneasy 
not always 
%possible 
%in general 
meaningful 
to 
%rigorously 
%define 
%they does not 
%a unique 
consider 
%discuss 
%mention 
soliton's width and velocity. 
%uniquely. 
%of each soliton 
%rigorously. 
%in a rigorous manner. 
%
%This is in contrast to other approaches 
%%already 
%%known 
%%reported 
%in the literature, wherein 
%the parameters in 
%the obtained 
%%reported 
%soliton solutions involve a constraint 

%For constructing 
%explicit 
%exact solutions of integrable partial differential equations, 
%the 
There exist 
two major methods of 
obtaining exact solutions of integrable 
%systems:
partial differential 
%equations:\ 
equations:\ 
%i.e., 
%are 
the inverse scattering method~\cite{AS81}
%[AS]
%, Novikov] 
and 
the Hirota bilinear method~\cite{Hirota04}.
%[Hirota]. 
%appear to the most efficient 
%fundamental. 
%Unexpectedly, 
However, 
%Unlike for the scalar NLS equation, 
%Unfortunately, 
%unlike for constructing scalar or 
%quasi-scalar NLS solitons,  
these two 
%both 
methods 
%turn out to be 
%do not appear to be 
are
%, in general, 
not 
%promising
%suitable 
%efficient and 
%useful 
effective 
for constructing general 
multicomponent 
%NLS 
solitons 
%with internal degrees of freedom
%on a general plane-wave background, which have 
under the 
%general 
plane-wave boundary conditions. 
%with 
Indeed, 
%the 
%Note that 
explicit expressions for 
%a 
the vector/matrix NLS 
%one-soliton 
solitons 
%solutions 
with 
%essential 
internal degrees of freedom
%, i.e., a polarization, 
%have to 
%is 
are 
%rather complicated and 
%have to satisfy 
%involve 
accompanied 
%with 
by a set of 
%highly nontrivial and 
complicated constraints 
%relations 
%among 
on the parameters, 
%(cf.~\cite{Dub88,Park00}), 
%in the solution. 
%parametric constraints. 
%The reason 
%We adapted 
which is quite 
%are 
%appear to be very 
difficult to 
%derive 
obtain and resolve 
%directly 
using these methods. 
%the inverse scattering method or the Hirota bilinear method. 
The method based on 
B\"acklund--Darboux transformations 
%turns our 
%appears to be 
is more 
%the 
%most 
appropriate for this purpose. 
%efficient and useful than the more universal 
%these methods. 
%It is also 
%In addition, 
%It is 
%interesting 
%%important 
%to note 
%that 
Interestingly, 
in contrast to 
the 
common belief, 
even the most fundamental 
%simplest 
B\"acklund--Darboux transformation, 
%i.e., 
%namely, 
%other approaches 
%the existing literature 
%that
%the simplest one among the three transformations, i.e., 
called an elementary 
B\"acklund--Darboux transformation, 
can 
%be 
generate 
%used to obtain 
%applied to generate 
%construct 
%their 
%some 
nontrivial 
dark-soliton solutions such as (\ref{m-one-dark}) 
and (\ref{gene-dark}). 
%; a
A more elaborate transformation, 
the limiting case of the binary B\"acklund--Darboux transformation, 
is indispensable
%becomes necessary 
%in some cases more efficient 
%to obtain 
for obtaining 
%construct 
vector 
dark-soliton solutions 
with internal degrees of freedom, 
such as (\ref{v-soliton1}) and (\ref{v-soliton2}). 
%Using the binary B\"acklund--Darboux transformation, 
%we can obtain multicomponent bright-soliton solutions on a general plane-wave 
%background, such as (\ref{matrix-bright}) and (\ref{one-vector-bright}). 
%Nevertheless, 
%Although 
%%these soliton solutions are not 
%%obtainable by 
%the conventional Hirota bilinear method 
%is not applicable for , 
%for obtaining formal expressions 
%of soliton solutions using the matrix exponential 
%functions
%and/or 
%and checking their correctness, 
It 
would be interesting 
%is a promising direction of research 
%possible 
to 
consider 
%develop 
a 
%matrix-valued 
%analog 
matrix 
generalization 
of the Hirota bilinear method, 
which 
%can be used 
%it 
%should allow 
%enables 
allows 
us to obtain 
%to check 
%that can be used to 
%and to confirm 
%for obtaining 
formal 
expressions 
of 
%soliton 
%for 
%exact 
solutions 
%using 
%based on 
%involving the matrix exponential function, 
%check the correctness
%the latter two expressions are indeed formal 
%involve 
involving 
%the 
%a matrix 
the exponential 
%function, 
of a 
%constant 
non-diagonal matrix, 
such as 
%(\ref{m-one-dark}), 
(\ref{gene-dark}) and (\ref{matrix-bright}). 
%function,. 

%In fact, such a generalized version of 
%the Hirota bilinear method 
%would be useful to construct 
%%vector/matrix 
%formal expressions for 
%%
%multicomponent 
%multisoliton
%%dark
%solutions on the plane-wave background. 
In this paper, we 
%discussed 
%constructed only 
focused on 
%presented 
the construction of 
%the 
one-soliton solutions 
of the 
%vector/matrix 
multicomponent 
NLS equations
%; the ``one-soliton" solution meand 
%in the sense of a single 
%that can be obtained by 
%applying 
%using 
%by a single application of a 
by applying a 
%single 
%using a 
``one-step" 
%``one-fold" 
B\"acklund--Darboux transformation. 
%only 
%once. 
It is
%, 
in principle
%, 
%also 
possible 
%straightforward 
to apply 
%``one-step" 
%B\"acklund--Darboux transformations
it 
%can be applied 
%consecutively 
%repeatedly 
iteratively
at (generally) different values of the B\"acklund parameter; 
%leading 
%to 
%which 
this 
defines 
%results in 
%is equivalent to 
a multifold
%``multi-fold" 
%``multi-step" 
B\"acklund--Darboux transformation, 
%and to 
%by which 
%and we can 
which 
%allows us to obtain
can be used to obtain 
%provide  
%more explicit 
%which 
%provides 
%formal expressions for the 
%
multicomponent 
multisoliton 
%two-soliton 
%dark
solutions 
%on the plane-wave background. 
under 
%satisfying 
the plane-wave boundary conditions. 
%; 
%such formal expressions for 
%construct 
%multisoliton solutions; 
%Indeed, 
%Note that the B\"acklund--Darboux transformations 
%describe how the linear eigenfunction of the Lax pair 
%is transformed.  
However, 
%to obtain their explicit expressions 
to 
write 
%express 
%rewrite 
%express 
them 
out
%in 
%more 
%explicit forms 
%for solutions 
explicitly, 
%without using the matrix exponential, 
%function, 
we need to 
%this does provide a 
consider and 
%deal with and 
resolve 
%cope with 
a rather 
%too 
complicated system 
of algebraic 
%constraints on 
equations 
%for 
satisfied by the 
%solution 
parameters 
%in the solution. 
in the solutions. 
%which 
%equations
%large number 
%does not provide a useful concrete 
%is not 
%very 
%generally appears to be 
%appropriate 
%unsuitable 
%for 
%to deal with 
%computation 
%and investigation 
%by hand. 
%by a hand calculation. 
%Thus, 
%as a practical problem, 
In practice, 
such explicit expresssions 
%based on 
%obtained by brute force 
%%calculations 
%computations 
%are 
become 
too 
cumbersome 
even for the two-soliton solutions, 
%and 
%matter of fact, 
%we 
%can express 
%should obtain 
%express 
%formal expressions
%for 
%should be satisfied with 
%the multisoliton solutions 
%only 
%no more useful than 
so they 
are not 
%very 
useful compared with 
%formally 
formal expressions 
%using 
involving 
%the matrix exponentials. 
a matrix exponential function. 

%The method based on 
%B\"acklund--Darboux transformations 
%turns out to be more efficient and useful than 
%the more universal 
%%conventional 
%methods 
Before closing 
%ending the 
this paper, 
we 
would like to 
remark 
%comment on some relevant literature. 
%mention 
%stress 
that 
soliton solutions of 
the 
%vector/matrix 
multicomponent 
NLS equations under  
%some 
simpler 
%special 
%decaying or non-generic 
%special 
%nonvanishing 
%plane-wave 
boundary conditions 
have 
%already 
been 
%solved 
reported 
in the 
%existing 
literature. 
%under  
%some 
%simpler 
%special 
%decaying or non-generic 
%special 
%nonvanishing 
%plane-wave 
%boundary conditions. 
%simpler than
%% that 
%the general plane-wave boundary conditions 
%considered in this paper. 
%have been reported in the literature. 
The list of references 
on the multicomponent NLS solitons 
%equations 
is too long 
(see, {\it e.g.}, 
%references in 
\cite{DM2010}), 
so we only mention 
%some 
%the most both 
%what
the original and 
most 
relevant 
%and original 
%papers. 
ones. 
%to supplement the introduction. 
%on the vector NLS equation (\ref{rvNLS}). 
%(cf.~(\ref{rvNLS})). 
%In the simplest case of 
%Under the decaying 
%%vanishing 
%boundary conditions 
%at spacial infinity, 
The bright 
%vector  
%multisoliton 
$N$-soliton solution 
of the 
%%focusing 
vector NLS equation 
under the 
%decaying 
vanishing 
boundary conditions 
%at spacial infinity 
%were obtained 
was 
%presented 
derived 
in~\cite{Dub88} (also see~\cite{Steu88,Date75})
and the corresponding Hirota 
bilinear form 
and tau functions were 
%was 
%given 
%discussed 
%clarified 
presented 
in~\cite{Zhang94}. 
%(also see~\cite{Steu88}, [Date75?]). 
%Degenerate dark-soliton
%%-like 
%solutions 
%and the so-called dark-bright soliton solutions of 
The two-component vector NLS equation 
was solved 
under 
special 
%non-generic 
%partially 
%the 
nonvanishing boundary conditions
%, 
such that 
%, i.e., 
%{\it e.g.}, 
%when 
%the wavenumbers of 
the 
two 
background 
plane waves 
%in 
%the two 
%%both 
%components 
%coincide
have the same wavenumber and frequency
%~\cite{Mak83,Mak84} 
%(up to 
%(cf.~the 
(note the 
%a 
Galilean 
%transformation
invariance)~\cite{Mak83,Mak84}
%up to a Galilean transformation 
or 
%one of the two 
%components decays at infinity
one of 
them 
%the two background plane waves 
%in one 
%%of the two components 
%component 
%vanishes
disappears~\cite{Dub88,Park00}; 
%the two components decays as \mbox{$x \to \pm \infty$}~\cite{Mak83,Dub88}. 
%Thus, 
%Consequently, 
%Correspondingly, 
%Accordingly, 
%In such a simpler case, 
%the so-called 
%Thus, 
thus, the 
%corresponding 
%solutions, i.e., 
%namely, 
%``degenerate" 
degenerate dark-soliton solutions (cf.~\cite{Ohta11})  
and 
%the 
so-called 
dark-bright soliton 
solutions
%, 
were obtained. 
%by Makhankov 
%{\em et al.\ }
%in these 
%%those 
%papers. 
%the 1980s. 
%in these papers. 
%; 
%~\cite{}
%such that one 
%was solved 
%it is possible to 
%Note that 
%identify 
%these 
%
%and 
%note that 
%the relationship between 
%Moreover, 
In addition, 
%the 
these 
two classes of 
%soliton 
solutions 
%they 
can be 
%explained 
related 
%with each other 
%cases 
%identified 
%with the aid of 
through 
%using 
%by 
the symmetry group 
of the two-component 
%vector 
NLS equation, 
such as $SU(2)$ or 
%$U(1,1)$
$SU(1,1)$~\cite{Park00,MMP81}. 
Surprisingly,
most of the 
recent papers 
%publications 
%on this subject 
on 
the 
%soliton 
solutions of 
the vector NLS equation 
under nonvanishing boundary conditions  
%in this research area 
%do 
did not 
%even 
%cite 
%rarely 
refer to 
%the work of 
%their work 
%these 
%relevant 
%the above-mentioned 
%references 
%appropriately 
the pioneering work 
%important contributions 
of 
%V.\ G.\ 
Makhankov 
%and coworkers 
{\em et al.\ }in the 1980s 
%~\cite{Mak} 
%in particluar~\cite{Dub88},  
(see~\cite{Dub88} and references therein)  
%and 
or the 
subsequent 
%important 
contribution 
%work of 
by 
Park and 
%H.\ J.\ 
Shin~\cite{Park00,Park02}. 
%in 2000~\cite{Park00}. 
%at all.  
%have not 
%What is more, 
%Unfortunately, 
%Actually, 
%In fact, 
As far as we could check, 
%``new"  
%results 
%result on the 
%soliton solutions 
%of the vector NLS equation 
%presented 
such 
%recent 
%those 
papers 
%usually 
generally 
%%appear to 
%tend to 
%contain 
%report 
%find 
%actually 
%turn out to 
present 
no 
%essentially 
new 
%results 
%result on the soliton 
solutions 
of the vector NLS equation. 
%and only present the already known results as the
%rather, they are usually 
%%so we do not discuss which paper is exemplify 
%rediscoveries of 
%already known results, 
%so we 
%dare not 
%this is why we dare not 
%do not think it necessary to 
%cite them 
%and 
%not demonstrate 
%discuss 
%that they are just rediscoveries of 
%already known solutions. 
%in this paper. 

%
%parametric constraints~\cite{Dub88}
%
%Readers interested in 
%the special case wherein some of the components 
%of the background planewaves vanish are referred to 
%the work of~\cite{Dub88,Park00} etc.\

%Each soliton with internal degrees of freedom 
%has a time-dependent shape 
%because of beating effects with the background plane waves 

%dark-bright soliton solutions correspond to the case 
%wherein one (or more) background plane waves vanish or 
%the wavenumbers of some background plane waves coincide, 
%see [Makhankov 1983, Park--Shin]

\section*{Acknowledgments}
The author is indebted to 
%thanks
Professor 
%Folkert M\"uller-Hoissen and
%Dr.\ 
Ken-ichi
%K.\ 
Maruno
for his 
kind help. 
%in completing this paper. 
%useful comments. 
%and discussions. 
He also thanks Professor Takeshi Iizuka 
for 
%giving 
sending him 
a copy 
%reprint 
of~\cite{Iizuka91}.

\end{document}